\documentclass[12pt]{elsarticle}                                          
\usepackage{axodraw4j}
\usepackage{hyperref}
\usepackage{graphicx}                                                         
\usepackage{a41}                                                                
\usepackage{xcolor}                     
\usepackage{slashed}
\usepackage[rflt]{floatflt}
\usepackage{float}
\usepackage{hyperref}
\hypersetup{colorlinks=true, 
linkcolor=blue, filecolor=blue, urlcolor=mygreen}
\usepackage{breakurl} 
\usepackage{times}     %  
\usepackage{array}
\usepackage[english]{babel}
\usepackage[latin1]{inputenc}
\usepackage[T1]{fontenc}
\usepackage{ae}
\usepackage{url}
\usepackage{amsmath, amsthm, amssymb}
\usepackage{slashed}

\usepackage{rotating}
\usepackage{graphicx}
\usepackage{comment}
\newcounter{mmacnt}
\def\restartmma{\setcounter{mmacnt}{0}}
\restartmma \catcode`|=\active
\def|#1|{\mathrm{#1}}
\catcode`|=12
\newenvironment{mma}{
\par\smallskip
\catcode`|=\active
\parskip=0pt\parindent=0pt % locally
\small
\def\In##1\\{%
\def\linebreak{\hfill\break\null\qquad}%
\refstepcounter{mmacnt}
\hangindent=2.5em\hangafter=0
\leavevmode
\llap{\tiny\sffamily In[\arabic{mmacnt}]:=\kern.5em}%
\mathversion{bold}\footnotesiVe$
\displaystyle##1$\normalsiVe
\mathversion{normal}\par
 }%
\def\Print##1\\{%
\def\linebreak{\hfill\break}%
\hangindent=2.5em\hangafter=0
\leavevmode ##1\par}%
\def\Out##1\\{%
\def\linebreak{$\hfill\break\null\hfill$}%
\kern\abovedisplayskip\par
\hangindent=2.5em\hangafter=0
\leavevmode
\llap{\tiny\sffamily Out[\arabic{mmacnt}]=\kern.5em}
\footnotesiVe$\displaystyle##1$
\normalsiVe\hfill\null\par
\kern\belowdisplayskip
}%
\def\Warning##1##2\\{%
\def\linebreak{\hfill\break}%
\hangindent=2.5em\hangafter=0
\leavevmode
{\scriptsiVe##1 : ##2}\par}%
}{%
\par\smallskip
}

\usepackage{color}
\newenvironment{fshaded}{%
\MakeFramed {\FrameRestore}
}%
{\endMakeFramed}

\makeatletter
\def\ps@pprintTitle{%
\let\@oddhead\@empty
\let\@evenhead\@empty
\def\@oddfoot{\reset@font\hfil\thepage\hfil}
\let\@evenfoot\@oddfoot
}
\makeatother
\usepackage{tcolorbox}
\usepackage{slashed}
\usepackage{comment}
\begin{document}  %%%%
%%%%%%%%%%%%%%%%%%%%%%
\begin{frontmatter}%%%
%%%%%%%%%%%%%%%%%%%%%%%%%%%%%%%%%%%%%%%%%%%%%%%
\title{
\textbf{One-loop contributions for
$A^0 \rightarrow \ell \bar{\ell} V$
with $\ell \equiv e, \mu$ and
$V\equiv \gamma, Z$
in Higgs Extensions of the Standard Model}
}
%%%%%%%%%%%%%%%%%%%%%%%%%%%%%%%%%%%%%%%%%%%%%%
\author[1,2]{Khiem Hong Phan}
\ead{phanhongkhiem@duytan.edu.vn}
\author[1,2]{Dzung Tri Tran} 
\author[3]{Thanh Huy Nguyen}
%%%%%%%%%%%%%%%%%%%%%%%%%%%%%%%%%%%%%%%%%%%%%%
\address[1]{\it Institute of Fundamental 
and Applied Sciences, Duy Tan University, 
Ho Chi Minh City $700000$, Vietnam}
\address[2]{Faculty of Natural Sciences, 
Duy Tan University, Da Nang City $550000$, 
Vietnam}
\address[3]
{\it VNUHCM-University of Science, 
$227$ Nguyen Van Cu, District $5$, 
Ho Chi Minh City $700000$, Vietnam}
\pagestyle{myheadings}
\markright{}
%%%%%%%%%%%%%%%%%%%%%%%%%%%%%%%%%%
\begin{abstract} %%%
%%%%%%%%%%%%%%%%%%%%
We present one-loop formulas for
the decay of CP-odd Higgs
$A^0 \rightarrow \ell \bar{\ell} V$
with $\ell \equiv  e, \mu$ and
$V\equiv \gamma, Z$
in Higgs Extensions of the Standard Model,
considering two higgs doublet model with
a complex (and real) scalar, two higgs
doublet model as well as triplet higgs
model. Analytic results for one-loop
amplitudes are expressed in terms of
Passarino-Veltman functions following the
standard notations of {\tt LoopTools}. As
a result, physical results
can be generated numerically by using the
package. In phenomenological results, the
total decay widths and the
differential decay rates with respect
to the invariant mass of lepton pair
are analyzed for two typical models
such as two higgs doublet model
and triplet higgs model.
\end{abstract}
%%%%%%%%%%%%%%%%%%%%%%%%%%%%%%%%%%%%%%%%
\end{frontmatter}
%%%%%%%%%%%%%%%%%%%%%%%%%%
\section{Introduction}%%%%
%%%%%%%%%%%%%%%%%%%%%%%%%%
Discovering the scalar Higgs
potential, subsequently answering
the nature of dynamic of
the electroweak spontaneous
symmetry breaking (EWSB),
is one of the priority tasks at
future colliders, e.g. High-luminosity
Large Hadron Collider
(HL-LHC)~\cite{Liss:2013hbb,
CMS:2013xfa} as well as future
Lepton Colliders
(LC)~\cite{Baer:2013cma}.
It is well-known that
the scalar potential
is extended by including
scalar singlets or scalar
multiplets in many of beyond
the standard models (BSM).
As a result, there exist many
new heavy scalar particles,
for examples,
neutral CP-even and CP-odd
Higges, singly charged Higgses
as well as doubly charged Higgses
in many of BSMs. The precise
measurements for decay widths
and the production cross-sections
of the scalar particles are
important for the indirect and direct
searches for new physic signals
at future colliders. From the
measured data, we can therefore
verify the nature of scalar Higgs
potential and understand deeply
the dynamic of EWSB. Recently,
direct production of
a light CP-odd Higgs boson
has been performed  at the Tevatron
and LHC~\cite{Dermisek:2009fd},
search for a CP-odd Higgs boson 
decaying to $Zh$ in $pp$ collisions
has performed at the
LHC~\cite{ATLAS:2015kpj,
ATLAS:2023zkt}. Probing for
a light pseudoscalar Higgs boson
in $\mu\mu\tau\tau$ events at the LHC
in~\cite{CMS:2020ffa} and
in the di-muon decay channels
in $pp$ collisions at
$\sqrt{s}=7$ TeV~\cite{CMS:2012fgd}
has reported.

From theoretical views, the
detailed evaluations for one-loop
radiative corrections to the
decay rates and the production
cross-sections for the standard
model-like Higgs boson
(SM-like Higgs) as well as for
all new scalar particles in many
of BSMs play a crucial role for
probing new physics signals at future
colliders. One-loop contributing
to the decay and production processes
of SM-like Higgs, CP-even Higgses
have computed in many Higgs
Extensions of the SM (HESM). It is
worth to refer to typical works
in this paper, for examples, in
the works of following
papers~\cite{Krause:2018wmo,
Athron:2021kve,Denner:2019fcr,Kanemura:2017gbi,
Kanemura:2019slf,Kanemura:2022ldq,Aiko:2023xui,
Phan:2021xwc,VanOn:2021myp,Kachanovich:2020xyg,
Hue:2023tdz,Chiang:2012qz,Benbrik:2022bol}
and the references
in therein. One-loop radiative
corrections to the CP-odd Higgs ($A^0$)
production processes in the HESM
have evaluated at LHC~\cite{Akeroyd:1999xf, 
Akeroyd:2001aka, Yin:2002sq}
and at future LC~\cite{Arhrib:2002ti,
Farris:2003pn, Sasaki:2017fvk, Abouabid:2020eik}. 
Furthermore, the decay channels
of the CP-odd Higgs including
one-loop and beyond one-loop
corrections have also computed
in many of BSMs as
in~\cite{Bernreuther:2018ynm,
Accomando:2020vbo, Aiko:2022gmz,
Akeroyd:2023kek,Esmail:2023axd, 
Biekotter:2023eil}. 
In this work, we present the
first calculations of one-loop
contributing for decay channels
$A^0 \rightarrow \ell \bar{\ell} V$
with $\ell \equiv e,\mu$
and $V\equiv \gamma, Z$
within many of HESM frameworks,
including two higgs doublet model with
a complex (and real) scalar,
two higgs doublet model
as well as triplet higgs model.
Analytic results for one-loop
form factors are expressed in
terms of Passarino-Veltman scalar
functions (PV-functions) following 
the standard notations of
{\tt LoopTools}~\cite{Hahn:1998yk}. 
Subsequently, physical results
can be computed numerically
by using the package. In phenomenological
results, the decay rates of CP-odd Higgs
and its differential decay widths with respect
to the invariant mass of lepton pair
are examined for two typical models
such as two higgs doublet model (THDM)
and triplet higgs model (THM).

Overview of the
paper is as follows. 
We first review
two specific classifications
of HESMs in detail in this work
such as two higgs doublet model
with a complex 
scalar field and
triplet higgs model
in the section $2$.
We then present in concrete
the evaluations for one-loop
contributions to the decay 
amplitudes 
$A^0\rightarrow \ell
\bar{\ell} V$ in the section $3$.
Phenomenological results
for the HESMs are shown
in section $4$.
Conclusion and outlook
are devoted in section $5$.
In appendices $A, B$ we derive
all related couplings 
to the processes under 
consideration in the
above-mentioned models.
Proving one-loop
mixings of $A^0$ with
scalar CP-even Higgs $\phi$,
of with $Z$ boson as well as
the mixings of $\phi$ with
$Z$ boson are vanished in
appendices $C, D$.
%%%%%%%%%%%%%%%%%%%%%%%%%
\section{Higgs Extension 
Standard Models}%%
%%%%%%%%%%%%%%%%%%%%%%%%%%
In this section, we review the
Higgs Extensions of the
Standard Models, examining
two higgs doublet model
with a complex scalar field
(noted as THDMS, or STHDM
hereafer)~\cite{Jiang:2019soj},
with soft breaking $Z_2$ symmetry 
and triplet higgs models.
From the general STHDM, we can
reduce to the next minimum
two higgs doublet models
(NTHDM)~\cite{Muhlleitner:2016mzt,
Glaus:2022rdc}
and THDM~\cite{Branco:2011iw, Nie:1998yn,
Kanemura:1999xf,Akeroyd:2000wc, 
Ginzburg:2005dt,Kanemura:2015ska,
Bian:2016awe, Xie:2018yiv}.
We then turn our attention to
the second classification of
HESM, triplet higgs model (THM),
in the
subsection $2.2$.
%%%%%%%%%%%%%%%%%%%%%%%
\subsection{STHDM} %%%%
%%%%%%%%%%%%%%%%%%%%%%%
In this subsection, we arrive
at STHDM, the two higgs doublet model
with adding a complex 
scalar field $S$,
following the soft
breaking $Z_2$ symmetry. 
In this model, 
two scalar fields $\Phi_1$ 
and $\Phi_2$ are doublets of $SU(2)_L$ 
with hypercharge $Y=+1/2$. The additional 
complex scalar $S$ is a scalar singlet of 
$SU(2)_L$ with with hypercharge $Y=0$. 
In this paper, we only concern the 
CP-conservating case for the scalar sector.
As a result, all parameters in scalar 
potential are considered to be real
parameters. Furthermore, the scalar potential
follows the soft breaking term of $Z_2$
symmetry, e.g. $\Phi_i \rightarrow -\Phi_i$
for $i=1,2$. Under the above assumptions,
the most generalized gauge invariant
formulation in accordance with the
renormalizable condition for the scalar
potential is given by:
%%%%%%%%%%%%%%%%%%%%%%%%%%%%%
\begin{eqnarray}
\mathcal{V}_{\text{STHDM}} 
&=&
m_{11}^2|{\Phi _1}|^2 
+ m_{22}^2|{\Phi _2}|^2 
- m_{12}^2(\Phi _1^\dag {\Phi _2} 
+ \Phi _2^\dag {\Phi _1}) 
+ \frac{{{\lambda _1}}}{2}|{\Phi _1}{|^4} 
+ \frac{{{\lambda _2}}}{2}|{\Phi _2}{|^4}
\\
&& + {\lambda _3}|{\Phi _1}|^2|{\Phi _2}|^2 
+ {\lambda _4}|\Phi _1^\dag {\Phi _2}|^2 
+ \frac{{{\lambda _5}}}{2}
[{(\Phi _1^\dag {\Phi _2})^2}
+ {(\Phi _2^\dag {\Phi _1})^2}]
\nonumber\\
&& - m_S^2|S|^2 
+ \frac{{{\lambda _S}}}{2}|S{|^4} 
+ \lambda_{S\Phi_1}|{\Phi _1}|^2|S|^2 
+ \lambda_{S\Phi_2} |{\Phi _2}|^2|S|^2 
\nonumber\\
&&
+ \Big\{ 
-\frac{{m'^2_S}}{4}{S^2} 
+ \mathrm{H.c.}
\Big\}. 
\nonumber
\label{scalarSTHDM}
\end{eqnarray} 
The potential includes a scalar sector
from THDM, a singlet complex Higgs part
and the mixing term of THDM with singlet
complex scalar $S$. In the above scalar
potential, the last term is broken softly
the $U(1)$ symmetry.

For the electroweak spontaneous
symmetry breaking, the scalar
fields can be parameterized
as follows:
\begin{equation}
\Phi _1 =
\begin{pmatrix}
\phi _1^ +  \\
({v_1} + {\rho _1}
+ i{\eta _1})/\sqrt 2
\\
 \end{pmatrix},
 \quad
 \Phi _2 =
 \begin{pmatrix}
\phi _2^ +
\\
({v_2} + {\rho _2}
+ i{\eta _2})/\sqrt 2
\\
\end{pmatrix},
\quad S =
\frac{v_s + \rho_3 + i\chi }
{\sqrt 2 }.
\end{equation}
By minimizing the potential, we find
the following system equations:
\begin{eqnarray}
m_{11}^2
&=& \frac{v_2}{v_1}{m}_{12}^2
- \frac{1}{2} \lambda _1 v_1^2 
- \frac{1}{2}\lambda_{345}v_2^2 
- \frac{1}{2} 
\lambda_{S\Phi_1}v_s^2, \\
m_{22}^2
&=&
\frac{v_1}{v_2}{m}_{12}^2 
- \frac{1}{2} \lambda_2 v_2^2 
- \frac{1}{2}\lambda_{345}v_1^2 
- \frac{1}{2}\lambda_{S\Phi_2}v_s^2, 
\\
m_S^2
&=&  - \frac{1}{2}m'^2_S
+ \frac{1}{2} \lambda_S v_s^2 
+ \frac{1}{2}
\lambda_{S\Phi_1} v_1^2 
+ \frac{1}{2} \lambda_{S\Phi_2} 
v_2^2,
\label{Vmin}
\end{eqnarray}
where the notation
$\lambda_{ijk\cdots} = \lambda_i
+ \lambda_j + \lambda_k+\cdots$ has used.
In this model, $\chi$ becomes
a stable dark matter (DM) candidate.
After the EWSB, the mass of $\chi$
is given by:
\begin{eqnarray}
m_\chi ^2 &=&
m'^2_S =
- m_S^2 + \frac{1}{2}m'^2_S
+ \frac{1}{2} \lambda_S v_s^2 
+ \frac{1}{2}\lambda_{S\Phi_1}v_1^2 
+ \frac{1}{2}\lambda_{S\Phi_2} 
v_2^2.
\end{eqnarray}
In scenario of $m'^2_S=0$,
it means there isn't
soft breaking term in this case.
As a result, $\chi$ becomes
a massless Nambu-Goldstone boson.
For the case of $m'^2_S>0$,
$\chi$ has non-zero masss,
$\chi$ is a pseudo-Nambu-Goldstone
boson. In the scope of this paper,
the role of $\chi$ isn't related
to the processes under concern.
Therefore, we skip discussing
$\chi$ in the rest of this paper.

The charged scalars part
can be collected in the form of
\begin{eqnarray}
\mathcal{L}_{\mathrm{mass}}^{\phi^{\pm} }
&=& -
\left(
m_{12}^2 
- \frac{1}{2}\lambda_{45}{v_1}{v_2}
\right)
\begin{pmatrix}
\phi_1^- , 
& 
\phi_2^- 
\\
\end{pmatrix}
\begin{pmatrix}
v_2/v_1 &  - 1  \\
- 1     & v_1/v_2  \\
 \end{pmatrix}
 \begin{pmatrix}
\phi_1^+  \\
\phi_2^+  \\
 \end{pmatrix}.
\end{eqnarray}
The $CP$-odd scalar sector
is presented as the same form
as follows:
\begin{equation}
\mathcal{L}_{\mathrm{mass}}^{\eta}
=
-\frac{1}{2}( m_{12}^2 - \lambda_5v_1 v_2 )
\begin{pmatrix}
\eta _1,
&
\eta _2   \\
\end{pmatrix}
\begin{pmatrix}
v_2/v_1 & -1         \\
-1      &  v_1/v_2   \\
\end{pmatrix}
\begin{pmatrix}
\eta_1   \\
\eta_2   \\
\end{pmatrix}.
\end{equation}
Physical mass terms can be
obtained by diagonalying 
the above matrices.
This can be done by
applying the following 
rotation matrices: 
\begin{equation}
\begin{pmatrix}
\phi_1^+   \\
\phi_2^+   \\
\end{pmatrix} 
= R(\beta)
\begin{pmatrix}
G^+  \\
H^+  \\
\end{pmatrix},
\quad
\begin{pmatrix}
\eta _1  \\
\eta _2  \\
 \end{pmatrix} 
 = R(\beta )
 \begin{pmatrix}
G^0  \\
A^0  \\
 \end{pmatrix},
 \quad
R(\beta ) 
= \begin{pmatrix}
c_{\beta} & -s_{\beta}  \\
s_{\beta} & c_{\beta}  \\
\end{pmatrix},
\end{equation}
where the rotation
angle $\beta$ is given by:
\begin{equation}
t_{\beta} = \frac{v_2}{v_1}.
\end{equation}

We know that 
$G^\pm$ and $G^0$ are being 
massless Nambu-Goldstone bosons
giving the masses for the
weak gauge bosons $W^\pm$ and $Z$,
respectively. The remaining
physical states $H^\pm$ and $A^0$ 
become charged Higgs and CP-odd Higgs.
Their masses are then given by
\begin{eqnarray}
M_{{H^ + }}^2
&=& \frac{v_1^2 + v_2^2}{v_1 v_2}
\Big[
m_{12}^2 
- \frac{1}{2}\lambda_{45} v_1 v_2 
\Big],
\\
M_{A^0}^2 &=&
\frac{v_1^2 + v_2^2}{v_1 v_2 }
(m_{12}^2 - \lambda_5 v_1 v_2 ).
\label{MHPMA0}
\end{eqnarray}

As the same procedure,
the $CP$-even scalars 
$\rho_1$, $\rho_2$, and $\rho_3$
can be first collected in the
from of
\begin{equation}
 \mathcal{L}^{\rho}_{\mathrm{mass}}
 =
 -\frac{1}{2}
 \begin{pmatrix}
{{\rho _1,}}
& {{\rho _2,}} & \rho_3
\\
\end{pmatrix}
\mathcal{M}_{\rho}^2
\begin{pmatrix}
{{\rho _1}}  \\
{{\rho _2}}  \\
\rho_3 \\
\end{pmatrix}.
\end{equation}
Where the elements 
of the matrix 
$\mathcal{M}_{\rho}^2$ 
are shown as follows:
\begin{eqnarray}
(\mathcal{M}_{\rho}^2)_{11} 
&=& 
\lambda _1 v_1^2 
+ \frac{v_2}{v_1}m_{12}^2,
\quad
(\mathcal{M}_{\rho s}^2)_{22}
=
\lambda_2 v_2^2 
+ \frac{v_1}{v_2}  m_{12}^2,
\\
(\mathcal{M}_{\rho}^2)_{33}
&=& \lambda_S v_s^2,
\hspace{2cm}
(\mathcal{M}_{\rho}^2)_{12} 
= \lambda_{345}{v_1}{v_2} 
- {m}_{12}^2,
\\
(\mathcal{M}_{\rho}^2)_{13}
&=& \lambda_{S\Phi_1} {v_1}{v_s},
\hspace{1.25cm}
 (\mathcal{M}_{\rho}^2)_{23}
 = {\lambda_{S\Phi_2} }{v_2}{v_s}.
\end{eqnarray}
The matrix
$\mathcal{M}_{\rho}^2$ can be 
diagonalized by applying
the rotation matrix $\mathcal{O}$
as follows:
\begin{eqnarray}
\mathrm{diag} (M_{H_1}^2,
M_{H_2}^2, M_{H_3}^2)
=
\mathcal{O}^{\mathrm{T}} 
\mathcal{M}_{\rho}^2 \mathcal{O}.
\label{Omatrix}
\end{eqnarray}
As a result, the mass eigenstates 
$H_i$ are then related to the flavor bases
$\rho_i$ for $i=1,2,3$ via the
rotation matrix $\mathcal{O}$ as
\begin{equation}
\begin{pmatrix}
   \rho_1  \\
   \rho_2  \\
   \rho_3  \\
 \end{pmatrix} = \mathcal{O}
 \begin{pmatrix}
   H_1  \\
   H_2  \\
   H_3  \\
 \end{pmatrix}.
\end{equation}
Detail the form of $\mathcal{O}$
is given explicitly in Appendix $A$.
Finally, we get the physical masses
CP-even Higgses. One of $H_i$
becomes the SM-like Higgs boson.
In this paper, we note that
$h^0$ is the SM-like Higgs boson
and $H_j$ for remaining CP-even
Higges in the model under
consideration
for the later computation.

We turn our attention
to kinetic terms of
the above Lagrangian.
The terms are expressed
as follows:
\begin{eqnarray}
\mathcal{L}_{\mathrm{kin}} 
= 
(D^{\mu}\Phi_1)^{\dag} 
D_{\mu}\Phi_1 
+ (D^{\mu} \Phi_2)^{\dag} 
D_{\mu} \Phi_2.
\label{LkinemSTHDM}
\end{eqnarray}
Masses for the weak gauge bosons
can be derived
from expanding the kinetic terms. 
In detail, mass terms can be
collected in the form of
\begin{eqnarray}
\mathcal{L}_{\mathrm{mass}}^{V}
= \frac{g^2}{4}
(v_1^2 + v_2^2)W^{-,\mu }W_\mu^+  
+ \frac{1}{2}\frac{g^2}{4c_{\mathrm{W}}^2}
(v_1^2 + v_2^2)Z^\mu Z_\mu.
\end{eqnarray}
Here, we use $c_\mathrm{W}
=\cos\theta_\mathrm{W}$
which is cosine of Weinberg's angle.
In this equation, we fix
$v =\sqrt{v_1^2 + v_2^2} \sim 246$ GeV at 
electroweak scale. The masses of $W$
and $Z$ bosons then read
\begin{eqnarray}
M_W = \frac{gv}{2},
\quad 
M_Z = \frac{gv}{2c_{\mathrm{W}}}.
\end{eqnarray}

Finally, we consider the Yukawa sector.
In order to avoid Tree-level
Flavor-Changing Neutral Currents (FCNCs),
four independent types of Yukawa
couplings are listed as follows:
\begin{eqnarray}
&& \text{Type I:}
\quad\;\;
\mathcal{L}^{(I)}_{\mathrm{Y}}
=  
- y_{\ell_i} 
\bar{L}_{i\mathrm{L}} 
\ell_{i{\mathrm{R}}} 
\Phi_2
- 
\tilde{y}_d^{ij}
\bar{Q}_{i{\mathrm{L}}}
d'_{j{\mathrm{R}}} \Phi_2
%%%%%%%%%%
- 
\tilde{y}_u^{ij} 
\bar{Q}_{i{\mathrm{L}}}
u'_{j{\mathrm{R}}} 
\tilde{\Phi}_2 
+ \mathrm{H.c.}
\\
%%%%%%%%%%%%%%%%%%%%%%%%%%%%
&& \text{Type II:}
\quad
\mathcal{L}^{(II)}_{\mathrm{Y}}
=  
-y_{\ell_i} \bar{L}_{i \mathrm{L}}
\ell_{i{\mathrm{R}}} \Phi_1 
- 
\tilde{y}_d^{ij} \bar{Q}_{i{\mathrm{L}}}
d'_{j{\mathrm{R}}} \Phi_1 
- 
\tilde{y}_u^{ij} \bar{Q}_{i{\mathrm{L}}} 
u'_{j{\mathrm{R}}} \tilde{\Phi}_2
+ \mathrm{H.c.}
%%%%%%%%%%%%%%%%%%%%%%%
\\
&& \text{Type X:}
\quad
\mathcal{L}^{(X)}_{\mathrm{Y}}
= 
- y_{\ell_i}
\bar{L}_{i \mathrm{L}} 
\ell_{i \mathrm{R} } \Phi_1
- 
\tilde{y}_{d^{ij}} 
\bar{Q}_{i\mathrm{L}}
d'_{j{\mathrm{R}}} 
\Phi_2
- 
\tilde{y}_{u^{ij}} 
\bar{Q}_{i\mathrm{L}} 
u'_{j{\mathrm{R}}} 
\tilde{\Phi}_2 
+ \mathrm{H.c.}
\\
%%%%%%%%%%%%%%%%%%%%%%%%%
&& \text{Type Y:}\quad
\mathcal{L}^{(Y)}_{\mathrm{Y}}
= 
- y_{\ell _i}
\bar{L}_{i\mathrm{L}}
\ell_{i\mathrm{R}}
\Phi_2 
- \tilde{y}_d^{ij}
\bar{Q}_{i{\mathrm{L}}}
d'_{j\mathrm{R}}\Phi_1 
- \tilde{y}_u^{ij}
\bar{Q}_{i{\mathrm{L}}}
u'_{j\mathrm{R}}
\tilde{\Phi}_2 
+ \mathrm{H.c.}
\end{eqnarray}
Here 
$\tilde{\Phi}_2= 
i\sigma^2 \Phi_2^*$, 
lepton doublet $L_{i\mathrm{L}} 
=(\nu_{i\mathrm{L}}, 
\ell_{i\mathrm{L}})^\mathrm{T}$ 
for generation index 
$i=1,2,3$
and quark doublet 
$Q_{i\mathrm{L}} =
(u'_{i\mathrm{L}}, 
d'_{i\mathrm{L}})^\mathrm{T}$.
The mixing matrices 
$\tilde y_d^{ij}$ for down-quarks
and $\tilde y_u^{ij}$ for up-quarks
can be diagonalized by
following rotation matrix
$({U_d})_{ij}$. In detail, we
have 
\begin{eqnarray}
({U_d})_{ij}^\dag
\tilde y_d^{jk}{({U_d})_{kl}}
&=& {y_{{d_i}}}{\delta _{il}}, \\
({U_u})_{ij}^\dag
\tilde y_u^{jk}{({U_u})_{kl}}
&=& {y_{{u_i}}}{\delta _{il}}.
\end{eqnarray}
The quarks basis 
$u'_i$ and $d'_i$ are related 
to their mass basis 
$u_i$ and $d_i$ through
\begin{eqnarray}
d'_i = ({U_d})_{ij}{d_j}, 
\quad 
u'_i= ({U_u})_{ij}{u_j}. 
\end{eqnarray}
The Cabibbo-Kobayashi-Maskawa 
matrix is then identified as
\begin{eqnarray}
\textrm{CKM}_{ij} 
= ({U_u})_{ik}^\dag {({U_d})_{kj}}.
\end{eqnarray}
Since we aren't interested in
neutrino physics in this work, we assume the
lepton sector is the same as in the SM.
After the EWSB,
the four types of Yukawa interactions
can be written in the form of
\begin{eqnarray}
\mathcal{L}_{\mathrm{Y}}
&\supset& -
\sum\limits_{f = \ell_j,d_j,u_j} 
\Big[
 m_f\bar{f}f
+
\sum\limits_{i = 1}^3
\kappa_{H_i}^f \frac{m_f}{v}
\; H_i\bar{f} f
+
\kappa_{A^0}^f \frac{m_f}{v}
\; A^0 \bar{f}
i \gamma_5 f
\Big].
\end{eqnarray}
The couplings of $H_i$ and
$A_0$ to fermion pair
are taken the form of
\begin{eqnarray}
 g_{H_i f\bar{f}} =
 -\kappa_{H_i}^{f}
 \frac{m_f  }{v},
\quad
 g_{A_0f\bar{f}} =
 -i\kappa_{A^0}^{f}
 \frac{m_f}{v}\gamma_5.
\end{eqnarray}
The coefficients
$\kappa_{H_i}^f$
and  $\kappa_{A^0}^f$ are
are shown in detail in
Table~\ref{YukawaKappa}.
Here we note $h^0$ is SM-like
Higgs bososn
and $H_j$ for new heavy
CP-even Higges in the
corresponding models
in the following Table.
%%%%%%%%%%%%%%%%%%%%%%%%%%%%%%%%%%%
\begin{table}[H]
\centering
\begin{tabular}{l@{\hspace{1.5cm}}
l@{\hspace{1.5cm}}l@{\hspace{1.5cm}}
l@{\hspace{1.5cm}}l}
\hline
\hline
$\kappa$-factors
& Type I & Type II
& Type X & Type Y \\
\hline \hline
$\kappa_{H_i}^{\ell_j}$ 
& $\mathcal{O}_{2i}/s_{\beta}$ 
& $\mathcal{O}_{1i}/c_{\beta}$ 
& $\mathcal{O}_{1i}/c_{\beta}$ 
& $\mathcal{O}_{2i}/s_{\beta}$ \\
$\kappa_{H_i}^{d_j}$ 
& $\mathcal{O}_{2i}/s_{\beta}$ 
& $\mathcal{O}_{1i}/c_{\beta}$ 
& $\mathcal{O}_{2i}/s_{\beta}$ 
& $\mathcal{O}_{1i}/c_{\beta}$ \\
$\kappa_{H_i}^{u_j}$ 
& $\mathcal{O}_{2i}/s_{\beta}$ 
& $\mathcal{O}_{2i}/s_{\beta}$ 
& $\mathcal{O}_{2i}/s_{\beta}$ 
& $\mathcal{O}_{2i}/s_{\beta}$ \\
\hline
$\kappa_{A^0}^{\ell_j}$ 
& $1/t_{\beta}$ 
& $-t_{\beta}$ 
& $-t_{\beta}$ 
& $1/t_{\beta}$ \\
$\kappa_{A^0}^{d_j}$ 
& $1/t_{\beta}$ 
& $-t_{\beta}$ 
& $1/t_{\beta}$ 
& $-t_{\beta}$ \\
$\kappa_{A^0}^{u_j}$ 
& $-1/t_{\beta}$ 
& $-1/t_{\beta}$ 
& $-1/t_{\beta}$ 
& $-1/t_{\beta}$ \\
\hline
\hline
\end{tabular}
\caption{\label{YukawaKappa} 
The factors $\kappa_{H_i}^f$ 
and $\kappa_{A^0}^f$ are 
are listed for four
types of STHDM. The
elements of the rotation matrix
$\mathcal{O}_{ij}$ are
given explicitly
in Appendices A, B. It is noted
that $h^0$ is noted for the SM-like Higgs boson
and $H_j$ are for remaining CP-even Higgeses
in the mentioned models.}
\end{table}
%%%%%%%%%%%%%%%%%%%%%%%%%%%%%%%%%
Other couplings relating to the processes
under consideration are listed in
Table~\ref{AHZ1}. In the Table~\ref{AHZ1},
a general form of the couplings are shown
in the second column. In the third column,
we present the couplings in STHDM. While
the couplings in THDM can be reduced
from the third column and are presented
in the last column. For the vertices of
scalar particle with vector boson pair,
we express the couplings in terms of
$\kappa$-factor of the SM's couplings.
The couplings for SM-like Higgs to
vector boson pair in the SM are given
$g_{h^0ZZ}^{\textrm{SM}}=
\frac{e M_W}{c_W^2 s_W}$,
$g_{h^0WW}^{\textrm{SM}}=
\frac{e M_W}{s_W}$.
%%%%%%%%%%%%%%%%%%%%%%%%%%%%%%%
\begin{table}[H]
\begin{center}
\begin{tabular}{l@{\hspace{0.3cm}}
l@{\hspace{0.3cm}}l@{\hspace{0.3cm}}l}
\hline \hline 
Vertices
& HESM
& STHDM
& THDM
\\
\hline
\hline
%%%%%%%%%%%%%%%%%%%%
$A^0 Z_{\mu} \, \phi$
& 
$g_{A^0 Z \, h^0}
\cdot
(p^{h^0}-p^{A^0})^{\mu}
$
&
$\dfrac{e}{s_{2W}}
(-s_{\beta}\mathcal{O}_{12}
+c_{\beta}\mathcal{O}_{22})
\cdot
(p^{h^0}-p^{A^0})^{\mu}$
&
$
\dfrac{e\;
c_{\beta - \alpha}}{s_{2W} }
(p^{h^0}-p^{A^0})^{\mu}
$
\\ 
&
$g_{A^0 Z \, H_j}
\cdot
(p^{H_j}-p^{A^0})^{\mu}
$
&
$\dfrac{-e}{s_{2W}}
(c_{\beta}\mathcal{O}_{2j}
-\mathcal{O}_{1j}s_{\beta})
\cdot
(p^{H_j}-p^{A^0})^{\mu}$
&
$
\dfrac{-e
\; s_{\beta - \alpha}}{s_{2W} }

(p^{H}-p^{A^0})^{\mu}
$
\\ \hline
%%%%%%%%%%%%%%%%%%%
$\phi Z_{\mu}Z_{\nu}$
& 
$ - i \,
\kappa_{h^0 Z Z}
\cdot
g_{h^0ZZ}^{\textrm{SM}}
\cdot
g^{\mu\nu}
$
&
$- i \, [c_{\beta}\mathcal{O}_{11}
+s_{\beta}\mathcal{O}_{21}]\cdot
g_{h^0ZZ}^{\textrm{SM}}
\cdot
g^{\mu\nu}
$
&
$ i \,
s_{\beta-\alpha}
\cdot
g_{hZZ}^{\textrm{SM}}
\cdot
g^{\mu\nu}
$
\\ 
&
$ - i \,
\kappa_{H_j Z Z}
\cdot
g_{h^0ZZ}^{\textrm{SM}}
\cdot
g^{\mu\nu}
$
&
$- i \, [c_{\beta}\mathcal{O}_{1j}
+s_{\beta}\mathcal{O}_{2j}]\cdot
g_{h^0ZZ}^{\textrm{SM}}
\cdot
g^{\mu\nu}
$
&
$ i \,
c_{\beta-\alpha}
\cdot
g_{h^0ZZ}^{\textrm{SM}}
\cdot
g^{\mu\nu}
$
\\
\hline
%%%%%%%%%%%%%%%%%%%
$\phi W^\pm_{\mu}
W^\mp_{\nu}$
& 
$ - i \,
\kappa_{h^0WW}
\cdot
g_{h^0 WW}^{\textrm{SM}}
\cdot
g^{\mu\nu}
$
&
$
- i \,
[c_{\beta}\mathcal{O}_{11}
+s_{\beta}\mathcal{O}_{21}]
\cdot
g_{h^0 WW}^{\textrm{SM}}
\cdot
g^{\mu\nu}
$
&
$ i \,
s_{\beta-\alpha}
\cdot
g_{h^0WW}^{\textrm{SM}}
\cdot
g^{\mu\nu}
$
\\ 
&
$ - i \,
\kappa_{H_jWW}
\cdot
g_{h^0WW}^{\textrm{SM}}
\cdot
g^{\mu\nu}
$
&
$
- i \,
[c_{\beta}\mathcal{O}_{1j}
+s_{\beta}\mathcal{O}_{2j}]
\cdot
g_{h^0WW}^{\textrm{SM}}
\cdot
g^{\mu\nu}
$
&
$ i \,
c_{\beta-\alpha}
\cdot
g_{h^0 WW}^{\textrm{SM}}
\cdot
g^{\mu\nu}
$
\\ \hline   
\hline      
\end{tabular}
\caption{\label{AHZ1} All related
couplings with
$V \equiv \gamma, Z$ and $\phi 
\equiv h^0, H_j$
to the processes under consideration.
$H_j$ is for remaining
CP-even Higgses
in the mentioned models.
Here, we have the
SM couplings as
$g_{h^0ZZ}^{\textrm{SM}}=
\frac{e M_W}{c_W^2 s_W}$,
$g_{h^0WW}^{\textrm{SM}}=
\frac{e M_W}{s_W}$. In THDM,
one has $H_j = H$ and
we have the appropriate
couplings in the last column.
}
\end{center}
\end{table}
We note that we can derive the
corresponding couplings for NTHDM
by considering the singlet scalar
$S$ being real scalar field. In order
to arrive at the respective couplings
in THDM, we take the limits for the rotation
matrix $\mathcal{O}$ in Appendix $A$
as follows: $c_{12} \rightarrow c_{\alpha}$
and $s_{23} = s_{13}=0$.
%%%%%%%%%%%%%%%%%%%%%%%%%%%%%%%%%%%%%%%%%%%%%%%%
\subsection{Triplet Higgs Models} %%%
%%%%%%%%%%%%%%%%%%%%%%%%%%%%%%%%%%%%%
We turn our attention to another type of
the HESM which is triplet higgs model
(THM), a classification of the SM with
adding a real Higgs triplet,
denoted as $\Delta$ with
hypercharge $Y_{\Delta}=2$~\cite{Chun:2012jw,
Chen:2013dh, Arhrib:2011uy, Arhrib:2011vc,
Akeroyd:2012ms,Akeroyd:2011zza,Akeroyd:2011ir,
Aoki:2011pz,Kanemura:2012rs,Chabab:2014ara,
Han:2015hba,Chabab:2015nel,Ghosh:2017pxl,
Ashanujjaman:2021txz,Zhou:2022mlz}.
The most general form of
the scalar potential of THM with obeying
the renormalizable condition and the gauge 
invariance is taken the form of:
\begin{eqnarray}
\mathcal{V}_{THM}(\Phi, \Delta)
&=& -m_{\Phi}^2{\Phi^{\dagger}\Phi}
+\frac{\lambda}{4}(\Phi^{\dagger}\Phi)^2 
+ M_{\Delta}^2 \textrm{Tr}
(\Delta^{\dagger} \Delta)
+[\mu(\Phi^T{i}\sigma^2\Delta^{\dagger}\Phi)
+{\rm H.c}]\nonumber\\
&&+
\lambda_1(\Phi^{\dagger}\Phi) 
\textrm{Tr}(\Delta^{\dagger}{\Delta})
+\lambda_2( \textrm{Tr}\Delta^{\dagger}{\Delta})^2
+\lambda_3 \textrm{Tr}(\Delta^{\dagger}{\Delta})^2
+\lambda_4{\Phi^\dagger\Delta\Delta^{\dagger}\Phi}.
\end{eqnarray}
Here, $\sigma^2$ is Pauli matrix.
All Higgs self couplings
$\lambda_i$ ($i=\overline{i,4}$)
are considered as real parameters. 
For the EWSB, two Higgs multiplets
are parameterized as follows: 
\begin{eqnarray}
\Delta &=& \frac{1}{\sqrt{2} }
\begin{pmatrix}
\delta^+  
&  \sqrt{2} \; \delta^{++} \\
v_{\Delta} +\eta_{\Delta} 
+i \chi_{\Delta}
& 
-\delta^+ 
\end{pmatrix}
\quad 
{\rm and}
\quad 
\Phi=\frac{1}{\sqrt{2} } 
\begin{pmatrix}
\sqrt{2}\; \phi^+ \\
v_{\Phi} +\eta_{\Phi} 
+i \chi_{\Phi}
\end{pmatrix},
\label{representa-htm}
\end{eqnarray}
where vacuum expectation values
(VEV) the two neutral Higgs 
are corresponding to
$v_{\Phi}$ and $v_{\Delta}$. 
The electroweak scale is fixed at 
$v= \sqrt{v_{\Phi}^2+2v_{\Delta}^2}
\sim 246$ GeV for agreement 
with the SM case. In order 
to obtain the masses of physical
scalar bosons, one rotates
the flavor bases into
the physics states.
The relations
are shown as follows:
\begin{eqnarray}
\begin{pmatrix}
\phi^{\pm}\\
 \delta^{\pm}
\end{pmatrix}
&=&
\begin{pmatrix}
c_{\beta^{\pm}} & 
-s_{\beta^{\pm}} \\
s_{\beta^{\pm}}
& c_{\beta^{\pm}}
\end{pmatrix}
\begin{pmatrix}
G^{\pm}\\
H^{\pm}
\end{pmatrix},
%%%%%%%%
\\\begin{pmatrix}
\eta_{\Phi}\\
 \eta_{\Delta}
\end{pmatrix}
&=&
\begin{pmatrix}
c_{\alpha}  &  
-s_{\alpha} \\
s_{\alpha}  & 
c_{\alpha}
\end{pmatrix}
\begin{pmatrix}
h^0\\
H
\end{pmatrix},
\end{eqnarray}
and 
\begin{eqnarray}
\begin{pmatrix}
\chi_{\Phi}\\
 \chi_{\Delta}
\end{pmatrix}
=
\begin{pmatrix}
c_{\beta^{0}} & 
-s_{\beta^{0}} \\
s_{\beta^{0}}
& c_{\beta^{0}}
\end{pmatrix}
\begin{pmatrix}
G^{0}\\
A^0
\end{pmatrix},
\end{eqnarray}
where the rotation angles are given
$t_{\beta^{\pm}}
= \frac{\sqrt{2} v_{\Delta}}
{v_{\Phi}}$, $t_{\beta^0} 
= \sqrt{2} t_{\beta^{\pm}}$ 
and the mixing angle $\alpha$
between two neutral Higgs is 
taken into account. After the EWSB, 
charged Nambu-Goldstone bosons $G^{\pm}$
and neutral Nambu-Goldstone bosons 
$G^0$ are eaten by charged gauge bosons
$W^\pm$ and neutral gauge boson $Z$, 
respectively. As a result, all gauge 
bosons gain their masses. The remaining 
fields become the physical Higgs states. 
The Higgs spectrum of the THM includes 
pair of doubly charged Higgs $H^{\pm\pm}$, 
two singly charged $H^\pm$, 
a neutral CP-odd $A^0$, and two CP-even 
$H$ and $h^0$ being the SM-like Higgs boson.
The masses of Higgses are expressed in 
terms of the Higgs self-coupling parameters
and $\mu$ as follows: 
\begin{eqnarray}
&& M_{H^{\pm\pm}}^2  = 
\frac{\sqrt{2}\mu{v_{\Phi}^2}
-\lambda_4v_{\Phi}^2v_\Delta
-2\lambda_3v_\Delta^3}{2v_\Delta},
\\
%%%%%%%%%%%%%%
&& M_{H^{\pm}}^2 
=  \frac{(v_{\Phi}^2
+2v_\Delta^2)\,[2\sqrt{2}\mu
-\lambda_4v_\Delta]}{4v_\Delta},\\
%%%%%%%%%%%%%%
&& M_{A^0}^2 
=  \frac{\mu(v_{\Phi}^2
+4v_\Delta^2)}{\sqrt{2}v_\Delta},
\\
%%%%%%%%%%%%%%%
&& M_H^2 = 
\frac{1}{2}\Big\{\lambda
v_{\Phi}^2 s_\alpha^2
+ c_\alpha^2 
\Big[\sqrt{2}\mu
\frac{v_{\Phi}^2}{v_\Delta}
\big(1+4\frac{v_\Delta}{v_{\Phi}} 
t_{\alpha}\big) 
+4v_\Delta^2\big((\lambda_2+\lambda_3) 
- (\lambda_1+\lambda_4)
\frac{v_{\Phi}}{v_\Delta} 
t_{\alpha}\big) \Big]\Big\},
\\
%%%%%%%%%%%%%%
&& M_{h^0}^2 = \frac{1}{2}
\Big\{
\lambda v_{\Phi}^2 c_\alpha^2 
+ s_\alpha^2 \Big[ \sqrt{2}
\mu\frac{v_{\Phi}^2}{v_\Delta}
\big(1-4\frac{v_\Delta}{v_{\Phi} 
t_{\alpha}}\big) + 4v_\Delta^2
\big( (\lambda_1+\lambda_4) 
\frac{v_{\Phi}}{v_\Delta t_{\alpha}} 
+ 
(\lambda_2+\lambda_3)
\big)\Big]
\Big\}.
\end{eqnarray}
%%%%%%%%%%%%%%%%%%%%%%%%%%%%%%%%%%%%%%%
As same procedure,
the Yukawa Lagrangian is expressed in 
terms of the mass eigenstates as follows:
\begin{eqnarray}
\label{YukawaTHM}
{\mathcal{L}}_{\rm Yukawa} = 
{\mathcal{L}}^{\rm SM}_{\rm Yukawa}
- L^T_i y_{\nu}C (i \sigma^2 \Delta) L_i 
+ \textrm{H.c},
\end{eqnarray}
where $L_i^T=(L_e,L_{\mu},L_{\tau})$ 
are three left-handed lepton doublets, 
$y_{\nu}$ is the $3\times3$ Yukawa 
coupling matrix and $C$ is the charge 
conjugation operator. Expanding the
Yukawa sector, one arrives
the couplings of $A^0\bar{f}f$
and $\phi\bar{f}f$ for $\phi = h_0, H$.
The corresponding
couplings are shown in
Table~\ref{AffTHM}.
We also express the couplings
in terms of $\kappa$-factors
in the Table~\ref{AffTHM} as
\begin{eqnarray}
 g_{A^0\bar{f}f} = -i \kappa_{A^0}^f
 \frac{m_f}{v}\gamma_5,
 \quad
 g_{\phi \bar{f}f} = -\kappa_{\phi}^f
 \frac{m_f}{v}.
\end{eqnarray}

%%%%%%%%%%%%%%%%%%
\begin{table}[H]
\begin{center}
\begin{tabular}{l@{\hspace{1.5cm}}l
@{\hspace{1.5cm}} l@{\hspace{1.5cm}} l}
\hline\hline
\text{Vertices} & \text{HESM} & \text{THM}
& $\kappa$-factors
\\
\hline \hline
$A^0\bar{f}f$
& $g_{A^0\bar{\ell}\ell}$
& $i\dfrac{\sqrt{2}\;t_{\beta^\pm}}
{v}m_{\ell} \; \gamma_5 $
& $-\sqrt{2}\;t_{\beta^\pm}$
\\
& $g_{A^0\bar{d}d}$
& $i\dfrac{\sqrt{2}\;t_{\beta^\pm}}
{v}m_d\;
\gamma_5 $
& $-\sqrt{2}\;t_{\beta^\pm}$
\\
& $g_{A^0\bar{u}u}$ &
$-i\dfrac{\sqrt{2}\;t_{\beta^\pm}}
{v}m_u\; \gamma_5$
& $\sqrt{2}\;t_{\beta^\pm}$
\\
\hline
$\phi\bar{f}f$ &
$g_{h^0\bar{f}f}$
&
$\dfrac{m_f}{v}c_{\alpha}$
& $-c_{\alpha}$
\\
& $g_{H\bar{f}f}$ &
$-\dfrac{m_f}{v}\; s_{\alpha}$
& $s_{\alpha}$
\\ \hline \hline
\end{tabular}
\caption{\label{AffTHM}
The couplings
of $A^0\bar{f}f$ and $\phi\bar{f}f$
are shown in this Table.
We have used $\phi = h_0, H$ in the THM.
}
\end{center}
\end{table}
%%%%%%%%%%%%%%%%%%%%%%%%%%%%%%%%%%%%%%%%%%%%%%
Other couplings relating to
the processes under consideration are shown
in Table~\ref{THDMCOUPLINGS}. In this Table,
the notations
$V \equiv \gamma, Z$ and $\phi \equiv h^0, H$
are used.
%%%%%%%%%%%%%%%%%%%%%%%%%%%%%%%%%%%%%%%%%%%%%%
\begin{table}[H]
\begin{center}
\begin{tabular}{l@{\hspace{1.5cm}}
l@{\hspace{1.5cm}}l}
\hline\hline
\text{Vertices} 
& \text{HESM} & \text{THM}
\\ \hline \hline
$A^0Z_{\mu}\phi$ 
&
$g_{A^0Zh^0}
\cdot
(p_{\mu}^h-p_{\mu}^{A^0})
$
&$\dfrac{-ie}{s_{2W}}
(2s_{\alpha}c_{\beta^0} 
- s_{\beta^0}c_{\alpha})
\cdot
(p_{\mu}^h-p_{\mu}^{A^0})$   
\\
&$g_{A^0ZH}
\cdot
(p_{\mu}^H-p_{\mu}^{A^0})
$
&$\dfrac{-ie}{s_{2W}}
(2c_{\alpha}c_{\beta^0} 
+ s_{\beta^0}s_{\alpha})
\cdot
(p_{\mu}^H-p_{\mu}^{A^0})$    
\\
\hline
%%%%%%%%%%%%%%%%%%%%%%%%%
$\phi Z_{\mu}Z^{\mu}$ 
& $-i\kappa_{h^0ZZ} \cdot
g_{h^0ZZ}^{\textrm{SM}}
\cdot g^{\mu\nu}
$
& $-i
(c_{\beta^0}c_{\alpha} 
+ 2s_{\alpha}s_{\beta^0})
\cdot
g_{h^0ZZ}^{\textrm{SM}
}
\cdot g^{\mu\nu}
$
\\
&$-i\kappa_{HZZ} \cdot
g_{h^0ZZ}^{\textrm{SM}}
\cdot g^{\mu\nu}
$
&$-i
(-c_{\beta^0}s_{\alpha} 
+ 2c_{\alpha}s_{\beta^0})
\cdot
g_{h^0ZZ}^{\textrm{SM}}
\cdot g^{\mu\nu}
$
\\
\hline
%%%%%%%%%%%%%%%%%%%%%%
$\phi W^\pm_{\mu} W^\mp_{\nu}$
& $-i\kappa_{h^0WW} \cdot
g_{h^0WW}^{\textrm{SM}}
\cdot g^{\mu\nu}
$
& $-i(c_{\alpha}c_{\beta^\pm}
+\sqrt{2}s_{\alpha}s_{\beta^\pm})
\cdot
g_{h^0WW}^{\textrm{SM}}
\cdot g^{\mu\nu}
$ \\
& $-i
\kappa_{HWW} \cdot
g_{h^0WW}^{\textrm{SM}}
\cdot g^{\mu\nu}
$
& $ -i(-s_{\alpha}c_{\beta^\pm}
+\sqrt{2}c_{\alpha}s_{\beta^\pm})
\cdot
g_{h^0WW}^{\textrm{SM}}
\cdot g^{\mu\nu}
$
\\ \hline \hline
\end{tabular}
\caption{\label{THDMCOUPLINGS}
All related couplings with
$V \equiv \gamma, Z$ and $\phi \equiv h^0, H$
to the processes under consideration.
Again, the couplings for SM-like Higgs
vector boson pair in the SM are given
$g_{h^0ZZ}^{\textrm{SM}}=
\frac{e M_W}{c_W^2 s_W}$,
$g_{h^0WW}^{\textrm{SM}}=
\frac{e M_W}{s_W}$.
}
\end{center}
\end{table}
%%%%%%%%%%%%%%%%%%%%%%%%%%%%%%%%%%%%%%%%%%%%%%%%
\section{One-loop formulas for decay of 
$A^0 \rightarrow \ell \bar{\ell} V$ in HESM} 
\begin{figure}[H]
\centering
\includegraphics[width=12cm, height=4cm]
 {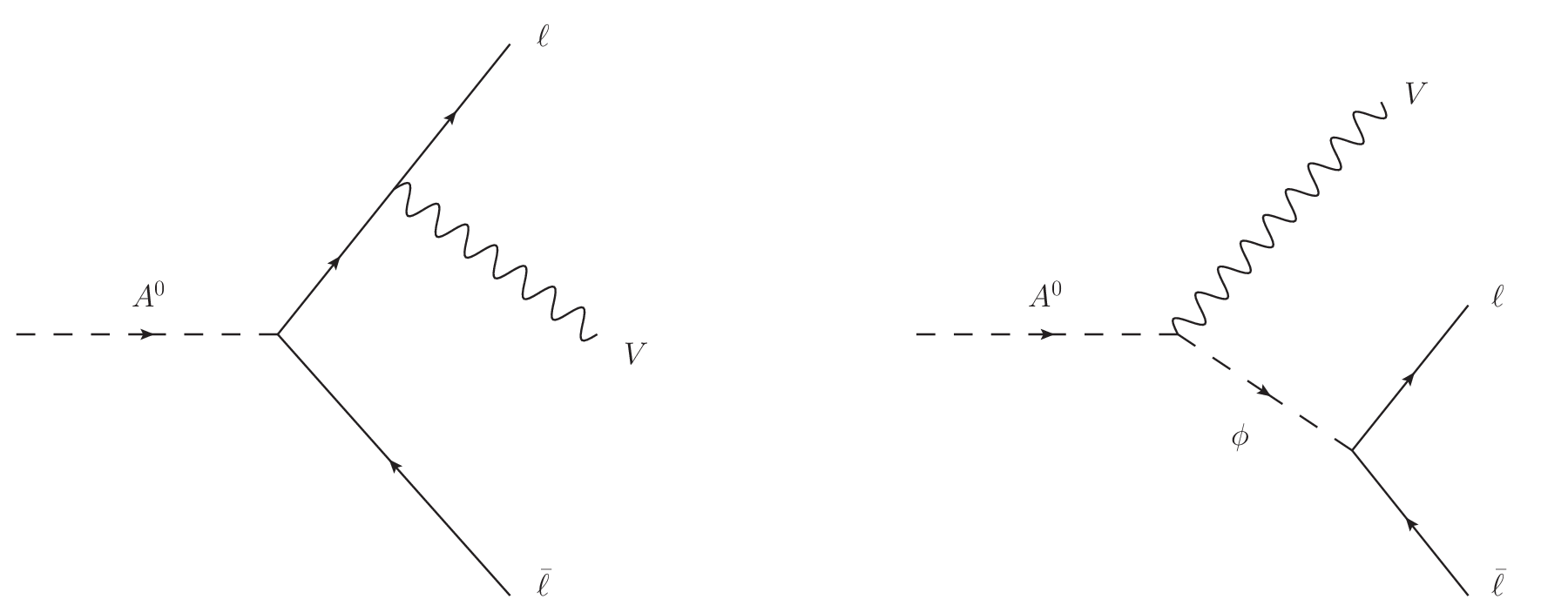}
\caption{\label{treediag} Tree-level
topologies
contributing to the decay processes.
In the plots, $\phi = h^0$ for
the SM-like Higgs
boson and $\phi=H_j$
for remaining CP-even Higgses in
the considered models.}
\end{figure}

\begin{figure}[H]
\centering
\begin{tabular}{cc}
\includegraphics[width=6cm, height=5cm]
{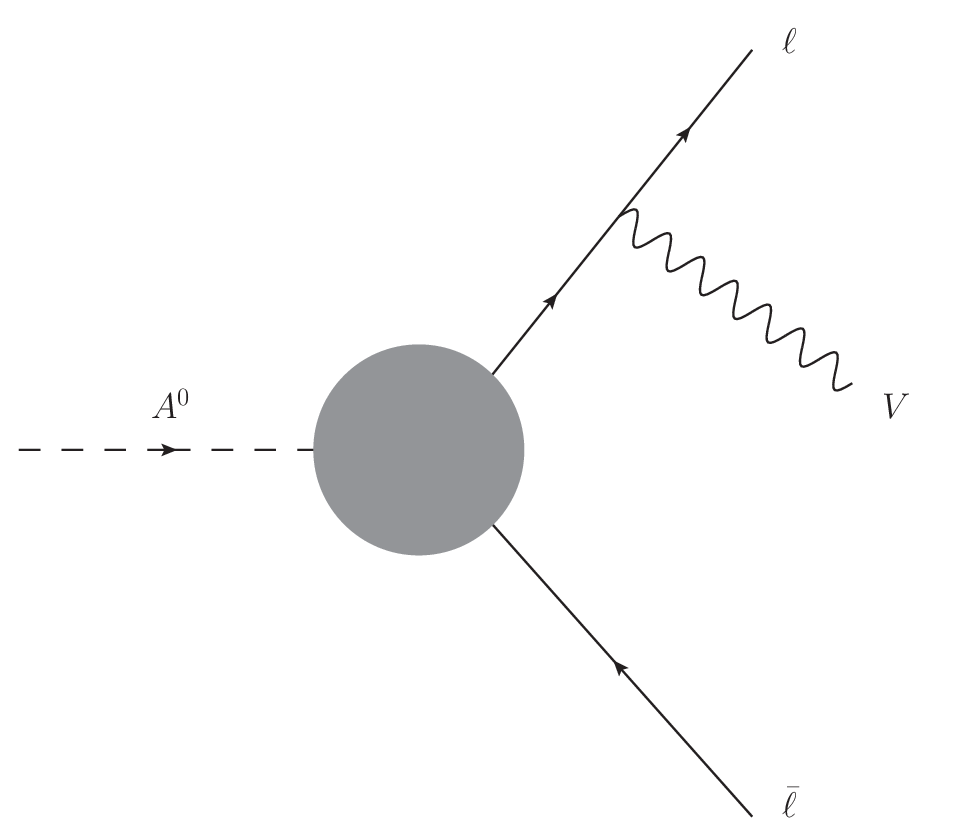}&
\includegraphics[width=6cm, height=5cm]
{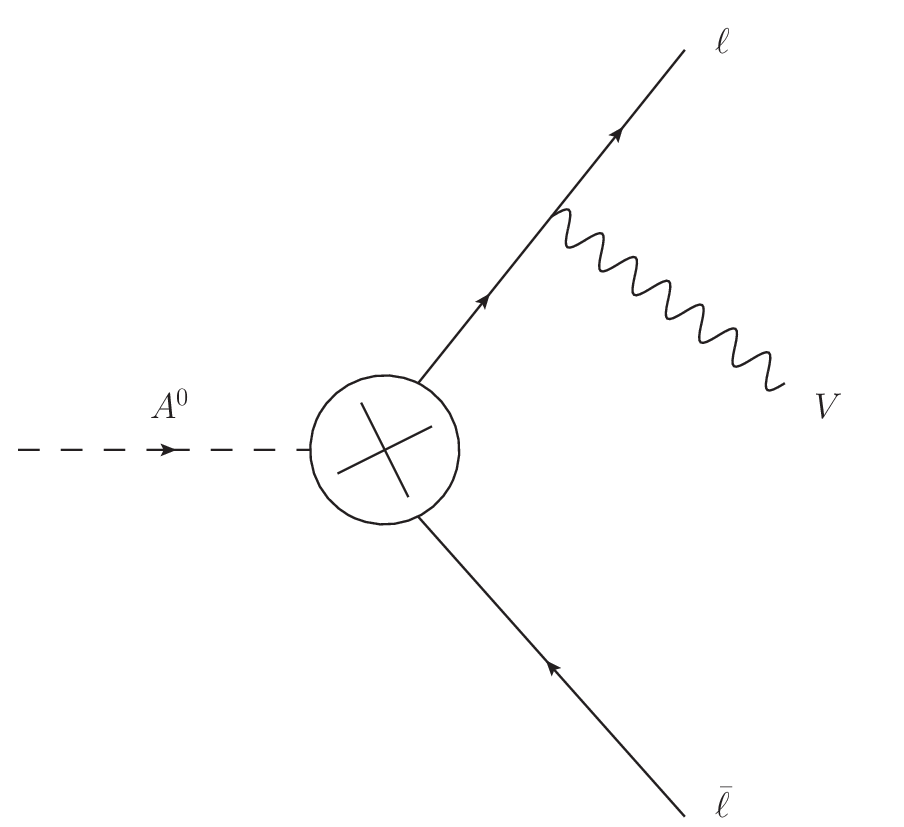}\\
(a) & (b) \\
\includegraphics[width=6cm, height=5cm]
{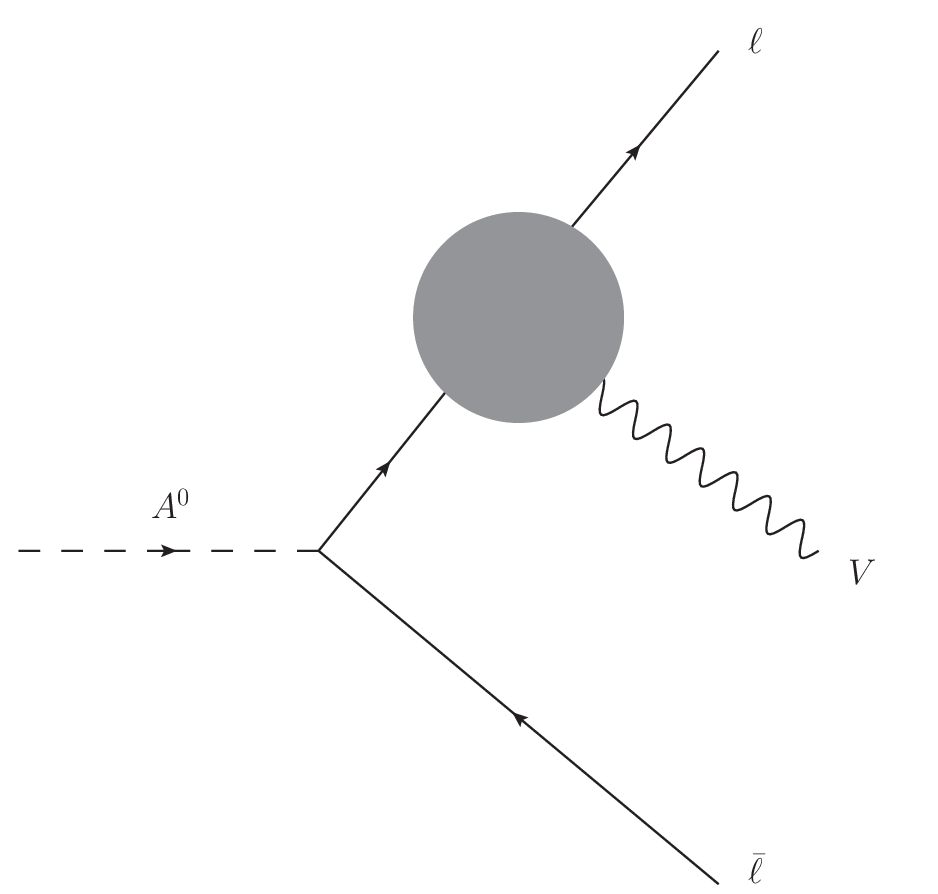}&
\includegraphics[width=6cm, height=5cm]
{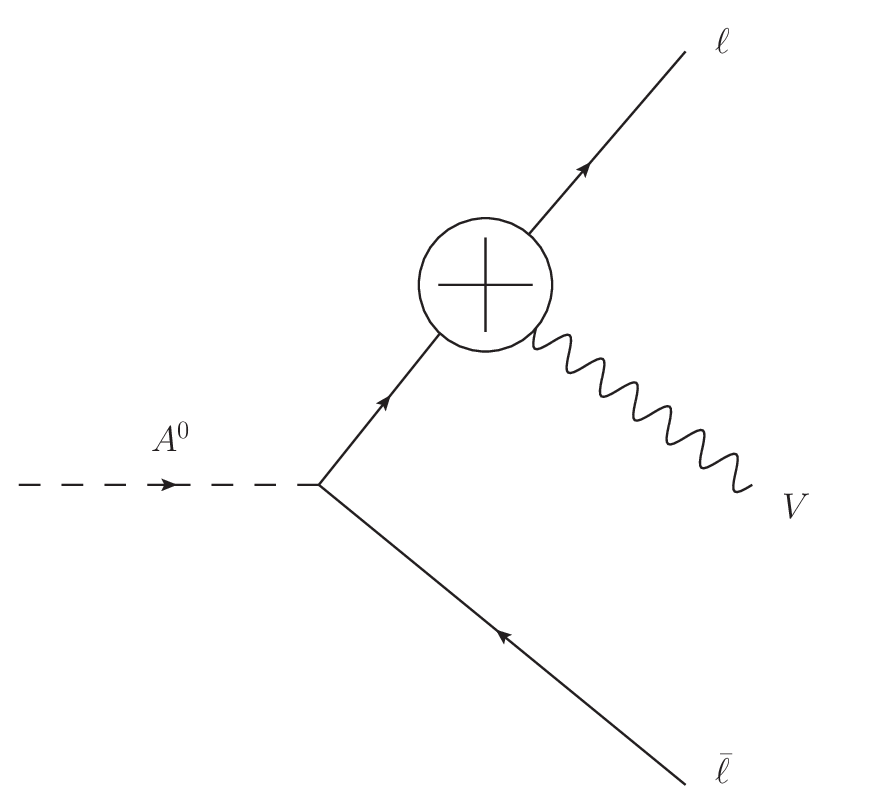}\\
(c) & (d) \\
\end{tabular}
\caption{\label{loopdiag1} One-loop
diagrams
contributing to the decay processes
$A^0 \rightarrow \ell \bar{\ell} V$ in HESM
within the HF gauge.}
\end{figure}

\begin{figure}[H]
\centering
\begin{tabular}{cc}
\includegraphics[width=6cm, height=5cm]
{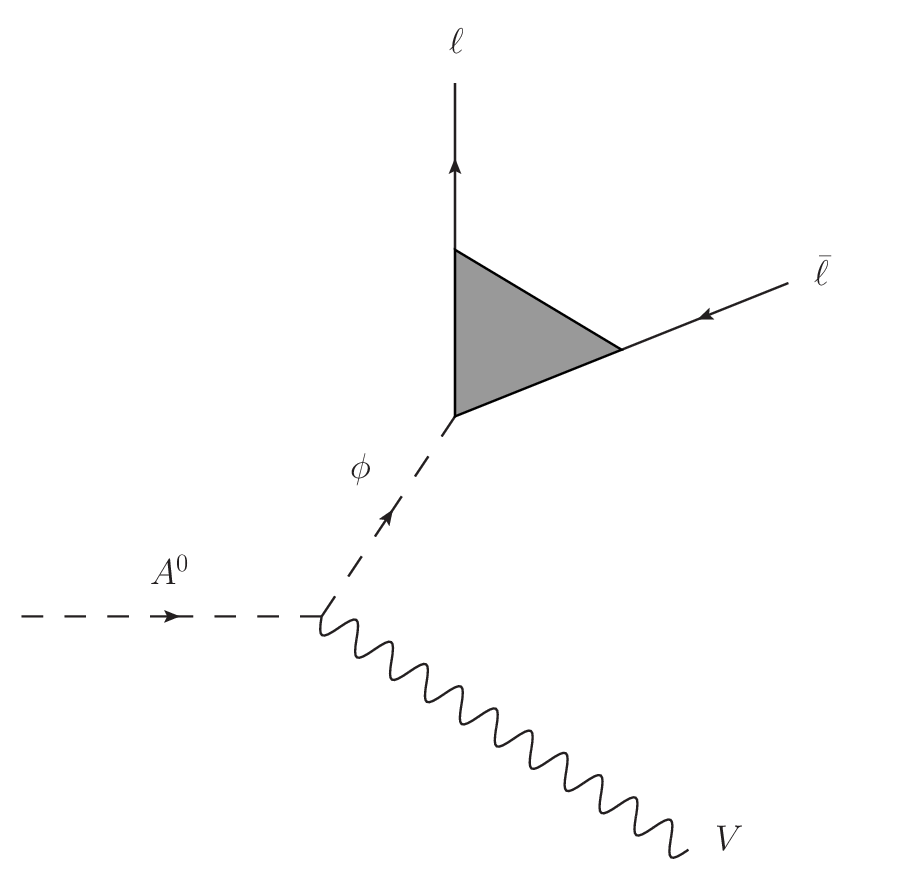}&
\includegraphics[width=6cm, height=5cm]
{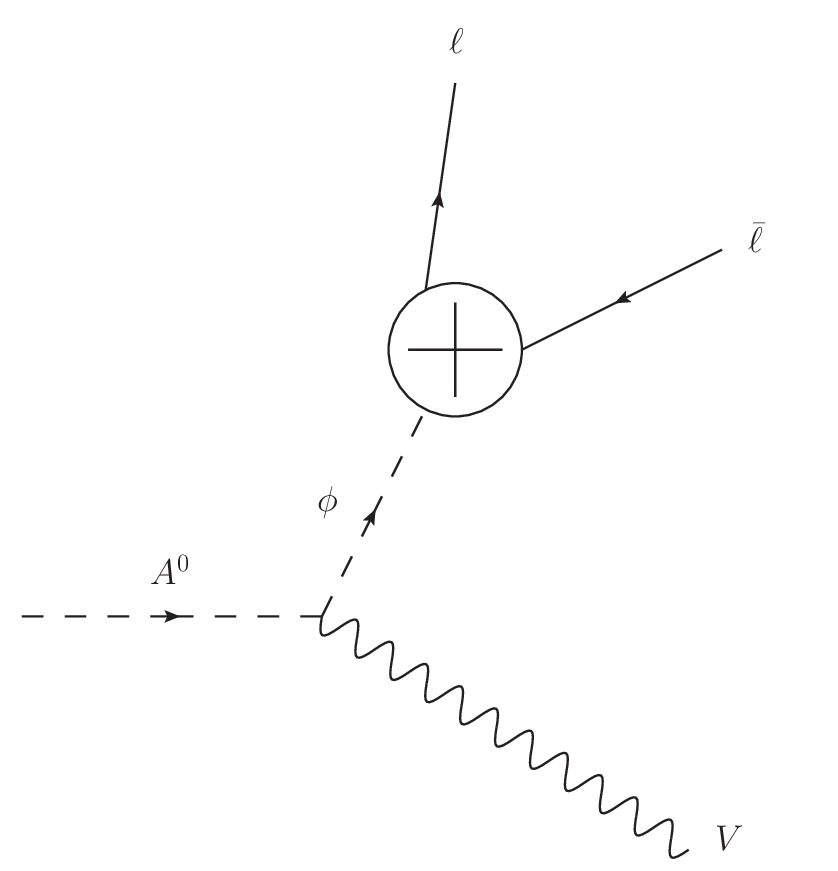}\\
(e) & (f) \\
\includegraphics[width=6cm, height=5cm]
{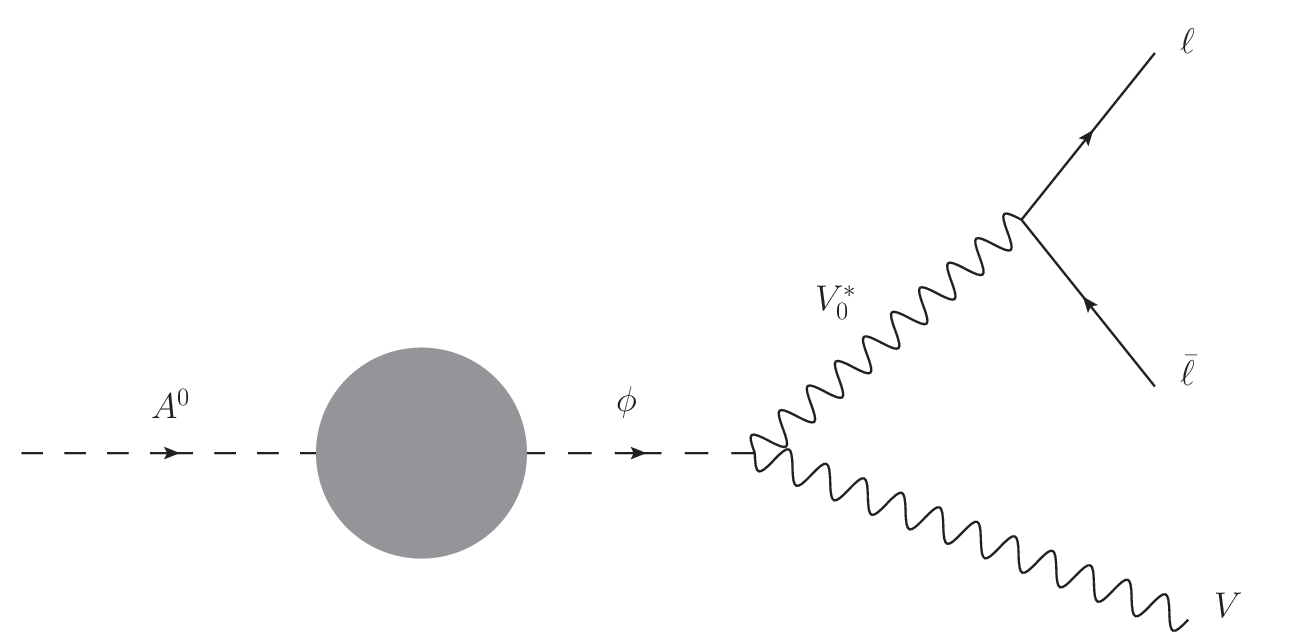}&
\includegraphics[width=6cm, height=5cm]
{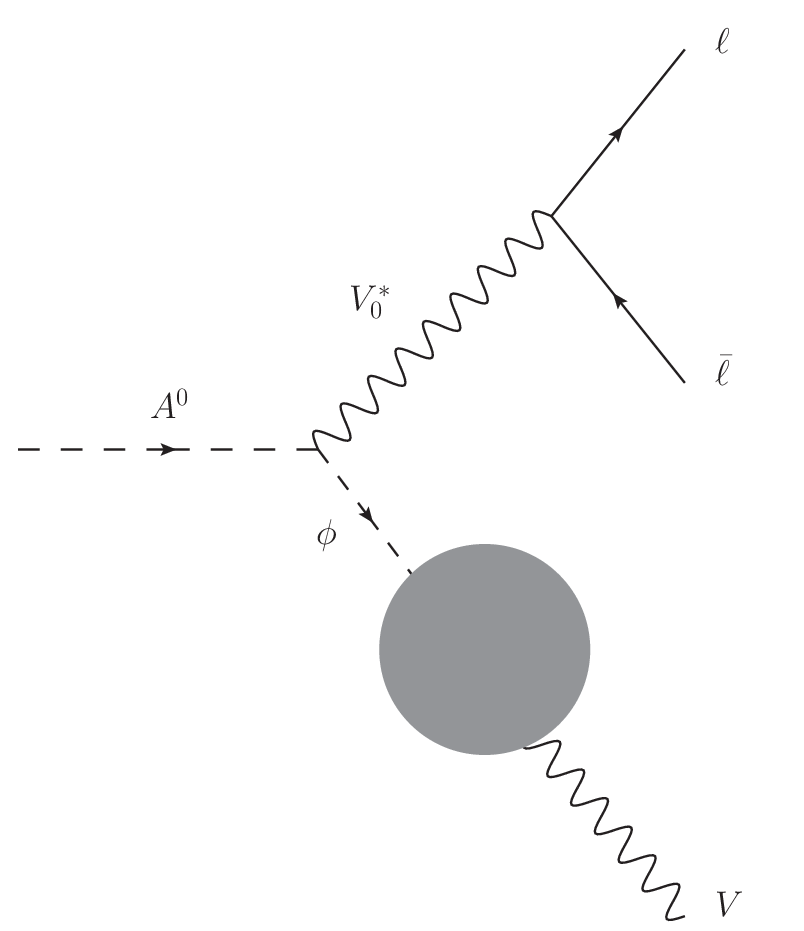}\\
(g) & (h) \\
\end{tabular}
\caption{\label{loopdiag2} One-loop
diagrams
contributing to the decay processes
$A^0 \rightarrow \ell \bar{\ell} V$
in HESM within the HF gauge.}
\end{figure}

\begin{figure}[H]
\centering
\begin{tabular}{ccc}
\includegraphics[width=4.5cm,height=5cm]
{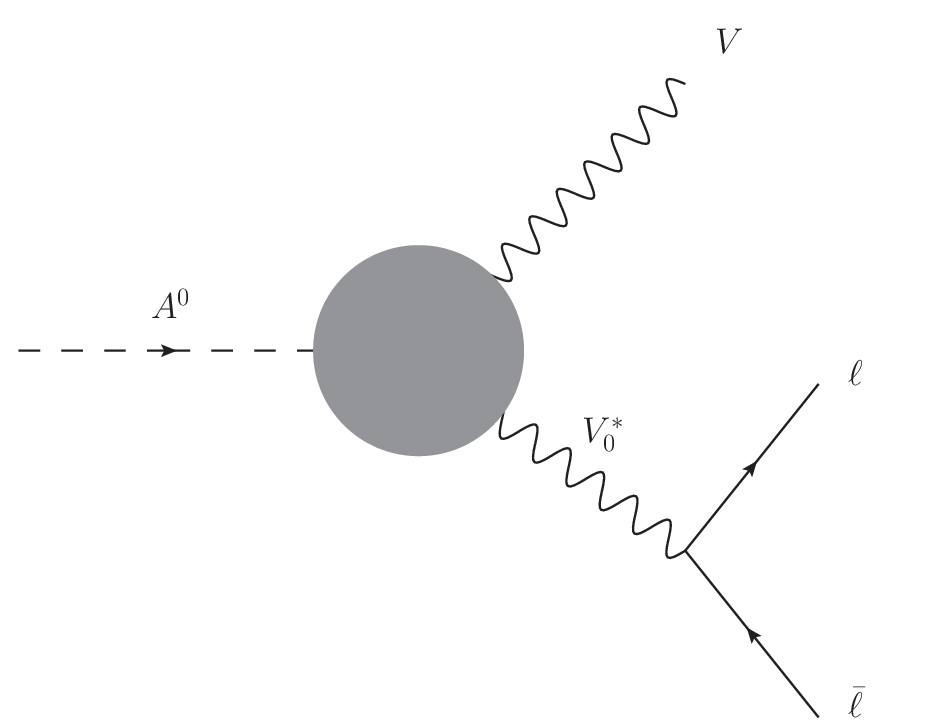}&
\includegraphics[width=6cm,height=5cm]
{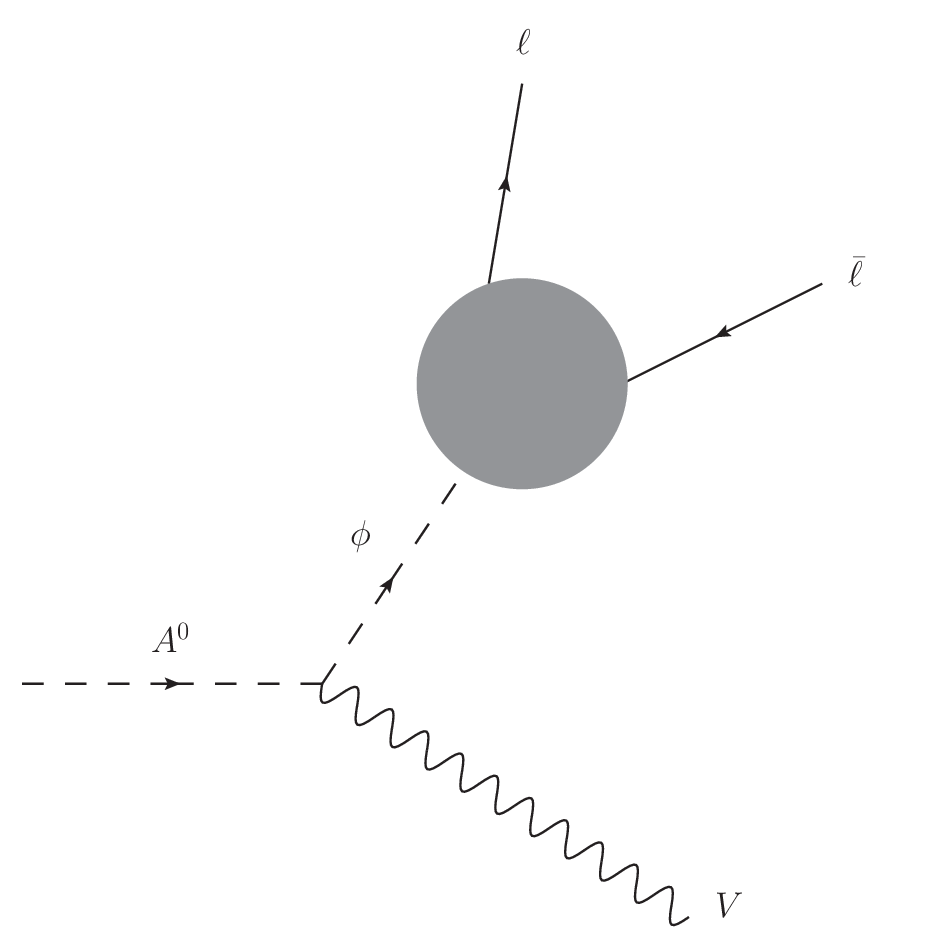}&
\includegraphics[width=4.5cm,height=5cm]
{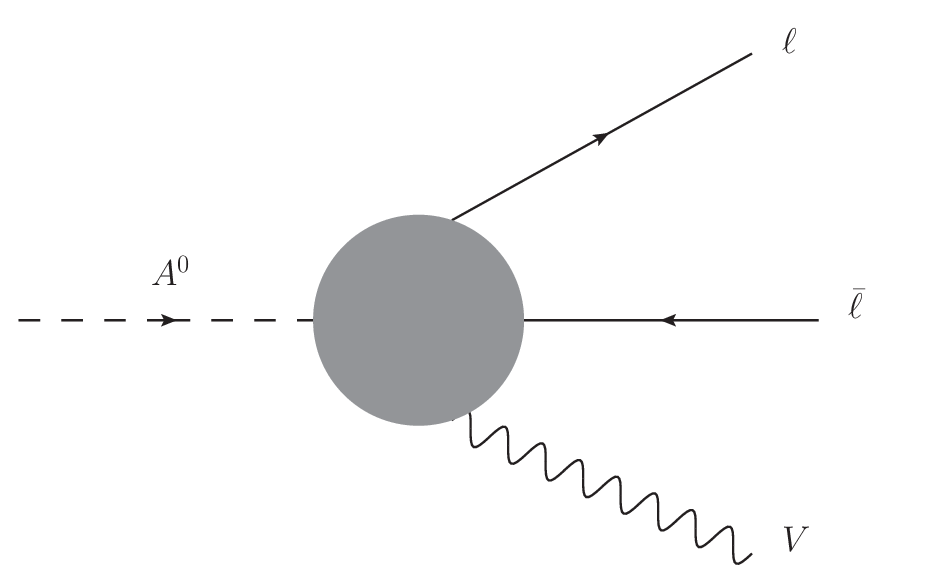}\\
(o) & (p) & (q)
\end{tabular}
\caption{\label{loopdiaglast}
One-loop
diagrams
contributing to the decay processes
$A^0 \rightarrow \ell \bar{\ell} V$ in HESM
within the HF gauge. They are
considered to be dominant
contributions to the processes. }
\end{figure}

\begin{figure}[H]
\centering
\includegraphics[width=8cm, height=6cm]
{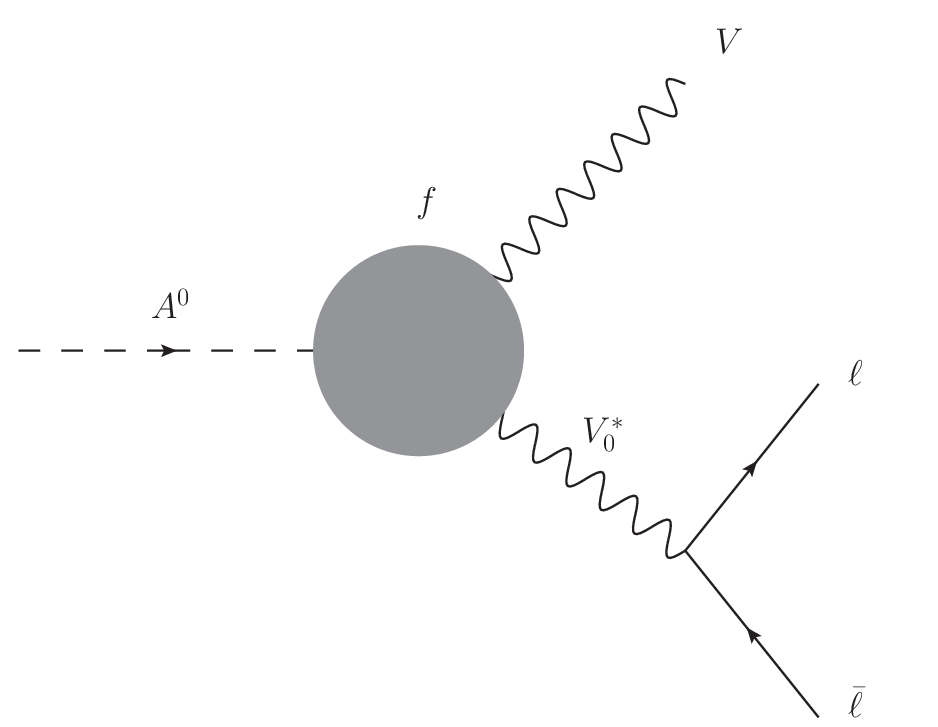}
\caption{\label{G1a} Group $1-$One-loop
Feynman diagrams $V_{0}^{*}$-poles contributing
to the processes. We denote that
$V_{0}^{*}$ can be $\gamma^*, Z^*$ in this
calculation.}
\end{figure}

\begin{figure}[H]
\centering
\includegraphics[width=8cm, height=6cm]
{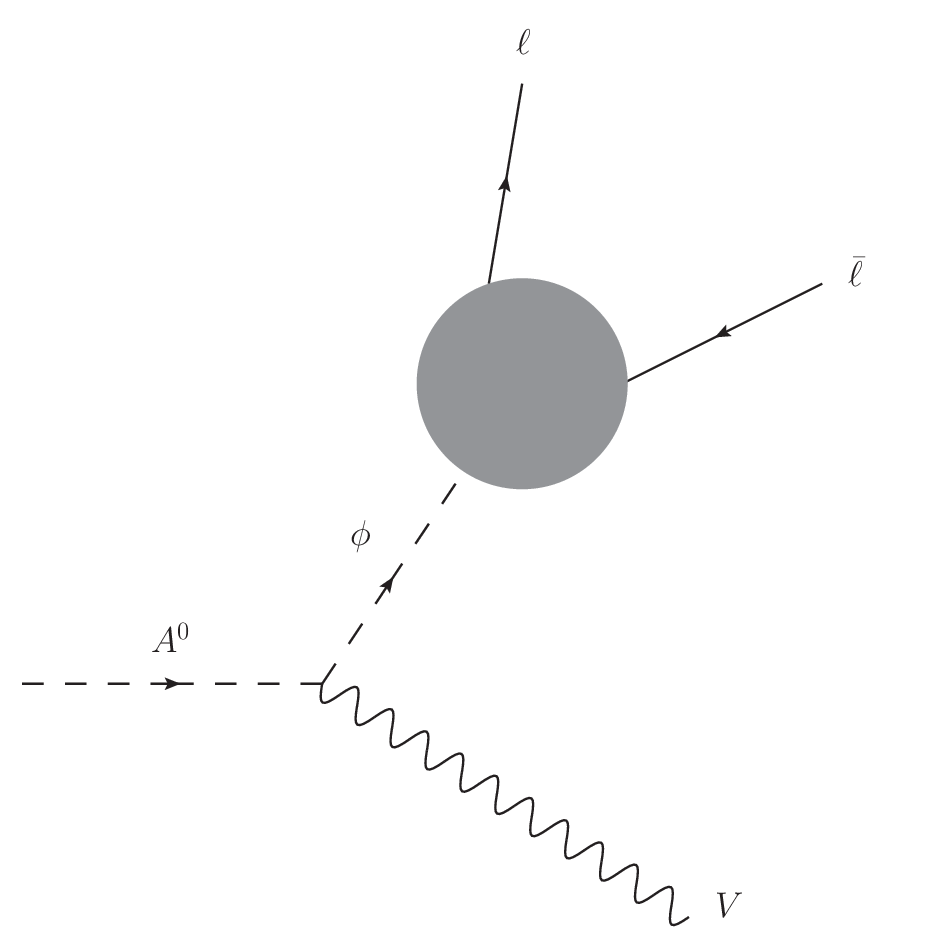}
\caption{\label{G2b} Group $2-$ One-loop
Feynman diagrams with $\phi$-poles including
the $W^\pm$ bosons and $Z$ boson exchanging 
in the loop.}
\end{figure}

\begin{figure}[H]
\centering
\includegraphics[width=15cm, height=5cm]
{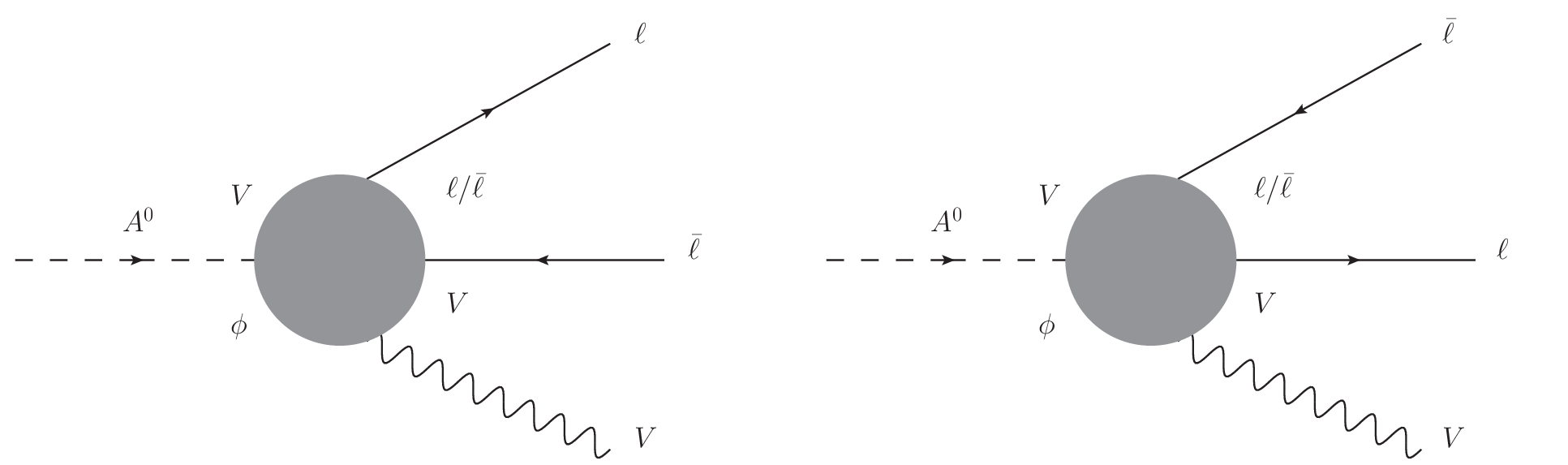}
\caption{\label{G3c} Group $3-$
One-loop box diagrams with exchanging 
$Z$ boson and $\phi$ in the loop.}
\end{figure}

%%%%%%%%%%%%%%%%%%%%%%%%%%%%%%%%%%%%%%%%%%%%%%%%%
In this section, the detailed evaluations for
one-loop contributing the decay
processes $A^0 \rightarrow \ell \bar{\ell} V$
in the HESM frameworks are presented.
At tree-level, one has two following
topologies contributing to the processes,
as plotted in Fig.~\ref{treediag}.
When $V$ becomes external photon, we
have only the first topology relating
to the decay channel. Noting that photon
can attach to both lepton and anti-lepton.
Therefore, we have two Feynman
diagrams accordingly in this case.
When $V$ becomes $Z$ boson, we have two
additional diagrams showing in the second
topology in Fig.~\ref{treediag}. Here,
$\phi$ can be the SM-like Higgs
as well as other CP-even Higgeses
in the HESM.
%%%%%%%%%%%%%%%%%%%%%%%%%%%%%%%%%%
%%%%%%%%%%%%%%%%%%%%%%%%%%%%%%%%
Working in on-shell
renormalization scheme,
we have listed
all topologies appear in
the decay channels within
't Hooft-Feynman (HF) gauge
at one-loop level.
In all the below topologies,
the first one-loop diagram
in Fig.~\ref{loopdiag1}~(a)
can be ignored, since there
isn't the tree-level couplings of
CP-odd Higgs to vector $W$, $Z$ bosons.
Within on-shell renormalization
conditions, the counterterm
diagram~Fig.~\ref{loopdiag1}~(b)
is included all field strength
renormalizablized constants for
external lines, the renormalizablized
coupling constant for $A^0 f\bar{f}$,
the renormalizabled mixing angle of
$A^0, G_0$, etc. However, this term is
expressed in terms of tree-level
amplitude and one-loop self-energies
from the renormalizabled constants.
Thus, the contribution is proportional
to $y_{\ell}
\times \mathcal{O}(\alpha)$
($y_{\ell}=m_{\ell}/v$ is the
Yukawa coupling).
As a result, this contribution
is assumed
to be small in comparison
with other ones and it can be omitted
in the present paper, as same course of
the works in Ref.~\cite{VanOn:2021myp,
Kachanovich:2020xyg,
Hue:2023tdz,Tran:2023vgk}.
The next diagrams in
Fig.~\ref{loopdiag1}~($c,d$)
as well as diagrams ($e,f$)
in Fig.~\ref{loopdiag2}
also can be
ignored in our current calculation
because their contributions are
proportional to
$y_{\ell}\times \mathcal{O}(\alpha)$.
Mixing of $A^0$ with $\phi$ (as plotted in
diagram $(g)$) and mixing of
$\phi$ with $V\equiv Z$ (depicted in diagram $(h)$)
are vanished as shown in the appendices $C, D$.
Self-energy diagrams in Fig.~\ref{loopdiag3}~($k, l$)
give zero contributions to the decay processes
(see appendix $D$ for detail). Furthermore,
mixing of $A^0$ with $V\equiv Z, \gamma$
will be canceled due to Slavnov-Taylor
identity~\cite{Aiko:2022gmz}.
%%%%%%%%%%%%%%%%%%%%%%%%%%%%%%%%
%%%%%%%%%%%%%%%%%%%%%%%%%%%%%%%%%%%%%%%%%%%%%%%%%%
%%%%%%%%%%%%%%%%%%%%%%%%%%%%%%%%%%%%%
\begin{figure}[H]
\centering
\begin{tabular}{c}
\includegraphics[width=11.5cm,height=5.5cm]
{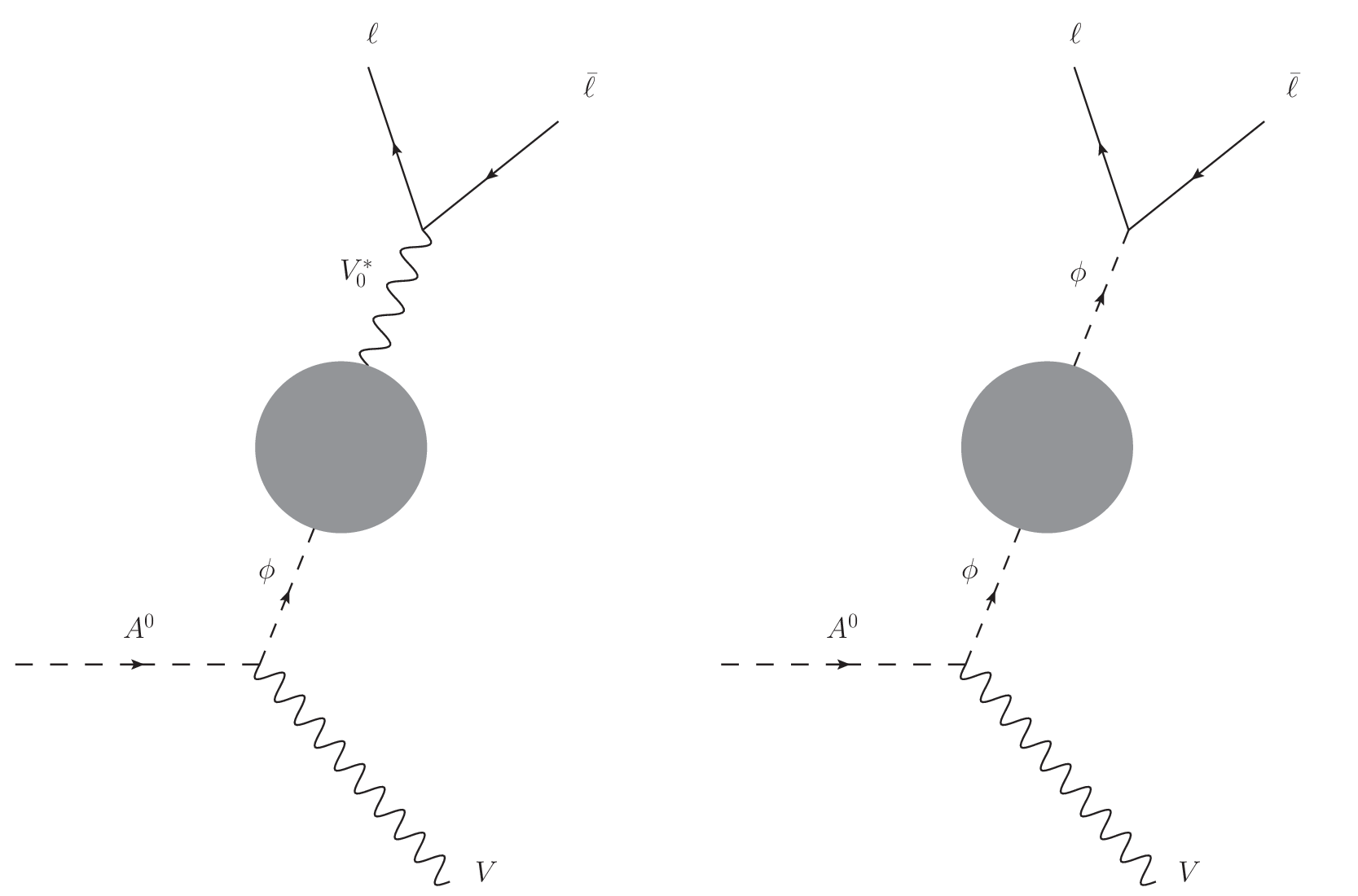}\\
(k) \hspace{5cm}(l)\\
\includegraphics[width=12cm,height=5cm]
{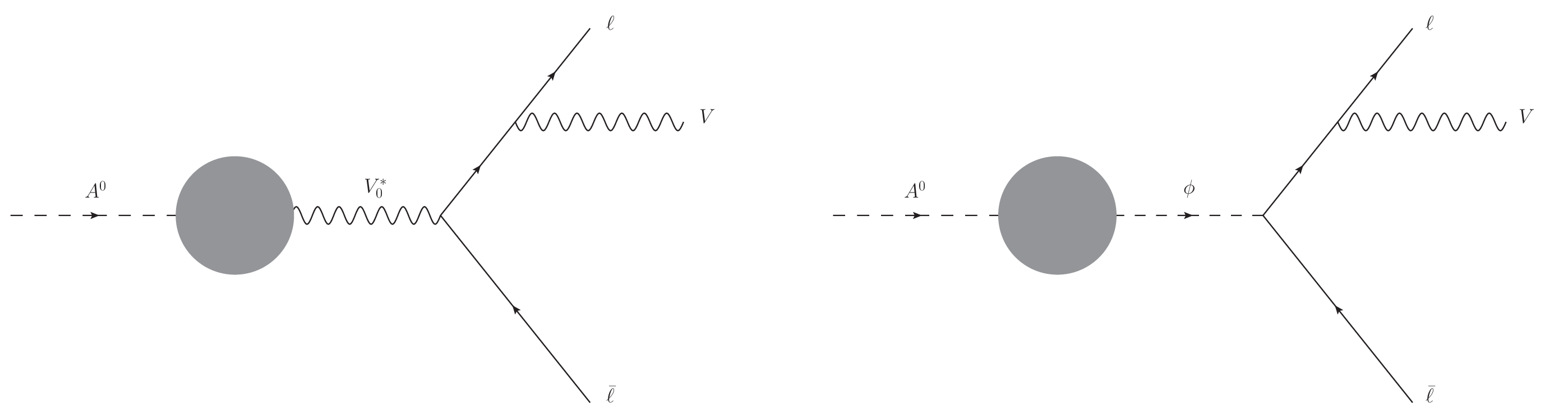}\\
(m) \hspace{5cm}(n)
\end{tabular}
\caption{\label{loopdiag3} One-loop
diagrams
contributing to the decay processes
$A^0 \rightarrow \ell \bar{\ell} V$ in HESM
within the HF gauge.}
\end{figure}
%%%%%%%%%%%%%%%%%%%%%%%%%%%%%%%%%%%%%
Lastly, we list all Feynman diagrams
contributing
to the processes under consideration
in Fig.~\ref{loopdiaglast}.  They are
considered to be dominant
contributions to the processes.
In the first topology ($o$),
CP-odd Higgs only couple to fermions,
we hence have fermion loop in this case.
The dominant contributions are
from top, bottom quarks and tau lepton
exchanging in the loop. The diagram
($p$), one takes into
account $W, Z$ propagating in the loop.
Finally, we have box diagram ($q$)
with $Z$ boson
exchanging in the loop
contributing to the processes.
%%%%%%%%%%%%%%%%%%%%%%%%%%%%%%%%%%%%%%
%%%%%%%%%%%%%%%%%%%%%%%%%%%%%%%%%
The total amplitude for the decay 
$A^0 (p) \rightarrow \ell (q_1) 
\bar{\ell} (q_2) V_\mu (q_3)$ is
expressed as the sum of
each group $G_i$ for $i=1, 2, 3$
listed in the following paragraphs
\begin{eqnarray}
\mathcal{A}_{\text{Total}} =
\mathcal{A}_{\text{Tree}}^{
A^0 \rightarrow \ell \bar{\ell} V }
+
\mathcal{A}_{\text{1-loop}}^{A^0 
\rightarrow \ell \bar{\ell} V}. 
\end{eqnarray}
The related kinematic variables 
for the processes are 
included as follows:
\begin{eqnarray}
p^2 = M_{A^0}^2,
\quad
q_1^2 = q_2^2 = m_\ell^2,
\quad
q_3^2 = M_V^2,
\quad
q_{jk} = (q_j + q_k)^2
\quad \text{for}
\quad j, k=1,2,3.
\end{eqnarray}
The above kinematic
variables follow
the relation: 
\begin{eqnarray}
q_{12} + q_{13} + q_{23}=
M_{A^0}^2
+
M_V^2
+ 
2 m_\ell^2
\backsimeq 
M_{A^0}^2
+
M_V^2.
\end{eqnarray}
First, tree-level amplitude for
the decay processes 
$A^0 \rightarrow \ell \bar{\ell} V$
can be expressed as follows:

%%%%%%%%%%%%%%%%%%%%%%%%%%%%%%%%%%%%
\begin{eqnarray}
\mathcal{A}_{\text{tree}
}^{A^0 \rightarrow \ell \bar{\ell} V}
&=&
\frac{
i\kappa_{A^0}^{\ell}
}{
(q_{13} - m_{\ell}^2)
(q_{23} - m_{\ell}^2)
}
\cdot
\frac{m_{\ell} }{v}
\cdot
u (q_1)
\Bigg\{
2
\big(
q_{13}
-
m_{\ell}^2
\big)
\Big(
\boldsymbol{a}_{V \ell \bar{\ell}}
-
\boldsymbol{v}_{V \ell \bar{\ell}}
\,
\gamma_5
\Big)
\;
p
\cdot
\epsilon^* (q_3)
\nonumber
\\
&&\hspace{2.4cm}
+
\Big[
\boldsymbol{a}_{V \ell \bar{\ell}}
\,
\big(
q_{23}
-
q_{13}
\big)
+
\boldsymbol{v}_{V \ell \bar{\ell}}
\,
\big(
q_{13}
+
q_{23}
-
2 m_{\ell}^2
\big)
\gamma_5
\Big]
\slashed p
\slashed \epsilon^*(q_3)
\Bigg\}
\bar{v} (q_2)
\nonumber
\\
&&
+
\sum\limits_{\phi = h^0, H_j}
i
\dfrac{2
m_{\ell}
}{q_{12} - M_\phi^2}
\,
g_{\phi A^0 V}
\;
\Big\{
u (q_1)
\Big[
\boldsymbol{v}_{\phi \, \ell \bar{\ell}}
+
\boldsymbol{a}_{\phi \, \ell \bar{\ell}}
\,
\gamma_5
\Big]
\bar{v} (q_2)
\Big\}
p
\cdot
\epsilon^* (q_3).
\end{eqnarray}
For CP-even Higgs boson
$\phi \equiv h^0, H_j$,
the axial-vector components
$\boldsymbol{a}_{\phi \,
\ell \bar{\ell}} =0$ for the
CP-conserving case. In the above
formulas, when $V$ or $V_0^*$ being
$Z$-boson, one has $M_V^2 = M_Z^2$,
$\Gamma_{V_0}=\Gamma_Z$ and
the involved couplings are replaced
by
\begin{eqnarray}
\boldsymbol{v}_{Z f \bar{f}}
=\frac{e}{s_{2W}} ( I_{3,f}
- 2 s_W^2 Q_f),
\quad
%%%%%%%%%
\boldsymbol{a}_{Z f \bar{f}} =
-\frac{e}{s_{2W}} I_{3,f}.
\end{eqnarray}
In the case of $V$ or $V_0^*$
being photon, one has $M_V^2 = 0$,
$\Gamma_{V_0}=0$ and
the corresponding couplings
read as
\begin{eqnarray}
\boldsymbol{v}_{\gamma f \bar{f}} &=& e Q_f,
\quad
%%%%%%%%%
\boldsymbol{a}_{\gamma f \bar{f}} = 0.
\end{eqnarray}
% %%%%%%%%%%%%%%%%%%%%%%%%%%%%%%%%%
% \subsection{One-loop amplitude}
% %%%%%%%%%%%%%%%%%%%%%%%%%%%%%%%%%
We next consider one-loop amplitude
which can be evaluated from
Fig.~\ref{loopdiaglast}.
These diagrams are divided into
three following groups in
Figs.~\ref{G1a},
\ref{G2b},~\ref{G3c}. In detail,
one-loop amplitude is calculated
as follows:
\begin{eqnarray}
\mathcal{A}_{\text{1-loop}}
=
\sum \limits_{i = 1}^3
\mathcal{A}_{G_i}^V.
\end{eqnarray}
Where $V$ can be external photon $\gamma$
and $Z$ boson. Each one-loop amplitude
can be derived in the following paragraphs.
It is noted that $V$ being  $\gamma$, we have
only one diagram in Fig.~\ref{G1a} attributing to
the amplitude. When $V$ become  external
$Z$ boson, we take into account two more
topologies in Figs.~\ref{G2b},~\ref{G3c}
contributing to the amplitude.

We first arrive at group $G_1$
(as depicted in Fig.~\ref{G1a})
in which
one-loop Feynman triangle diagrams
with $V_0^*$-pole. We denote that
$V_{0}^{*}$ can be $\gamma^*, Z^*$ in this
calculation. Since, there is no the coupling
of CP-odd Higgs with vector boson at
tree-level. We have only fermion
loop in this case.
%%%%%%%%%%%%%%%%%%%%%%%%%%%%%%%%%%
%%%%%%%%%%%%%%%%%%%%%%%%%%%%%%
Considering all
fermions $f$ exchanging in loop, 
one-loop amplitude reads as follows:
\begin{eqnarray}
\mathcal{A}_{G_1}^{(V)}
&=&
\sum \limits_{V_0^*
= \gamma^*, Z^*}
\sum \limits_{f}
% \Big\{
u (q_1)
\Big[
(
\boldsymbol{v}_{V_0^* \ell \bar{\ell}}
-
\boldsymbol{a}_{V_0^* \ell \bar{\ell}}
\,
\gamma_5
\Big)
\gamma^\mu
\Big]
\bar{v} (q_2)
% \Big\}
\frac{
F_{G_1,f}^{(V)}
\cdot
\big(
\varepsilon_{\mu \nu \rho \sigma} \,
p^\rho
q_3^\sigma
\big)
\cdot
\epsilon^{*,\nu}(q_3)
}{
\big(
q_{12} - M_{V_0}^2 
\big)
+ i M_{V_0} \Gamma_{V_0}
}.
\end{eqnarray}
In this formulas,
$\varepsilon_{\mu \nu \rho \sigma} $
anti-symmetry tensor
Levi-civita is taken into account. While
one-loop form factor $F_{G_1,f}^{(V)}$
is written in terms of scalar one-loop 
integrals in the form of
\begin{eqnarray}
\label{formfactorsG1}
F_{G_1,f}^{(V)}
&=&
i\frac{\kappa_{A^0}^f}
{2\pi^2} \dfrac{
N^C_f \; m_f^2}
{[M_{A^0}^4 - 2 M_{A^0}^2
(M_V^2+q_{12} )
+
(M_V^2-q_{12} )^2 ]}
\cdot
\frac{m_{f} }{v}
\times
\\
&&\times
\Bigg\{
2\; \boldsymbol{a}_{V_0^* f \bar{f}}
\;
\boldsymbol{a}_{V f \bar{f}}
\Big[
\big(
q_{12}-M_{A^0}^2-M_V^2
\big) 
B_0 (M_V^2,m_f^2,m_f^2)
\nonumber \\
&&\hspace{3cm}
+
\big(
M_V^2-M_{A^0}^2-q_{12}
\big) 
B_0 (q_{12},m_f^2,m_f^2)
+
2 M_{A^0}^2 
B_0 (M_{A^0}^2,m_f^2,m_f^2)
\Big]
\nonumber \\
&&\hspace{0.4cm}
+ 
\Big[
\boldsymbol{v}_{V_0^* f \bar{f}}
\,
\boldsymbol{v}_{V f \bar{f}} 
\Big(
2 M_{A^0}^2 
\big(M_V^2+q_{12}\big)
-
\big(M_V^2-q_{12}\big)^2
- 
M_{A^0}^4
\Big)
\nonumber \\
&&\hspace{2.2cm}
+
\boldsymbol{a}_{V_0^* f \bar{f}}
\,
\boldsymbol{a}_{V f \bar{f}}
\Big(
M_{A^0}^4
-
(M_V^2-q_{12})^2
\Big)
\Big]
C_0 (M_V^2,q_{12},M_{A^0}^2,m_f^2,m_f^2,m_f^2)
\Bigg\}.
\nonumber
\end{eqnarray}
%%%%%%%%%
We next consider one-loop amplitudes with
$\phi$-poles including
the $W^\pm$ bosons and $Z$ boson exchanging
in the loop (seen Fig.~\ref{G2b}). In this
case, we note $\phi$ for $h^0$ or new
heavy CP-even Higges $H_j$ in the considered
models.
Evaluating the processes in THDM and THM,
we have $H_j=H$.
%%%%%%%%%%%%%%%%%%%%%%%%%%%%%%%%%%%%%%%%%%%%%%%%%%
%%%%%%%%%%%%%%%%%%%
The amplitude is decomposed in form of
\begin{eqnarray}
\mathcal{A}_{G_2}^{V}
&=&
\sum\limits_{\phi=h^0, H_j}
\Big[
u (q_1)
\,
\mathbf{1}
\,
\bar{v} (q_2)
\Big]
\times
\Big[
\sum\limits_{P=W/ Z}
F_{G_2, P}^{(V)}
\Big] \times 
\frac{
[p\cdot \epsilon^*(q_3) ]
}{
\big(
q_{12}
- 
M_{\phi}^2 
\big)
}.
\end{eqnarray}
One-loop form factors are 
expressed as follows
\begin{eqnarray}
F_{G_2,W}^{(V)}
&=& -
\frac{i\; \alpha
}{2\pi\; s_W^2}
\frac{m_\ell
}{q_{12} - 4 m_\ell^2}
\cdot
g_{\phi A^0 V}
\cdot
g_{\phi W W}
\times
\\
&&
\hspace{-0.4cm}
\times
\Bigg[
B_0 (m_\ell^2,0,M_W^2)
-
B_0 (q_{12},M_W^2,M_W^2)
+
( M_W^2-m_\ell^2 )
C_0 (m_\ell^2,q_{12}, m_\ell^2, 
0, M_W^2,M_W^2)
\Bigg],
\nonumber
\\
%%%%%%%%%%%%%%%%%%%%%%%%%%%%%%%%%%
F_{G_2, Z}^{(V)}
&=&-
\frac{i
}{2\pi^2}
\frac{
m_\ell
}
{q_{12} - 4 m_\ell^2}
\cdot
g_{\phi A^0 V}
\cdot
g_{\phi Z Z}
\times
\\
&&
\hspace{-0.4cm}
\times
\Bigg\{
\Big[
(
\boldsymbol{a}_{Z \ell \bar{\ell}}
)^2
+
(
\boldsymbol{v}_{Z \ell \bar{\ell}}
)^2
\Big]
\Big[
B_0 (m_\ell^2,m_\ell^2,M_Z^2)
-
B_0 (q_{12},M_Z^2,M_Z^2)
\Big]
\nonumber
\\
&&
\hspace{-0.4cm}
+
\Big[
(
\boldsymbol{a}_{Z \ell \bar{\ell}}
)^2
(M_Z^2+q_{12}-6 m_\ell^2)
+
(
\boldsymbol{v}_{Z \ell \bar{\ell}}
)^2
(M_Z^2-q_{12}+2 m_\ell^2)
\Big]
C_0 (m_\ell^2,q_{12},m_\ell^2,
m_\ell^2,M_Z^2,M_Z^2)
\Bigg\}.
\nonumber
\end{eqnarray}

Finally, we concern the group $G_3$
(as Fig.~\ref{G3c}),
taking all one-loop Feynman box diagrams
which have
$\phi \equiv h^0, H_j$ with $H_j$ being
new heavy CP-even Higges in the mentioned
models and $V\equiv Z$ boson
propagating in the loop.
%%%%%%%%%%%%%%%%%%%%%%%%%%%%%%%%
%%%%%%%%%%%%%%%%%%%%%%%%%%%%%%%%%%%
The corresponding one-loop
amplitude can be written
in the form of
\begin{eqnarray}
\mathcal{A}_{G_3}^V
&=&
\sum\limits_{\phi=h^0, H_j}
u (q_1)
\Bigg\{
(
\boldsymbol{a}_{Z \ell \bar{\ell}}
)^2
+
(
\boldsymbol{v}_{Z \ell \bar{\ell}}
)^2
-
2 \,
\boldsymbol{a}_{Z \ell \bar{\ell}}
\,
\boldsymbol{v}_{Z \ell \bar{\ell}}
\; 
\gamma_5
\Bigg\}
\times
\\
&&\hspace{1.77cm} \times
\Bigg\{
F_{0, G_3}^{(V)}
\times
\slashed
\epsilon^*(q_3)
+
\sum \limits_{k = 1, 2}
F_{k, G_3}^{(V)}
\cdot
[
q_k
\cdot
\epsilon^*(q_3)
]
\times
\slashed p
\Bigg\}
\bar{v} (q_2).
\nonumber
\end{eqnarray}
The one-loop form
factors are expressed 
in terms of scalar
one-loop functions
as follows:
\begin{eqnarray}
F_{0, G_3}^{(V)}
&=&
-
\frac{i\;g_{\phi A^0 Z}
\cdot g_{\phi Z V}
}{16\pi^2}
\Bigg\{
C_0 (m_\ell^2,M_V^2,q_{13},
m_\ell^2,M_Z^2,M_\phi^2)
%%%%%%%%
\\
&&
\hspace{2.7cm}
+
M_Z^2\; D_0
(m_\ell^2,q_{12},M_V^2,
q_{13},m_\ell^2,M_{A^0}^2,
m_\ell^2,M_Z^2,M_Z^2,M_\phi^2)
\nonumber
\\
&&
\hspace{2.7cm}
+ 2\Big[
q_{13} - M_{A^0}^2
\Big]
\;
D_1(m_\ell^2,q_{12},M_V^2,
q_{13},m_\ell^2,M_{A^0}^2,
m_\ell^2,M_Z^2,M_Z^2,M_\phi^2)
\nonumber
\\
&&
\hspace{2.7cm}
+
q_{12} \;
D_2 (m_\ell^2,q_{12},M_V^2,
q_{13},m_\ell^2,M_{A^0}^2,
m_\ell^2,M_Z^2,M_Z^2,M_\phi^2)
\nonumber
\\
&&
\hspace{2.7cm}
+
\Big[
q_{13} + M_{A^0}^2
\Big]\;
D_3
(m_\ell^2,q_{12},M_V^2,
q_{13},m_\ell^2,M_{A^0}^2,
m_\ell^2,M_Z^2,M_Z^2,M_\phi^2)
\Bigg\}
\nonumber
\\
&&\hspace{0cm}
-\{ q_{13} \leftrightarrow
q_{23}\}-\text{terms},
\nonumber
\\
%%%%%%%%%%%%%%%%%%%%%%%%%%%%%%%%%%
F_{1, G_3}^{(V)}
&=&-
\frac{i\; g_{\phi A^0 Z}
\cdot
g_{\phi Z V}}{4\pi^2}
D_2 (m_\ell^2,q_{12},M_V^2,
q_{13},m_\ell^2,M_{A^0}^2,m_\ell^2,
M_Z^2,M_Z^2,M_\phi^2),
\\
%%%%%%%%%%%%%%%%%%%%%%%%%%%%%%%%%%
F_{2, G_3}^{(V)}
&=&
-
F_{1, G_3}^{(V)}
\,
\Big\{ q_{13} \, 
\leftrightarrow \, 
q_{23} \Big\}.
\end{eqnarray}
After collecting the one-loop form factors, we
check numerically for the calculation by verifying
all factors are ultraviolet (UV) and infrared
(IR) finiteness. One finds the results are good
stability. In this paper, we skip showing numerical
results of the test. Since the information
is not important for our discussion
(see our previous works for example of
the numerical test~\cite{VanOn:2021myp,
Hue:2023tdz,Tran:2023vgk}).
Having all correctness  one-loop
amplitudes, the decay rates are then
evaluated as follows:
\begin{eqnarray}
\Gamma_{A^0 \rightarrow \ell \bar{\ell} V}
&=&
\dfrac{1}{256 \pi^3 M_{A^0}^3}
\int \limits_{q_{12}^\text{min}}^{q_{12}^\text{max}}
d q_{12}
\int \limits_{q_{13}^\text{min}}^{q_{13}^\text{max}}
d q_{13}
\,
\sum \limits_\text{pol.}
\big|
\mathcal{A}_\text{Total}
\big|^2.
\end{eqnarray}
Where the total amplitude is
expressed as follows:
\begin{eqnarray}
\label{ampTOT}
 \sum \limits_\text{pol.}
\big|
\mathcal{A}_\text{Total}
\big|^2
= \sum \limits_\text{pol.}
\big|
\mathcal{A}_\text{Tree}
\big|^2
+
2
\sum \limits_\text{pol.}
\mathcal{R}e\Big\{
\mathcal{A}^*_\text{Tree}
\mathcal{A}_\text{Loop}
\Big\}
+
\sum \limits_\text{pol.}
\big|
\mathcal{A}_\text{Loop}
\big|^2.
\end{eqnarray}
The limitations of integration
are expressed for
the general vector boson $V$ in
final state as follows:
\begin{eqnarray}
q_{12}^\text{min} &=&  4 m_\ell^2,
\\
%%%%%%%%%%%%%%%%%
q_{12}^\text{max}
&=& (M_{A^0} - M_V)^2,
\\
%%%%%%%%%%%%%%%%%
q_{13}^\text{max(min)}
&=&
\dfrac{
M_{A^0}^2
+
M_V^2
+
2 m_\ell^2
-
q_{12}
}{2}
\pm
\frac{1}{2}
\sqrt{
\Big(
1
-
\dfrac{4 m_\ell^2}{q_{12}}
\Big)
\Big[
(
M_{A^0}^2
+
M_V^2
-
q_{12}
)^2
-
4 M_{A^0}^2
M_V^2
\Big]
}.
\nonumber\\
\end{eqnarray}
We comment that the last term
in Eq.~(\ref{ampTOT}) belong to
order of two-loop corrections.
However, as we show in later since
the interference between tree
and one-loop amplitude to be
very small contributions.
This behavior also holds
for the interference between
tree and two-loop amplitude.
With light fermions in final states
in scalar particle decays,
the loop amplitude is dominant contributions
as many previous works~\cite{Kachanovich:2020xyg,
VanOn:2021myp, Hue:2023tdz,Tran:2023vgk}.
Therefore, we also take
the last term in Eq.~(\ref{ampTOT})
into account for our analysis.
%%%%%%%%%%%%%%%%%%%%%%%%%%%%%%%%%%%%
\section{Phenomenological results}%%
%%%%%%%%%%%%%%%%%%%%%%%%%%%%%%%%%%%%
In the phenomenological results, we focus
on two typical HESMs in this paper such as
THDM and THM as examples.
For numerical results in this work,
we apply the input parameters as follows.
For the masses of gauge bosons and
their decay widths, one takes
$M_Z = 91.1876$ GeV, $\Gamma_Z  = 2.4952$ GeV,
$M_W = 80.379$ GeV, $\Gamma_W  = 2.085$ GeV.
The mass of the SM-like Higgs boson and its
total decay width are $M_{h^0} =125.1$ GeV,
$\Gamma_{h^0} =4.07\cdot 10^{-3}$ GeV.
For fermion sector, we use the input
parameters as $m_e =0.00052$ GeV,
$m_{\mu}=0.10566$ GeV and
$m_{\tau} = 1.77686$ GeV.
In the quark sector, their masses
are taken as $m_u= 0.00216$ GeV,
$m_d= 0.0048$ GeV,
$m_c=1.27$ GeV, $m_s = 0.93$ GeV,
$m_t= 173.0$ GeV, and $m_b= 4.18$ GeV.
The $G_{\mu}$-scheme is taken into
account for the following numerical
investigations. In this scheme,
the Fermi constant is treated as an
input parameter with the value $G_{\mu}
=1.16638\cdot 10^{-5}$ GeV$^{-2}$.
The electroweak constant is then
obtained subsequently:
\begin{eqnarray}
\alpha = \sqrt{2}/\pi G_{\mu}
M_W^2(1-M_W^2/M_Z^2)
= 1/132.184.
\end{eqnarray}
We note that the other input
parameters for generating the
decay rate and its distributions
are scanned appropriately in
the parameter space
corresponding to the models under
consideration.
%%%%%%%%%%%%%%%%%%%%%
\subsection{THDM} %%%
%%%%%%%%%%%%%%%%%%%%%%%%%%%%%%%%%%%%%
We first arrive at the phenomenological
analysis for THDM. In detail, we are going
to examine the decay rates of CP-odd Higgs
decay into $\ell \bar{\ell} V$ with
$V$ being photon and $Z$ boson.
Besides that, we also interested in
the differential decay widths with respect
to the invariant mass of lepton pair.
Before studying the detailed physical
results, we first review briefly
the current parameter space for THDM, e.g.
the theoretical and experimental constraints
on physical parameter space for THDM.
It is well-known that theoretical
constraints reply on tree-level unitarity
of the theory, vacuum stability conditions
for the scalar Higgs potential
as well as the requirements of
perturbative regime. These subjects
have examined in
Refs.~\cite{Nie:1998yn, Kanemura:1999xf,
Akeroyd:2000wc, Ginzburg:2005dt,
Kanemura:2015ska}. Following the above
constraints, the theoretical bounds
on the parameters
$\lambda_i$ for $i=1,2, \cdots, 5$
and $m_{11}, m_{12}, m_{22}$ have
given in the above references. In the
aspect of the experimental limitations,
the electroweak precision tests (EWPT)
for THDM have implicated at
LEP~\cite{Bian:2016awe, Xie:2018yiv}.
Furthermore, from the direct and indirect
searching for the masses of scalar particles
in THDM have performed at the LEP,
the Tevaron and the LHC as
summarized in~\cite{Kanemura:2011sj}.
In addition, implicating for
one-loop induced decays
of $h\rightarrow \gamma\gamma$
and $h\rightarrow Z\gamma$ in THDM
have performed in Refs.~\cite{Chiang:2012qz,
Benbrik:2022bol} and the references
therein. Combining all the above
constraints,
we can take logically the parameter
space for our analysis in
THDM as follows. We select
$126$ GeV $\leq M_H \leq 1000$ GeV,
$60$ GeV $\leq M_{A^0} \leq 1000$ GeV
and $80$ GeV $\leq M_{H^{\pm}}
\leq 1000$ GeV in the type I
and type X of THDM.
For the Type-II and
Y, we can scan consistently
the physical parameters as follows:
$500$ GeV $\leq M_H \leq 1000$ GeV,
$500$ GeV $\leq M_{A^0} \leq 1000$ GeV and
$580$ GeV $\leq M_{H^{\pm} }  \leq 1500$ GeV.
In both types, one takes
$2 \leq t_{\beta} \leq 20$,
$0.95\leq s_{\beta-\alpha} \leq 1$ for
the alignment limit of the SM (we take
$s_{\beta-\alpha}= 0.995$ for all below
physical results)
and $m_{12}^2 =M_H^2 s_\beta c_\beta$.
Furthermore, the constraints on $t_{\beta}$,
$M_{H^{\pm}}$ from flavor experimental data
have also performed for the THDM with
the softly broken $Z_2$ symmetry
in Ref.~\cite{Haller:2018nnx}.
In Ref.~\cite{Haller:2018nnx}, the
small values of $t_{\beta}$ are favoured in
explaining the flavor experimental data.
To complete our
discussions, we are also interested in
considering the small values of
$t_{\beta}$ (scan it
reasonably over the
region of $2 \leq t_{\beta} \leq 10$
for examples) in our work.
In our analysis, the decay widths
of SM-like Higgs is taken from
experimental value. While
the total decay width
of CP-even Higgs
$H$ is calculated at LO
as in~\cite{Kanemura:2022ldq}
(see Appendix $B$ for more detail).
We are going to present the
phenomenological
results for all decay processes
$A^0 \rightarrow
\ell \bar{\ell} \gamma,
\ell \bar{\ell} Z$ in THDM
in the following subsections.
We emphasize that we only
study phenomenological results
for THDM with types I, II because
there are same Yukawa couplings
in types I and X as well as types
II and Y.
%%%%%%%%%%%%%%%%%%%%%%%%%%%%%%%%
\subsubsection{Decay processes
$A^0 \rightarrow
\ell \bar{\ell} \gamma$ }%%%%%%%
%%%%%%%%%%%%%%%%%%%%%%%%%%%%%%%%
First, the physical results
for decay channels $A^0 \rightarrow
\ell \bar{\ell} \gamma$ in THDM are
presented. The effects of one-loop
contributing to the decay rates
are examined. The total decay widths
as a function of $M_{A^0}$ at
$t_{\beta} =5, 10$ are shown
in Fig.~\ref{DecayrateMA0}.
In the plots, we change
$150$ GeV $\leq M_{A^0}\leq 600$ GeV.
In the left panel figures, the
physical results for the THDM with
type I are shown at $t_{\beta} =5$
(in the above figure)
and $t_{\beta} =10$
(in the below figure), respectively.
While the similar data for the THDM
with type II is plotted in the right
panel figures. In the plots, the
black-dotted points are for total
decay rates and the red-squared points
show for tree-level decay widths.
While the triangle points with
green color present for one-loop the
decay widths and the blue-squared points
are the decay rates calculated
from the interference between tree
and one-loop amplitudes.
It is interested to observe that
the one-loop decay widths are dominant
in this case. In general, the decay rates
are proportional to $M_{A^0}$ and
$1/t_{\beta}$.
In all scatter plots, the small figures
in the corner of each plot
show for the decay widths computing
from the absolute value of the
interference between tree and one-loop
amplitudes. It is indicated that these
terms give a very small contributions
in comparison with other ones.
As a result,the contributions can be
omitted as we discussed in previous
section. From the data, the one-loop
effects on the decay rates are significance
attributions and they should be taken
into account for physical analysis.
%%%%%%%%%%%%%%%%%%%%%%%%%%%%%%%%%%
\begin{figure}[H]
\centering
\begin{tabular}{cc}
\hspace{-5.5cm}
$\Gamma$[KeV]
&
\hspace{-5.5cm}
$\Gamma$[KeV]
\\
\includegraphics[width=8cm, height=6cm]
 {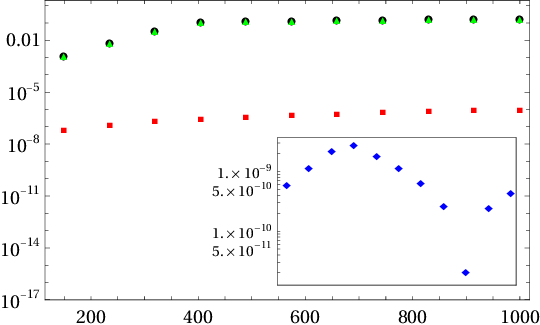}&
 \includegraphics[width=8cm, height=6cm]
 {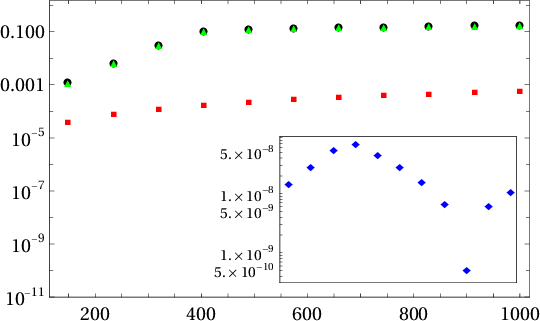}
 \\
\hspace{6cm}
$M_{A^0}$ [GeV]
&
\hspace{6cm}
$M_{A^0}$
[GeV]
\\
&
\\
\hspace{-5.5cm}
$\Gamma$[KeV]
&
\hspace{-5.5cm}
$\Gamma$[KeV]
\\
\includegraphics[width=8cm, height=6cm]
{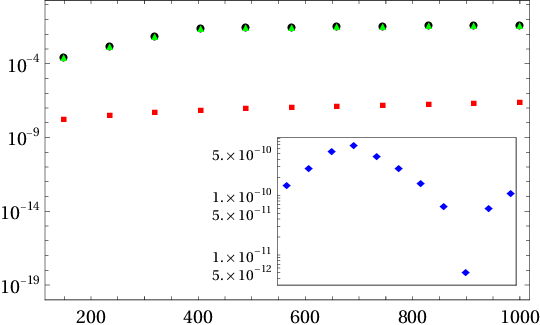}&
\includegraphics[width=8cm, height=6cm]
{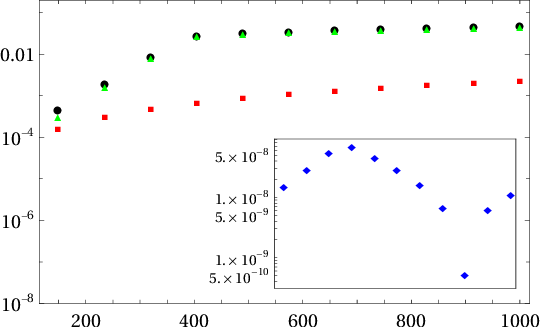}
\\
\hspace{6cm}
$M_{A^0}$ [GeV]
&
\hspace{6cm}
$M_{A^0}$ [GeV]
\end{tabular}
\caption{\label{DecayrateMA0}
Total decay rates for $A^0 \rightarrow
\ell \bar{\ell} \gamma$ in the THDM
with types I and II are presented.
In the left panel, we show the
decay rates at $t_{\beta}=5$
(the above figure)
and at $10$ (the below figure)
for the THDM with type I.
We present the decay rates at
$t_{\beta}=5, 10$ for the
THDM with type II in the right panel,
as same convention. In the plots,
we vary $150$ GeV $\leq M_{A^0}\leq 1000$
GeV.}
\end{figure}
%%%%%%%%%%%%%%%%%%%%%%%%%%%%%%%%

\begin{figure}[H]
\centering
\begin{tabular}{cc}
\hspace{-6cm}
$\frac{d\Gamma}{d m_{\ell\ell}}$
&
\hspace{-6cm}
$\frac{d\Gamma}{d m_{\ell\ell}}$
\\
\includegraphics[width=8cm, height=6cm]
{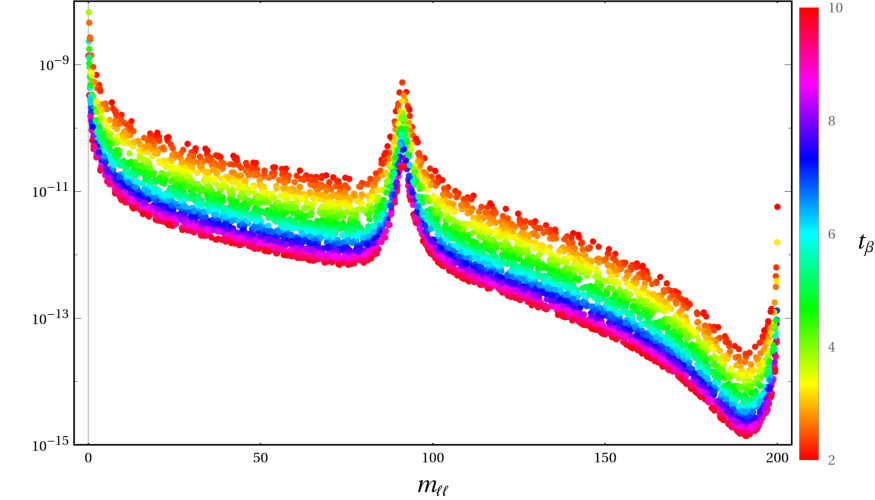}
&
\includegraphics[width=8cm, height=6cm]
{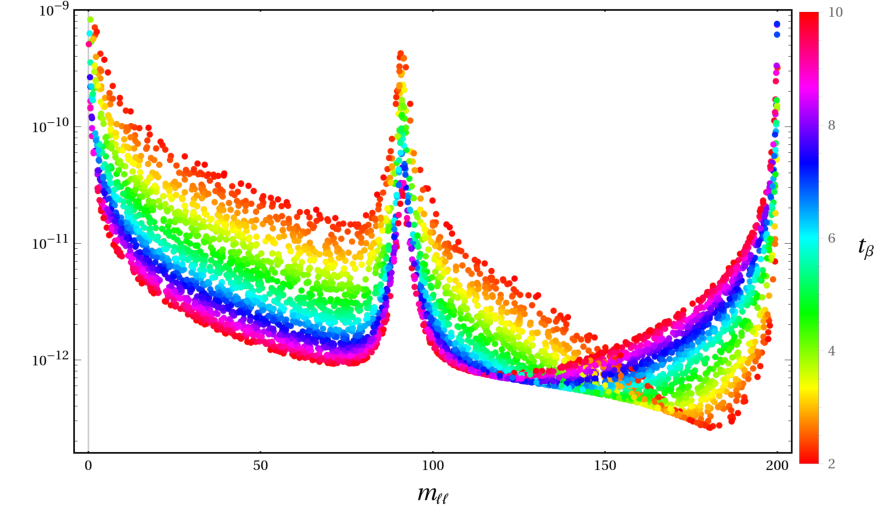}\\
\hspace{-6cm}
$\frac{d\Gamma}{d m_{\ell\ell}}$
&
\hspace{-6cm}
$\frac{d\Gamma}{d m_{\ell\ell}}$
\\
\includegraphics[width=8cm, height=6cm]
{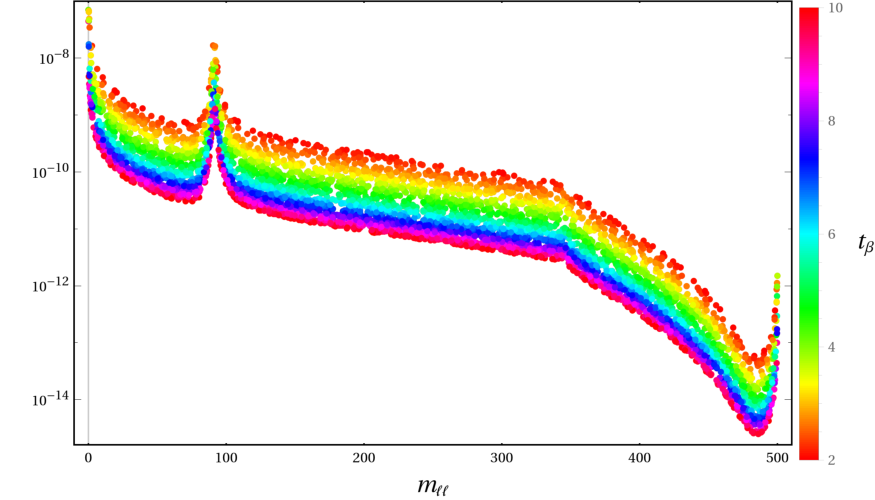}
&
\includegraphics[width=8cm, height=6cm]
{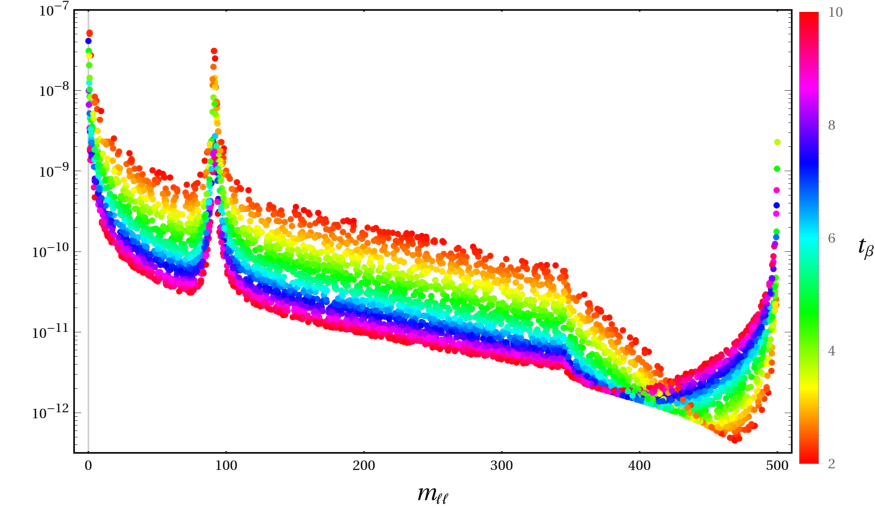}
\end{tabular}
\caption{\label{DecayMLL}
Differential decay rates with
respect to $m_{\ell\ell}$
for $A^0 \rightarrow
\ell \bar{\ell} \gamma$ as
function of $t_{\beta}$
in the THDM with types I and II.
In the left panel, we show the
decay rates at $M_{A_0} =200$
(above figure)
and at $500$ GeV (below figure)
for the THDM with type I.
We present the
decay rates at $M_{A_0} =200, 500$ GeV
for the
THDM with type II in the right panel,
as same convention.
In the plots, we vary
$2\leq t_{\beta} \leq 10$.}
\end{figure}

Differential decay rates with
respect to $m_{\ell\ell}$
for $A^0 \rightarrow
\ell \bar{\ell} \gamma$ as
function of $t_{\beta}$
in the THDM with types I and II
are shown in
Fig.~\ref{DecayMLL} for type I
(on the left panel) and for type II
(on the right panel). In the plots,
we set $M_{A_0} = 200$ GeV
for all the above figures and
$M_{A_0} = 500$ GeV for all
the below figures, respectively.
In the plots, we vary
$2\leq t_{\beta} \leq 10$.
Overall, we find that the decay decay
rates are decreased with $m_{\ell \ell}$.
There are two peaks observing
in the differential decay widths
which are
corresponding to $\gamma^*$-peak
and $Z^*$-peak. The decay rates
develop to the peaks
and are decrease rapidly
with $m_{\ell \ell}$ beyond the peaks.
Moreover, expression for one-loop
form factors in Eq.~(\ref{formfactorsG1})
indicates that the form factors are
largest at $q_{12}=m_{\ell \ell}^2
\rightarrow M_{A_0}^2$. In fact,
the one-loop factors
in Eq.~(\ref{formfactorsG1}) are taken
the form of
\begin{eqnarray}
F_{G_1,f}^{(\gamma)}
&\sim& \frac{1}{
[
M_{A^0}^4 - 2 M_{A^0}^2
(M_V^2+q_{12} )
+
(M_V^2-q_{12} )^2 ]|_{M_V^2=0}}
= \frac{1}{[q_{12} -M_{A^0}^2 ]^2}.
\end{eqnarray}
Subsequently,
one finds the peak of decay rates
at $m_{\ell \ell} \rightarrow M_{A_0}$.
The data is also shown that the decay
rates are inversely proportional to
$t_{\beta}$. At tree-level amplitude
in THDM with type II,
we find that the couplings of $A^0$ to
lepton pair
are proportional to $t_{\beta}$ as in Table
~\ref{YukawaKappa}.
At high mass region of $M_{A^0}$,
the decay rates of tree level
amplitude are more significance
contributions, as indicated in
previous data.
It explains for the reason that
decay rates are proportional
to $t_{\beta}$ in high mass
regions of $M_{A^0}$
in THDM with type II.
%%%%%%%%%%%%%%%%%%%%%%%%%%%%%%%%%%
%%%%%%%%%%%%%%%%%%%%%%%%%%%%%%%%
\subsubsection{Decay processes
$A^0 \rightarrow
\ell \bar{\ell} Z$ }%%%%%%%%%%%%
\begin{figure}[H]
\centering
\begin{tabular}{cc}
\hspace{-5.5cm}
$\Gamma$[KeV]
&
\hspace{-5.5cm}
$\Gamma$[KeV]
\\
\includegraphics[width=8cm, height=6cm]
 {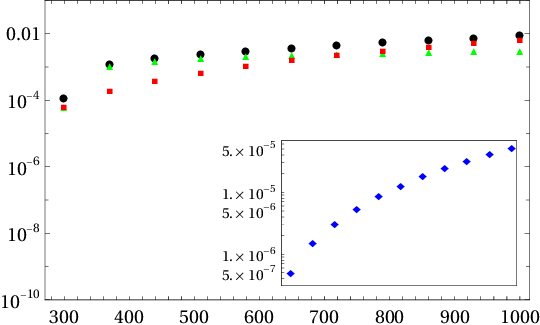}&
 \includegraphics[width=8cm, height=6cm]
 {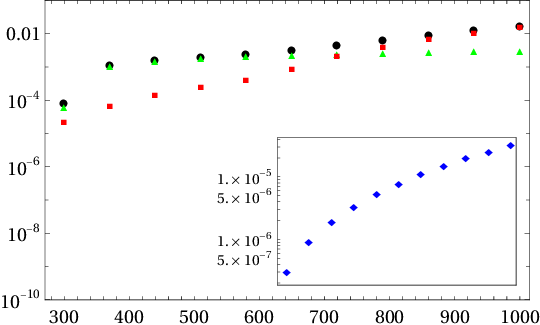}
 \\
\hspace{6cm}
$M_{A^0}$ [GeV]
&
\hspace{6cm}
$M_{A^0}$
[GeV]
\\
&
\\
\hspace{-5.5cm}
$\Gamma$[KeV]
&
\hspace{-5.5cm}
$\Gamma$[KeV]
\\
 \includegraphics[width=8cm, height=6cm]
 {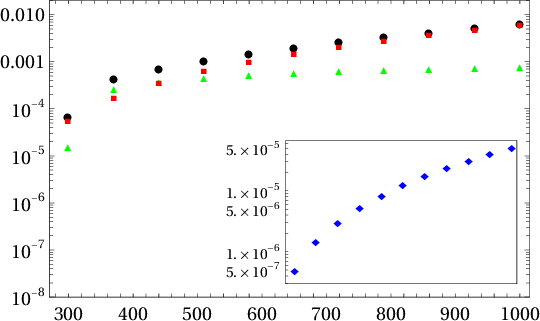}&
 \includegraphics[width=8cm, height=6cm]
 {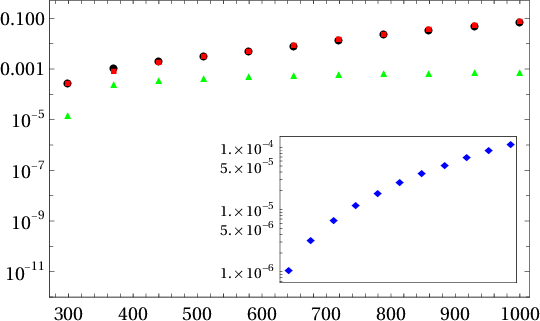}
 \\
\hspace{6cm}
$M_{A^0}$ [GeV]
&
\hspace{6cm}
$M_{A^0}$ [GeV]
\end{tabular}
\caption{\label{decayLLZ}
Total decay rates for $A^0 \rightarrow
\ell \bar{\ell} Z$ in the THDM with
types I and II are shown in the plots.
In the left panel, we show the
decay rates at
$t_{\beta}=5$ (above figure) and
at $10$ (below figure) for the
THDM with type I. We present the
decay rates at
$t_{\beta}=5, 10$ for the
THDM with type II in the right panel
as same notation.
In the plots, we vary
$300$ GeV $\leq M_{A^0}\leq 1000$ GeV
and take $M_H=500$ GeV.
To avoid the
numerical instability
of the results, one applies
a cut on
$q_{12}\geq q_{12}^{\text{min}}
= (M_{h^0}^2 + 1)$ GeV$^2$.}
\end{figure}

%%%%%%%%%%%%%%%%%%%%%%%%%%%%%%%%
We turn our attention to the
decay processes $A^0 \rightarrow
\ell \bar{\ell} Z$.
In Fig.~\ref{decayLLZ}, we present
the total decay rates of
$A^0 \rightarrow
\ell \bar{\ell} Z$ as functions of
$M_{A_0}$ at $t_{\beta}=5, 10$
for both types I and II of THDM.
The data points are presented
in these plots
following the same previous notations.
In all plots, we set $300$ GeV
$\leq M_{A^0} \leq 1000$ GeV.
We show the results for type I
of THDM on the left panel of
the figures at $t_{\beta}=5$
(the above plot) and  $t_{\beta}=10$
(the below plot). The same data
for type II of THDM is shown
in the right panel. In all plots,
the small figures in the right corner
of each figure,
we plot the decay rates of
the interference between tree amplitude
and one-loop amplitude. Again,
there contributions
are much smaller than other contributions
and can be ignored in this calculation.
To avoid the
numerical instability
of the results which are from the
narrow peak of SM-like Higgs boson, one
applies a cut on
$q_{12}\geq q_{12}^{\text{min}}
= (M_{h^0}^2 + 1)$ GeV$^2$.
Generally, one find that the decay rates
are proportional to $M_{A_0}$.
Difference from the channels
$A^0 \rightarrow
\ell \bar{\ell} \gamma$, in processes
$A^0 \rightarrow
\ell \bar{\ell} Z$,
we don't find that the decay rates depend
certainly on $t_{\beta}$ in type I
and are proportional to $t_{\beta}$ in
type II. We observe that one-loop
contributing for the decay rates are
significance and they should be taken
into account for physical analysis.
%%%%%%%%%%%%%%%%%%%%%%%%%%%%%%%%%%
%%%%%%%%%%%%%%%%%%%%%%%%%%%%%%%%

\begin{figure}[H]
\centering
\begin{tabular}{cc}
\hspace{-6cm}
$\frac{d\Gamma}{d m_{\ell\ell}}$
&
\hspace{-6cm}
$\frac{d\Gamma}{d m_{\ell\ell}}$
\\
\includegraphics[width=8cm, height=6cm]
 {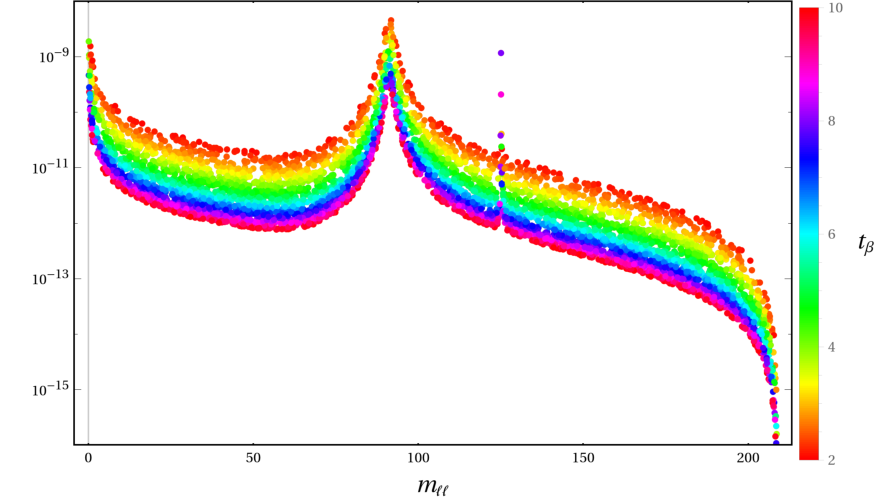}
 &
 \includegraphics[width=8cm, height=6cm]
 {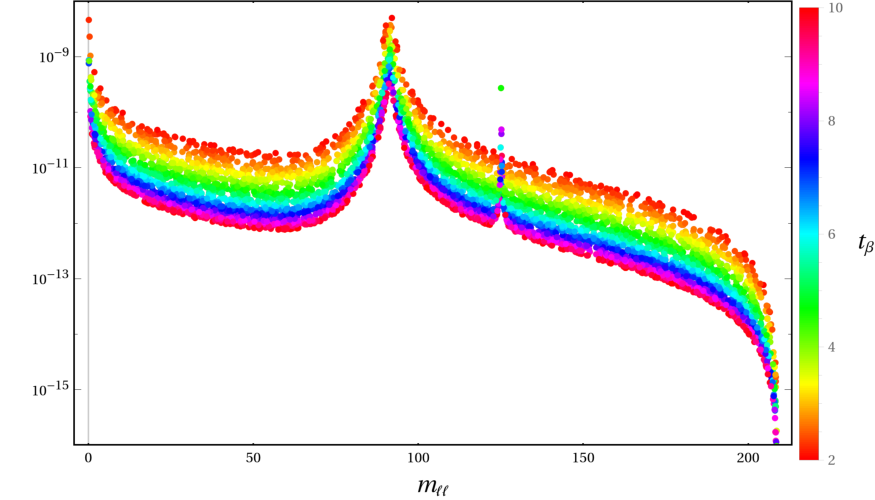}\\
 \hspace{-6cm}
$\frac{d\Gamma}{d m_{\ell\ell}}$
&
\hspace{-6cm}
$\frac{d\Gamma}{d m_{\ell\ell}}$
\\
 \includegraphics[width=8cm, height=6cm]
 {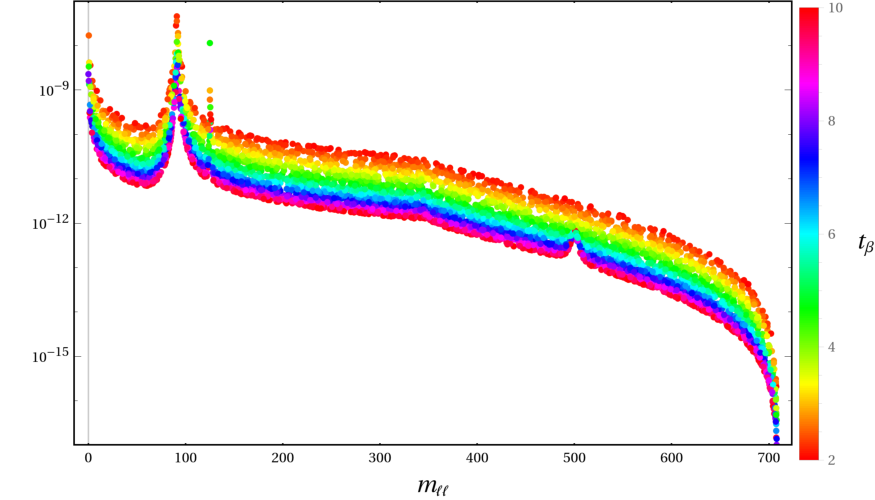}
 &
 \includegraphics[width=8cm, height=6cm]
 {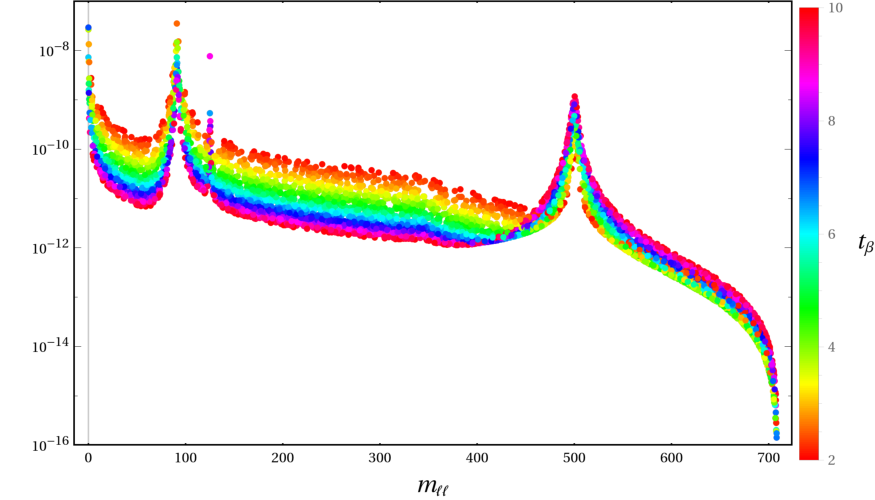}
\end{tabular}
\caption{\label{mllALLZ}
Differential decay rates with
respect to $m_{\ell\ell}$
for $A^0 \rightarrow
\ell \bar{\ell} Z$ as function of
$t_{\beta}$
in the THDM with types I and II
are shown in the plots.
In the left panel, we show the
decay rates at $M_{A_0} =300$
(above figure) and at $800$ GeV
(below figure)
for the THDM with type I.
We present the decay rates at
$M_{A_0} = 300, 800$ GeV for the
THDM with type II in the right
panel, as same convention.
In the plots, we vary
$2\leq t_{\beta} \leq 10$ and set
$M_H =500$ GeV.}
\end{figure}

We also interested in the
differential decay rates with
respect to $m_{\ell\ell}$
as shown in Fig.~\ref{mllALLZ}.
In the processes $A^0 \rightarrow
\ell \bar{\ell} Z$ we have two more
one-loop topologies
(Figs.~\ref{G2b},~\ref{G3c})
contributing to the channels.
In the left panel figures, the results
for THDM with type I are shown at
$M_{A_0}=300$ GeV (in above figure)
and at $M_{A_0}=800$ GeV (in below figure).
The corresponding results for THDM
with type II are presented in right
panel plots.
In general,
we observe
the same behave of the decay rates
in the case of $A^0 \rightarrow
\ell \bar{\ell} \gamma$, or
the decay rates
are generally decreased with
$m_{\ell \ell}$.
However, beyond the peaks of
$\gamma^*$-peak and $Z^*$-peak
we have two more peaks which are
corresponding to
$h^0$ and $H$ in this case.
In the case of
$M_{A_0}=300$ GeV, there aren't
peaks of $M_H$ since we set $M_H=500$ GeV
in the numerical analysis.
At $M_{A_0}=800$ GeV,
we observe very small peak at $m_{\ell\ell}
=500$ GeV in type I because the contribution
from $H$-pole is smaller than other ones.
In the case of type II, the attribution
from $H$-pole is enhanced by $t_{\beta}$.
As a result, this contribution is significance
in this case. Subsequently, we have a visible
peak of $m_{\ell\ell} = M_H=500$ GeV.

At high mass region of $M_{A^0}$,
the decay rates of tree level
amplitude are more significance
contributions which are proportional
to $t_{\beta}$ in type II.
At $M_{A^0}=800$ GeV,
one finds that
decay rates are proportional
to $t_{\beta}$ in high mass
regions for $m_{\ell\ell}
\geq \sim 450$ GeV
in THDM with type II.
%%%%%%%%%%%%%%%%%%%%%%%%%%%%%%%%%%
%%%%%%%%%%%%%%%%%%%
\subsection{THM}%%%
%%%%%%%%%%%%%%%%%%%
We next consider the phenomenological
results for the THM in this paper. The
constraints on the physical parameters
in the THM by including the theoretical
constraints as well as experimental
data are first reviewed and the subjects
have studied in
Refs.~\cite{Chun:2012jw,Chen:2013dh,
Arhrib:2011uy, Arhrib:2011vc,Akeroyd:2012ms,
Akeroyd:2011zza,Akeroyd:2011ir,Aoki:2011pz,
Kanemura:2012rs,Chabab:2014ara,Han:2015hba,
Chabab:2015nel,Ghosh:2017pxl,
Ashanujjaman:2021txz,Zhou:2022mlz}.
Combining all of the current constraints,
we take the parameters for
the THM logically as follows:
$c_{\alpha}=0.995$ for
the alignment limit of the SM,
varying $2\leq t_{\beta^{\pm}} \leq 10$
and setting $M_H =500$ GeV. The decay
widths of CP-even Higgs $H$ is calculated
as in
Eq.~(\ref{GMAH}) with
following coefficient couplings as
$\kappa_H^{f}= c_{\alpha}$,
$\kappa_H^{V}
= c_{\alpha}\; c_{\beta^{\pm}}
+\sqrt{2} s_{\alpha}\;
s_{\beta^{\pm}}$.
%%%%%%%%%%%%%%%%%%%%%%%%%%%%%%%%
\subsubsection{Decay processes
$A^0 \rightarrow
\ell \bar{\ell} \gamma$ }%%%%%%%
\begin{figure}[H]
\centering
\begin{tabular}{cc}
\hspace{-5.5cm}
$\Gamma$[KeV]
&
\hspace{-5.5cm}
$\Gamma$[KeV]
\\
\includegraphics[width=8cm, height=6cm]
{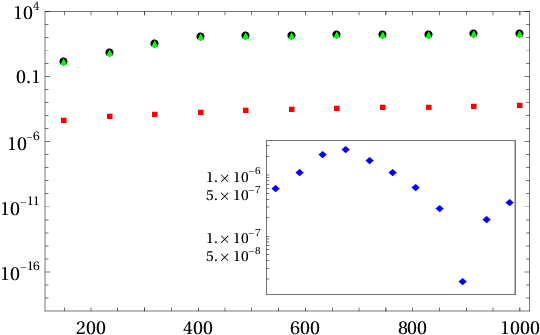}
 &
 \includegraphics[width=8cm, height=6cm]
 {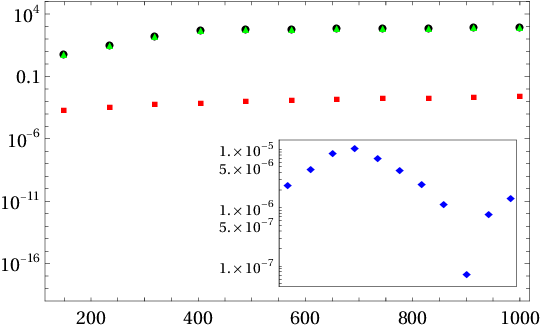}
\\
\hspace{6cm}
$M_{A^0}$ [GeV]
&
\hspace{6cm}
$M_{A^0}$ [GeV]
\end{tabular}
\caption{\label{DecayrateMA0THM}
Total decay rates for $A^0 \rightarrow
\ell \bar{\ell} \gamma$ in the THM.
In the left panel, we show the
decay rates at $t_{\beta}=5$.
We present the
decay rates at $t_{\beta}= 10$ in
the right panel. In the plots, we vary
$150$ GeV $\leq M_{A^0}\leq 1000$ GeV.}
\end{figure}

%%%%%%%%%%%%%%%%%%%%%%%%%%%%%%%%
Total decay rates for $A^0 \rightarrow
\ell \bar{\ell} \gamma$ are shown
as functions of $M_{A_0}$ in
Fig.~\ref{DecayrateMA0THM}.
In the left (and right) figure, the results
for $t_{\beta^\pm}=5$ (and $t_{\beta^\pm}=10$)
are plotted, respectively. In these plots,
the red-dotted points present for tree
decay rates and the blacked-dotted
points are for the total decay rates.
While the green-dotted points show
for the decay rates from one-loop
amplitude. The small figures in the corner of
the corresponding plots are the absolute value
of decay rates evaluating from the interference
tree and one-loop amplitudes.
The results of the interference tree and
one-loop amplitudes can be omitted in
this case.
In general, the total
decay rates are proportional to
$M_{A^0}$ and are
sightly proportional to
$t_{\beta^\pm}$.
%%%%%%%%%%%%%%%%%%%%%%%%%%%%%%%%%%
%%%%%%%%%%%%%%%%%%%%%%%%%%%%%%%%
We next concern the differential
decay widths with respect
to $m_{\ell \ell}$ for channels
$A^0 \rightarrow
\ell \bar{\ell} \gamma$ in the THM.
In the left figure, we show the results
for $M_{A_0}=300$ GeV. While the results
for $M_{A_0}=800$ GeV are plotted
in the right figure. In all cases,
we vary $2 \leq t_{\beta^{\pm}} \leq 10$.
In general, we find that the decay
rates are decreased with $m _{\ell \ell}$
and proportional to $t_{\beta^{\pm}}$.
We observe two peaks of decay rates
which are from $\gamma^*$-pole and
$Z^*$-pole. The last peak
of decay rates at
$m_{\ell \ell}= M_{A_0}$
because
one-loop form factors in
Eq.~(\ref{formfactorsG1}) are dominant
when $q_{12}=m_{\ell \ell}^2
\rightarrow M_{A_0}^2$.
%%%%%%%%%%%%%%%%%%%%%%%%%%%%%%%%%%
\begin{figure}[H]
\centering
\begin{tabular}{cc}
\hspace{-6cm}
$\frac{d\Gamma}{d m_{\ell\ell}}$
&
\hspace{-6cm}
$\frac{d\Gamma}{d m_{\ell\ell}}$
\\
\includegraphics[width=8cm, height=6cm]
 {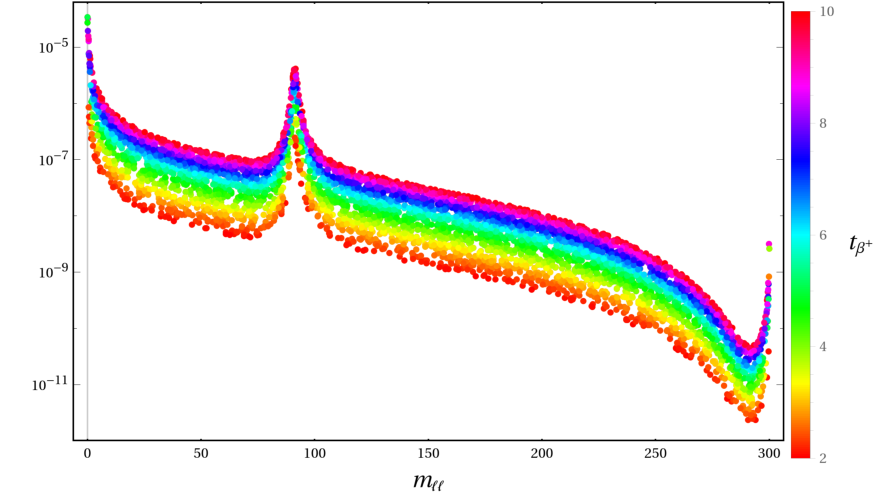}
 &
 \includegraphics[width=8cm, height=6cm]
 {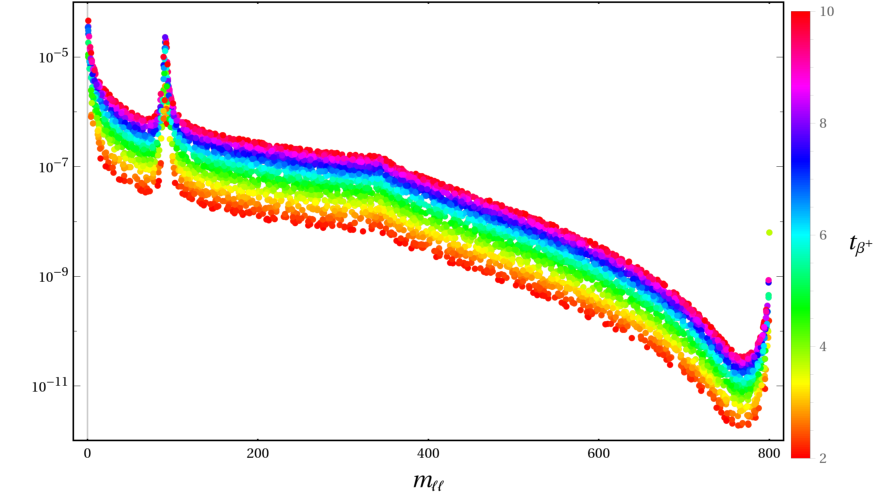}
\end{tabular}
\caption{\label{mllTHM}
Differential decay rates with
respect to $m_{\ell\ell}$
for $A^0 \rightarrow
\ell \bar{\ell} \gamma$ as
function of
$t_{\beta^{\pm}}$ in the THM.
In the left panel, we show the
decay rates at $M_{A_0} =300$ GeV.
We present the
decay rates at $M_{A_0} =800$ GeV
for the THM in the
right panel.
In the plots, we vary
$2\leq t_{\beta} \leq 10$.}
\end{figure}
%%%%%%%%%%%%%%%%%%%%%%%%%%%%%%%%
\subsubsection{Decay processes
$A^0 \rightarrow
\ell \bar{\ell} Z$ }%%%%%%%%%%%%
%%%%%%%%%%%%%%%%%%%%%%%%%%%%%%%%
Total decay rates for the processes
$A^0 \rightarrow
\ell \bar{\ell} Z$ in THM are presented.
In the left panel, we show the
decay rates at $t_{\beta^+}=5$.
We present the
decay rates at $t_{\beta^+}=10$ in
the right panel. In these plots, we vary
$150$ GeV $\leq M_{A^0}\leq 1000$ GeV.
To avoid the numerical instability
of the results which are from
the narrow peak of the SM-like
Higgs boson, one applies a cut on
$q_{12}\geq q_{12}^{\text{min}}
= (M_{h_0}^2 +1)$ GeV$^2$.
In all plots, we use the same notations
as previous cases.
Two small figures in the corner
of these plots are shown for decay
rates of the interference between
tree and one-loop amplitudes.
The results indicate that these
contributions are very small
in comparison with other attributions
and they can be ignored in this
computation. In general, the total
decay rates are proportional to
$M_{A^0}$ and
sightly proportional to
$t_{\beta^\pm}$.
%%%%%%%%%%%%%%%%%%%%%%%%%%%%%%%%%%
\begin{figure}[H]
\centering
\begin{tabular}{cc}
\hspace{-5.5cm}
$\Gamma$[KeV]
&
\hspace{-5.5cm}
$\Gamma$[KeV]
\\
\includegraphics[width=8cm, height=6cm]
{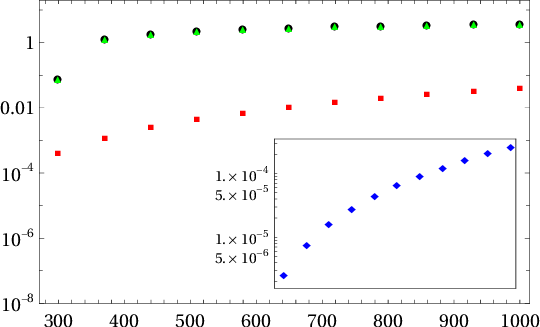}
 &
 \includegraphics[width=8cm, height=6cm]
 {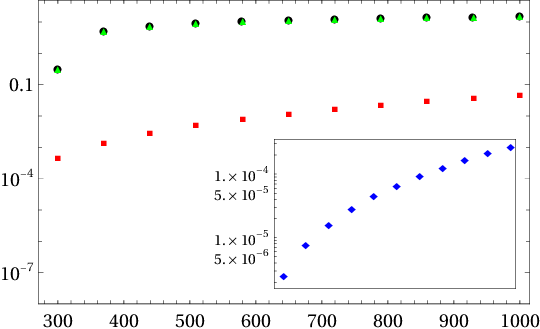}
 \\
\hspace{6cm}
$M_{A^0}$ [GeV]
&
\hspace{6cm}
$M_{A^0}$ [GeV]
\end{tabular}
\caption{\label{DecayrateMA0}
Total decay rates for $A^0 \rightarrow
\ell \bar{\ell} Z$ in the THM.
In the left panel, we show the
decay rates at $t_{\beta^+}=5$.
We present the
decay rates at $t_{\beta^+}=10$ in
the right panel. In the plots, we vary
$300$ GeV $\leq M_{A^0}\leq 1000$ GeV.
To avoid the numerical instability
of the results, one applies a cut on
$q_{12}\geq q_{12}^{\text{min}}
= (M_{h_0}^2 +1)$ GeV$^2$. }
\end{figure}
%%%%%%%%%%%%%%%%%%%%%%%%%%%%%%%%

Differential decay rates with respect
to $m_{\ell \ell}$ are presented
in Fig.~\ref{mllTHM}.
In the left panel, we show the
decay rates at $M_{A_0} =300$ GeV.
We present the
decay rates at $M_{A_0} =800$ GeV
for the THM in the
right panel.
In the plots, we vary
$2\leq t_{\beta^+} \leq 10$.
In general, we find that the decay
rates are decreased with $m _{\ell \ell}$
and proportional to $t_{\beta^{\pm}}$.
One find two peaks corresponding to
$\gamma^*$-peak and $Z^*$-peak in these plots.
Furthermore, we also observe a narrow peak
of $h^0$. The peak from $H$ give small
contribution
and it is invisible in these plots.
As same previous cases, the decay rates
develop to the peaks and decrease rapidly
beyond the peaks.
%%%%%%%%%%%%%%%%%%%%%%%%%%%%%%%%
\begin{figure}[H]
\centering
\begin{tabular}{cc}
\hspace{-6cm}
$\frac{d\Gamma}{d m_{\ell\ell}}$
&
\hspace{-6cm}
$\frac{d\Gamma}{d m_{\ell\ell}}$
\\
\includegraphics[width=8cm, height=6cm]
 {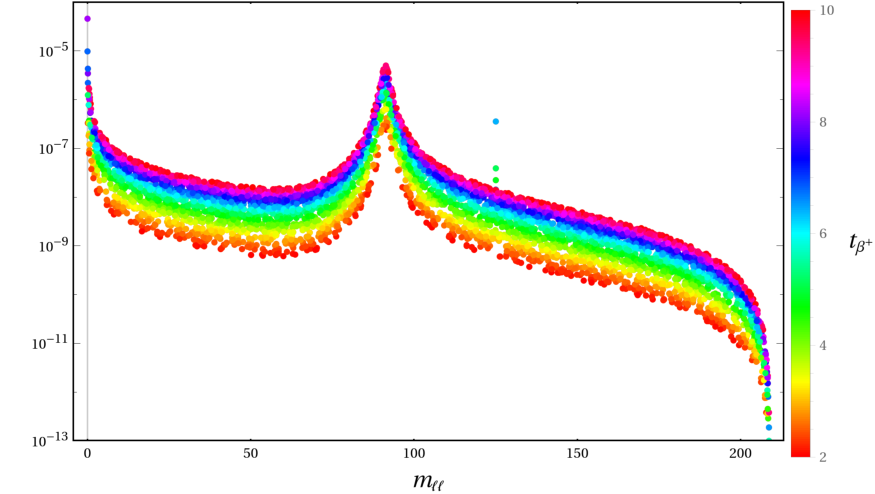}
 &
 \includegraphics[width=8cm, height=6cm]
 {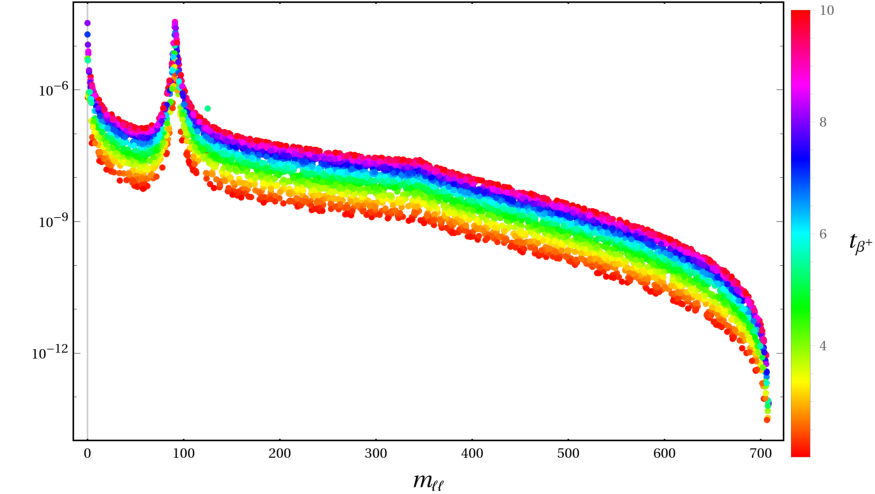}
\end{tabular}
\caption{\label{mllTHM}
Differential decay rates with
respect to $m_{\ell\ell}$
for $A^0 \rightarrow
\ell \bar{\ell} Z$ as
function of
$t_{\beta^{\pm}}$ in the THM.
In the left panel, we show the
decay rates at $M_{A_0} =300$ GeV.
We present the
decay rates at $M_{A_0} =800$ GeV
for the THM in the
right panel.
In the plots, we vary
$2\leq t_{\beta^+} \leq 10$.}
\end{figure}
%%%%%%%%%%%%%%%%%%%%%%%%%%%%%
\section{Conclusions} %%%%%%%
%%%%%%%%%%%%%%%%%%%%%%%%%%%%%
In this work, we have presented 
a general one-loop formulas for 
the decay of CP-odd Higgs $A^0
\rightarrow \ell \bar{\ell} V$
with $V\equiv \gamma, Z$
within Higgs Extension Standard Models,
including two higgs doublet model with
a complex scalar, two higgs doublet model
as well as triplet
higgs model. Analytic results are
expressed in terms of
PV-functions following the standard
notations of {\tt LoopTools}. As a result,
physical results can be generated
numerically by using the package.
In phenomenological results, we have
studied the decay rates of CP-odd Higgs
and the differential decay widths with
respect to the invariant mass of
lepton pair for all
decay channel
$A^0 \rightarrow \ell \bar{\ell} \gamma,
\ell \bar{\ell} Z$ in THDM and THM.
One finds
that one-loop contributing to the decay
rates and the differential decay rates
are significance contributions
and they should be taken into account
at future colliders.
\\

\noindent
{\bf Acknowledgment:}~
This research is funded by 
Vietnam National Foundation 
for Science and Technology 
Development (NAFOSTED) under 
the grant number
$103.01$-$2023.16$.
K. H. Phan and D. T. Tran 
express their gratitude to all 
the valuable support from 
Duy Tan University, for the 
30th anniversary of establishment 
(Nov. 11, 1994 - Nov. 11, 2024) 
towards "Integral, 
Sustainable and Stable Development".

%%%%%%%%%%%%%%%%%%%%%%%%%%%%%%%%%%%%
\section*{Appendix A: The effective 
Lagrangian for STHDM}
%%%%%%%%%%%%%%%%%%%%%%%%%%%%%%%%%%%%
In this Appendix, we derive effective
Lagrangian containing all the related
couplings to the processes under
consideration in STHDM.
From the kinetic terms, we have
%%%%%%%%%%%%%%%%%
\begin{eqnarray}
\mathcal{L}_{K}
&=& (D_\mu{\Phi_1})^\dagger(D_\mu{\Phi_1})
+ (D_\mu{\Phi_2})^\dagger(D_\mu{\Phi_2})
\\
&\supset&
-\frac{e}{2s_Wc_W}
(c_{\beta}\mathcal{O}_{2j}
- \mathcal{O}_{1j}s_{\beta})
(A^0Z_{\mu}\partial^{\mu}H_j
-H_jZ_{\mu}\partial^{\mu}
A^0)
\nonumber
\\
&&+\frac{e}{2s_Wc_W}
(-s_{\beta}\mathcal{O}_{12}
+ c_{\beta}\mathcal{O}_{22})
(A^0Z_{\mu}\partial^{\mu}h^0
- h^0Z_{\mu}\partial^{\mu}A^0)
\nonumber\\
&&
+
\frac{1}{2}\frac{e^2v}{s_W^2c_W^2}
[c_{\beta}\mathcal{O}_{1j}
+ s_{\beta}\mathcal{O}_{2j}]
Z^{\mu}Z_{\mu}H_j
+
\frac{eM_W}{s_W}
[c_{\beta}\mathcal{O}_{1j}
+s_{\beta}\mathcal{O}_{2j}]
W^{\pm}_{\mu}W^{\mp,\mu}H_j
\nonumber\\
&&
+
\frac{1}{2}\frac{e^2v}{s_W^2c_W^2}
[c_{\beta}\mathcal{O}_{12}
+ s_{\beta}\mathcal{O}_{22}]
Z^{\mu}Z_{\mu}h^0
+
\frac{eM_W}{s_W}
[c_{\beta}\mathcal{O}_{12}
+s_{\beta}\mathcal{O}_{22}]
W^{\pm}_{\mu}W^{\mp,\mu}h^0.
 \end{eqnarray}
%%%%%%%%%%%%%%%%%%%%%%%%%%%%%%%%%%
Where the rotation matrix of
mixing all neutral CP-even
Higgses is shown as below:
%%%%%%%%%%%%%%%%%%%%%%%%%%%%%%%%%%
\begin{eqnarray}
\mathcal{O}
&=& \left(
\begin{array}{ccc}
c_{13}c_{12}
& -(c_{23}s_{12}+s_{23}s_{13}c_{12})
& s_{23}s_{12}-c_{23}s_{13}c_{12}\\
c_{13}s_{12}
& c_{23}c_{12}-s_{23}s_{13}s_{12}
& -(s_{23}c_{12}+c_{23}s_{13}s_{12}) \\
s_{13}
& s_{23}c_{13}
& c_{23}c_{13} \\
\end{array}
\right).
\end{eqnarray}
%%%%%%%%%%%%%%%%%%%%%%%%%%%%%%%%%%%%%%%%%%%%%%%%
The generalized Yukawa Lagrangian of STHDM is
presented as follows:
\begin{eqnarray}
-\mathcal{L}_Y
&=& -\bar{Q}_L\frac{m_u}{v_i}\tilde{\Phi}_iu_R
-\bar{Q'}_L\frac{m_d}{v_j}{\Phi}_jd_R
-\bar{L}_L\frac{m_{l}}{v_k}{\Phi}_kl_R+H.c
\end{eqnarray}
The four Yukawa Lagrangian types as follows
%%%%%%%%%%%%%%%%%%%%%%%%%%%%%%%%%%%%%%%%%%%%%%%%
\begin{itemize}
 \item Type I:
 %%%%%%%%%%%%%%%%%%%%%%%%%%%%%%%%%%%%%%%%%%%%%%%%
\begin{eqnarray}
-\mathcal{L}_Y &=&
-\bar{Q}_L\frac{m_u}{v_2}\tilde{\Phi}_2u_R
-\bar{Q'}_L\frac{m_d}{v_2}{\Phi}_2d_R
-\bar{L}_L\frac{m_{l}}{v_2}{\Phi}_2l_R
+ H.c \\
%%%%%%%%%%%%%%%%%%%%
&\supset&
\frac{im_uc_{\beta}}{v\sqrt{2}s_{\beta}}
A^0\bar{u}^i\gamma_5 u^i
-\frac{im_dc_{\beta}}{v\sqrt{2}s_{\beta}}
A^0\bar{d}^i\gamma_5d^i
-\frac{im_{l}c_{\beta}}{v\sqrt{2}s_{\beta}}
A^0\bar{l}^i\gamma_5l^i
\nonumber\\
%%%%%%%%%%%%%%%%%%%%
&&-\frac{m_u}{v\sqrt{2}s_{\beta}}
\mathcal{O}_{2j}H_j\bar{u}^iu^i
-\frac{m_d}{v\sqrt{2}s_{\beta}}
\mathcal{O}_{2j}H_j\bar{d}^id^i
-\frac{m_{l}}{v\sqrt{2}s_{\beta}}
\mathcal{O}_{2j}H_j\bar{l}^il^i
\nonumber\\
%%%%%%%%%%%%%%%%%%%%%
&&-\frac{m_u}{v\sqrt{2}s_{\beta}}
\mathcal{O}_{22}h\bar{u}^iu^i
-\frac{m_d}{v\sqrt{2}s_{\beta}}
\mathcal{O}_{22}h\bar{d}^id^i
-\frac{m_{l}}{v\sqrt{2}s_{\beta}}
\mathcal{O}_{22}h\bar{l}^il^i.
\end{eqnarray}
\item Type II
\begin{eqnarray}
-\mathcal{L}_Y
&=&-\bar{Q}_L\frac{m_u}{v_2}\tilde{\Phi}_2u_R
-\bar{Q'}_L\frac{m_d}{v_1}{\Phi}_1d_R
-\bar{L}_L\frac{m_{l}}{v_1}{\Phi}_1l_R
+H.c
\\
&\supset&
\frac{im_uc_{\beta}}{v\sqrt{2}s_{\beta}}
A^0\bar{u}^i\gamma_5u^i
+\frac{im_ds_{\beta}}{v\sqrt{2}c_{\beta}}
A^0\bar{d}^i\gamma_5d^i
+\frac{im_{l}s_{\beta}}{v\sqrt{2}c_{\beta}}
A^0\bar{l}^i\gamma_5 l^i
\nonumber\\
&&
-\frac{m_u}{v\sqrt{2}s_{\beta}}
\mathcal{O}_{2j}H_j\bar{u}^iu^i
-\frac{m_d}{v\sqrt{2}c_{\beta}}
\mathcal{O}_{1j} H_j\bar{d}^id^i
-\frac{m_{l}}{v\sqrt{2}c_{\beta}}
\mathcal{O}_{1j} H_j\bar{l}^il^i
\nonumber\\
&&
-\frac{m_u}{v\sqrt{2}s_{\beta}}
\mathcal{O}_{22} h\bar{u}^iu^i
-\frac{m_d}{v\sqrt{2}c_{\beta}}
\mathcal{O}_{12} h\bar{d}^id^i
-\frac{m_{l}}{v\sqrt{2}c_{\beta}}
\mathcal{O}_{12} h\bar{l}^il^i.
\end{eqnarray}
%%%%%%%%%%%%%%%%%%%%%%%%%%%%%%%
\item Type X
%%%%%%%%%%%%%%%%%%%%%%%%%%%%%%%
\begin{eqnarray}
-\mathcal{L}_Y &=&
-\bar{Q}_L\frac{m_u}{v_2}\tilde{\Phi}_2u_R
-\bar{Q'}_L\frac{m_d}{v_2}{\Phi}_2d_R
-\bar{L}_L\frac{m_{l}}{v_1}{\Phi}_1l_R
+H.c \\
&\supset&
\frac{im_uc_{\beta}}{v\sqrt{2}s_{\beta}}
A^0\bar{u}^i\gamma_5u^i
-\frac{im_dc_{\beta}}{v\sqrt{2}s_{\beta}}
A^0\bar{d}^i\gamma_5 d^i
+\frac{im_{l}s_{\beta}}{v\sqrt{2}c_{\beta}}
A^0\bar{l}^i\gamma_5 l^i
\nonumber\\
&&
-\frac{m_u}{v\sqrt{2}s_{\beta}}
\mathcal{O}_{2j}H_j\bar{u}^iu^i
-\frac{m_d}{v\sqrt{2}s_{\beta}}
\mathcal{O}_{2j}H_j\bar{d}^id^i
-\frac{m_{l}}{v\sqrt{2}c_{\beta}}
\mathcal{O}_{1j}H_j\bar{l}^il^i
\nonumber\\
&&
-\frac{m_u}{v\sqrt{2}s_{\beta}}
\mathcal{O}_{22}h\bar{u}^iu^i
-\frac{m_d}{v\sqrt{2}s_{\beta}}
\mathcal{O}_{22}h\bar{d}^id^i
-\frac{m_{l}}{v\sqrt{2}c_{\beta}}
\mathcal{O}_{12}h\bar{l}^il^i.
\end{eqnarray}
\item Type Y:
\begin{eqnarray}
-\mathcal{L}_Y &=&
-\bar{Q}_L\frac{m_u}{v_2}\tilde{\Phi}_2u_R
-\bar{Q'}_L\frac{m_d}{v_1}{\Phi}_1d_R
-\bar{L}_L\frac{m_{l}}{v_2}{\Phi}_2l_R
+H.c \\
&\supset&
\frac{im_uc_{\beta}}{v\sqrt{2}s_{\beta}}
A^0\bar{u}^i\gamma_5 u^i
+\frac{im_ds_{\beta}}
{v\sqrt{2}c_{\beta}}A^0\bar{d}^i\gamma_5d^i
-\frac{im_{l}c_{\beta}}
{v\sqrt{2}s_{\beta}}A^0\bar{l}^i\gamma_5l^i
\nonumber\\
&&
-\frac{m_u}{v\sqrt{2}s_{\beta}}
\mathcal{O}_{2j}H_j\bar{u}^iu^i
-\frac{m_d}{v\sqrt{2}c_{\beta}}
\mathcal{O}_{1j}H_j\bar{d}^id^i
-\frac{m_{l}}{v\sqrt{2}s_{\beta}}
\mathcal{O}_{2j}H_j\bar{l}^il^i
\nonumber\\
&&
-\frac{m_u}{v\sqrt{2}s_{\beta}}
\mathcal{O}_{22}
h\bar{u}^iu^i
-\frac{m_d}{v\sqrt{2}c_{\beta}}
\mathcal{O}_{12}
h\bar{d}^id^i
-\frac{m_{l}}{v\sqrt{2}s_{\beta}}
\mathcal{O}_{22}
h\bar{l}^il^i.
\end{eqnarray}
\end{itemize}
%%%%%%%%%%%%%%%%%%%%%%%%%%%%%%%%%%%%%%%%%%%%%%%%%%
\section*{Appendix B: The effective 
Lagrangian for THM}
%%%%%%%%%%%%%%%%%%%%%%%%%%%%%%%%%%%%%%%%%%%%%%%%%%
We next derive the couplings relating
to the processes under consideration in the THM.
From the kinematic terms, one has
%%%%%%%%%%%%%%%%%%%%%%%%%%%%%%%%%%%%%%%%%%%%%%%%
\begin{eqnarray}
\mathcal{L}_{K} &=&
(D_\mu\Phi)^\dagger(D^\mu\Phi)
+
Tr[(D_\mu\Delta)^\dagger(D^\mu\Delta)]
\\
&\supset&
\frac{e(2s_{\alpha}c_{\beta^0}
- s_{\beta^0}c_{\alpha})}{s_{2W}}
(Z_{\mu}A^0\partial^{\mu}{h^0}
- Z^{\mu}h^0\partial_{\mu}{A^0})
% \nonumber\\
% &&
+\frac{e(2c_{\alpha}c_{\beta^0}
+s_{\beta^0}s_{\alpha})}{s_{2W}}
(Z_{\mu}A^0\partial^{\mu}{H}
-Z^{\mu}H\partial_{\mu}{A^0})
\nonumber\\
&&
+\frac{e^2v}{s_{2W}^2}(c_{\beta^{0}}c_{\alpha}
+2s_{\alpha}s_{\beta^{0}})
h^0Z^{\mu}Z_{\mu}
+\frac{ve^2}{s_{2W}^2}(-c_{\beta^{0}}s_{\alpha}
+2c_{\alpha}s_{\beta^{0}})
HZ^{\mu}Z_{\mu}
\nonumber\\
&&
+\frac{eM_W}{s_W}(c_{\alpha}c_{\beta^\pm}
+\sqrt{2}s_{\alpha}s_{\beta^\pm})
h^0W_{\mu}^{\pm}W^{\mp,\mu}
% \nonumber\\
% &&
+\frac{eM_W}{s_W}(-s_{\alpha}c_{\beta^\pm}
+\sqrt{2}c_{\alpha}s_{\beta^\pm})
HW_{\mu}^{\pm}W^{\mp,\mu}.
\end{eqnarray}

%%%%%%%%%%%%%%%%%%%%%%%%%%%%%%%%%%%%%%%%%%%%%%%%%
The Yukawa Lagrangian for THM can be expanded
as follows:
\begin{eqnarray}
\mathcal{L}_{Y}
&=& \mathcal{L}_{Y}^{SM}
- L_{i}^Ty_{\nu}C(i\sigma^2\Delta)L_i
+ H.c, \\
&\supset&
 i\frac{s_{\beta^0} Y_l}{\sqrt{2}}
(\bar{l}_{L}A^0l_R-\bar{l}_{R}A^0l_L)
+ i\frac{s_{\beta^0} Y_d}{\sqrt{2} }
(\bar{d}_{L}A^0d_R -\bar{d}_{R}A^0d_L)
% \nonumber\\
% &&
-i\frac{s_{\beta^0} Y_u}{\sqrt{2}
}(\bar{u}_{L}A^0u_R-\bar{u}_{R}A^0u_L)
\notag\\
&&-\frac{c_{\alpha
}Y_f}{\sqrt{2}
}(\bar{f}_{L}h^0f_R+\bar{f}_{R}hf_L)
+\frac{s_{\alpha}Y_f}{\sqrt{2}
}(\bar{f}_{L}Hf_R+\bar{f}_{R}Hf_L)
\\
&=&
i\frac{s_{\beta^0} Y_l}{\sqrt{2}}A^0\bar{l}\gamma_5{l}
+ i\frac{s_{\beta^0} Y_d}{\sqrt{2}}A^0\bar{d}\gamma_5{d}
- i\frac{s_{\beta^0} Y_u}{\sqrt{2} }A^0\bar{u}\gamma_5{u}
% \notag\\
% &&
-\frac{c_{\alpha}Y_f}{\sqrt{2} } h^0\bar{f}f
+\frac{s_{\alpha}Y_f}{\sqrt{2} }H\bar{f}f.
\end{eqnarray}
We can collect all couplings
of CP-even and CP-odd
Higgs to fermion pair
from the above Lagrangian.

The total decay width
of CP-even Higgs
$H$ is calculated at LO
as in~\cite{Kanemura:2022ldq}
\begin{eqnarray}
\label{GMAH}
\Gamma_H &\approx& \sum
\limits_{f}
\frac{N_C^f\; M_H}{8\pi}
\left(\kappa_H^{f}
\frac{m_f}{v} \right)^2
\left(
1 - \frac{4m_f^2}{M_H^2}
\right)^{3/2}
\\
&&
+\sum\limits_{V=Z, W}
\frac{M_H^3}{64\pi\; c_V \; M_V^4}
\left(\kappa_H^V
\frac{2M_V^2}{v}
\right)^2
\left(
 1-\frac{4M_V^2}{M_H^2} +
\frac{12M_V^4}{M_H^4}
\right)
\sqrt{1-\frac{4M_V^2}{M_H^2} }
.
\nonumber
\end{eqnarray}
Where $c_V=1(2)$ for $W (Z)$
respectively. The coefficient couplings
$\kappa_H^{f}
\sim c_{\beta -\alpha}$,
$\kappa_H^{V} =c_{\beta-\alpha}$ for THDM
and
$\kappa_H^{f}= c_{\alpha}$,
$\kappa_H^{V}
= c_{\alpha}\; c_{\beta^{\pm}}
+\sqrt{2} s_{\alpha}\; s_{\beta^{\pm}}$
for THM, respectively.
%%%%%%%%%%%%%%%%%%%%%%%%%%%%%%%%
\section*{Appendix C: Mixing
of $A^0$ with $\phi$
and $A^0$ with $V^*$}%%
%%%%%%%%%%%%%%%%%%%%%%%%%%%%%%%%%
In this Appendix, we consider one-loop 
mixing of $A^0$ with $\phi$ and the
mixing of $A^0$ with $V_0^*$. As we
mention in section 3, mixing of
$A_0$ with vector bosons $V_0^*$
will be vanished due to the
Slavnov-Taylor identity~\cite{Aiko:2022gmz}.

\begin{figure}[H]
\centering
\includegraphics[width=10cm, height=7cm]
{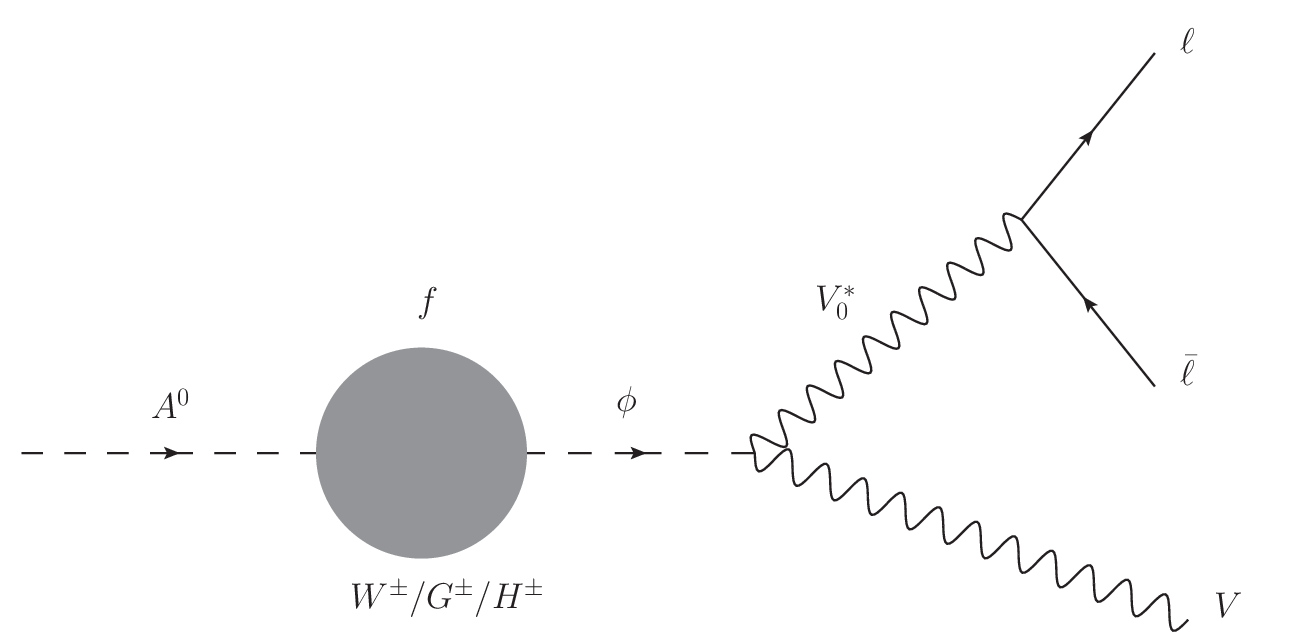}
\caption{\label{G4a} Group $4$ One-loop
self-energy Feynman diagrams mixing of 
$A^0$ with $\phi$ are plotted.
All fermions, $W$ boson,
Goldstone boson and charged Higgs
are exchanged in the loop.}
\end{figure}

All one-loop diagrams of
the mixing of $A^0$ with $\phi$
are shown in Fig.~\ref{G4a} in which
all fermions, $W$ boson,
Goldstone boson and charged Higgs
are exchanged in the loop.
%%%%%%%%%%%%%%%%%%%%%%%%%%%%%%%%%%%%%%%
%%%%%%%%%%%%%%%%%%%%%%%%%%%%%%%%%%%%%%%

The one-loop amplitude for the
mixing of $A_0$ with $\phi$ is
decomposed as follows:
\begin{eqnarray}
\mathcal{A}_{G_4}^V
&=&
\sum \limits_{\phi = h^0, H_j}
u(q_1)
\Big[
\boldsymbol{v}_{V_0^* \ell \bar{\ell}}
-
\boldsymbol{a}_{V_0^* \ell \bar{\ell}}
\,
\gamma_5
\Big]
\bar{v} (q_2)
\frac{
\Big[
F_{G_4,f}^{A^0\phi}
+
F_{G_4,(W,\cdots,
H^{\pm})}^{A^0\phi}
\Big]
\cdot
\slashed \epsilon^*(q_3)
}{
\big(
M_{A^0}^2 
- 
M_{\phi}^2 
\big)
\Big[
\big(
q_{12} - M_{V_0}^2 
\big)
+ i M_{V_0} \Gamma_{V_0}
\Big]
}.
\end{eqnarray}
One-loop form factors 
$F_{G_4,f}^{A^0\phi}$
are expressed as follows:
\begin{eqnarray}
%%%%%%%%%%%%%%%%%%%%%%%%%%%%%%%%%%%
F_{G_4,f}^{A^0\phi}
&=&
-i \frac{
N^C_f \; m_f^2
}{4\pi^2}
g_{\phi V_0^* V}
\Big\{
(
\boldsymbol{v}_{A^0 f \bar{f}}
\,
\boldsymbol{v}_{\phi f \bar{f}}
-
\boldsymbol{a}_{A^0 f \bar{f}}1
\,
\boldsymbol{a}_{\phi f \bar{f}}
)
A_0 (m_f^2)
\nonumber
\\
&&\hspace{0cm}
+
\Big[
\boldsymbol{a}_{A^0 f \bar{f}}
\, 
\boldsymbol{a}_{\phi f \bar{f}}
\,
M_{A^0}^2
-
\boldsymbol{v}_{A^0 f \bar{f}}
\,
\boldsymbol{v}_{\phi f \bar{f}}
\,
(
M_{A^0}^2
-
4 m_f^2
)
\Big]
B_0 (M_{A^0}^2,m_f^2,m_f^2)
\Big\}.
\end{eqnarray}
All related couplings in
THDM is given by
\begin{eqnarray}
\boldsymbol{v}_{h^0 f \bar{f}}
&=&
\frac{e}{2 s_W M_W}
\frac{c_\alpha}{s_\beta},
\quad
\boldsymbol{v}_{H f \bar{f}}=
\frac{e}{2 s_W M_W}
\frac{s_\alpha}{s_\beta},
\quad 
\boldsymbol{a}_{\phi f \bar{f}} = 0.
\end{eqnarray}
We verify easily that the form factor
$F_{G_4,f}^{A^0\phi}$ becomes zero
because we have
$\boldsymbol{v}_{A^0 f \bar{f}} = 0$ 
for CP-odd Higgs boson $A^0$.

We mention the second term of $G_4$
involving  all Feynman self-energy
diagrams contributing to the mixing $A^0$
with $\phi$ with considering all
vector bosons, Goldstone bosons,
charged scalar particles in the loop.
One-loop form factor
$F_{G_4,(W,\cdots H^{\pm})}^{A^0\phi}$
is expressed in terms of scalar
scalar one-loop functions as follows:
\begin{eqnarray}
F_{G_4,
(W,\cdots,H^{\pm})}^{A^0\phi}
&=&
-i \frac{
g_{\phi V_0^* V}
}{16 \pi^2}
\Bigg\{
(
g_{A^0 H^- W^+}
\,
g_{\phi H^+ W^-}
+
g_{A^0 H^+ W^-}
\,
g_{\phi H^- W^+}
)
\times
\\
&&\hspace{0.5cm} \times
\Big[
A_0 (M_{H^\pm}^2)
-
2 A_0 (M_W^2)
+
(
M_W^2
-
2M_{A^0}^2-2 M_{H^\pm}^2
) 
B_0 (M_{A^0}^2,M_{H^\pm}^2,M_W^2)
\Big]
\nonumber
\\
&&\hspace{0.5cm}
-
(
g_{A^0 H^- G^+}
\,
g_{\phi H^+ G^-}
+
g_{A^0 H^+ G^-}
\,
g_{\phi H^- G^+}
)
B_0 (M_{A^0}^2,M_{H^\pm}^2,M_W^2)
\Bigg\}.
\nonumber
\end{eqnarray}
We note that one-loop form factors
$F_{G_4,
(W,\cdots,H^{\pm})}^{A^0\phi}$
tend zero by replacing
the corresponding couplings
in THDM:
\begin{eqnarray}
g_{A^0 H^- W^+}
& \equiv &
g_{A^0 H^+ W^-}
=
\frac{e}{2 s_W}, \\
%%%%%%%%%%%%%%%%%%
g_{h^0 H^+ W^-}
& \equiv &
- g_{h^0 H^- W^+}
=
- \frac{e}{2 s_W}
c_{\beta - \alpha},
\\
%%%%%%%%%%%%%%%%%%
g_{H H^+ W^-}
& \equiv &
- g_{H H^- W^+}
=
\frac{e}{2 s_W}
s_{\beta - \alpha},
\\
g_{A^0 H^- G^+}
& \equiv &
- g_{A^0 H^+ G^-}
=
\frac{e}{2 s_W M_W}
\Big(
M_{A^0}^2
-
M_{H^\pm}^2
\Big), 
\\
%%%%%%%%%%%%%%%%%%
g_{h^0 H^+ G^-}
& \equiv &
g_{h^0 H^- G^+}
=
\frac{e}{2 s_W M_W}
\Big(
M_{h^0}^2 
-
M_{H^\pm}^2
\Big)
c_{\beta - \alpha},
\\
%%%%%%%%%%%%%%%%%%
g_{H H^+ G^-}
& \equiv &
g_{H H^- G^+}
=
- 
\frac{e}{2 s_W M_W}
\Big(
M_{H}^2
-
M_{H^\pm}^2
\Big)
s_{\beta - \alpha}.
\end{eqnarray}
%%%%%%%%%%%%%%%%%%%%%%%%%%%%%
\section*{Appendix D:
Mixing of $V$ with $\phi$}%%%
\begin{figure}[H]
\centering
\begin{tabular}{cc}
\includegraphics[width=8cm, height=6cm]
{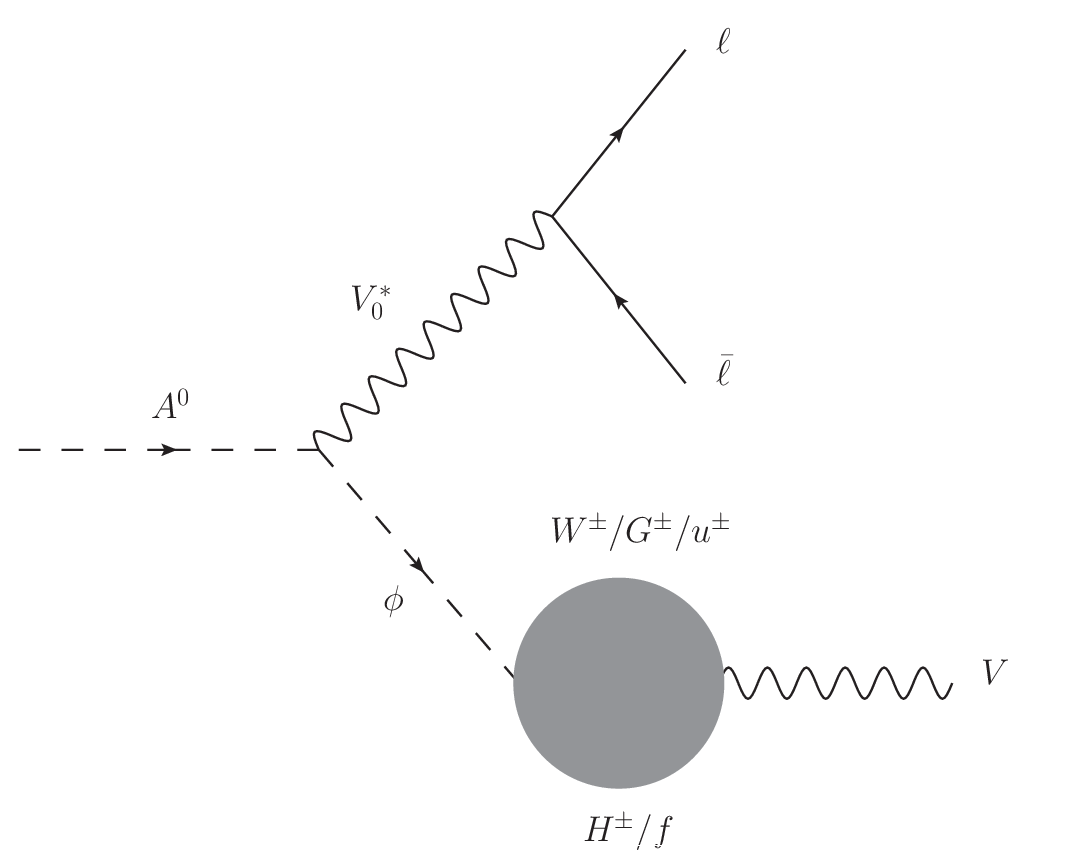}
&
\includegraphics[width=8cm, height=6cm]
{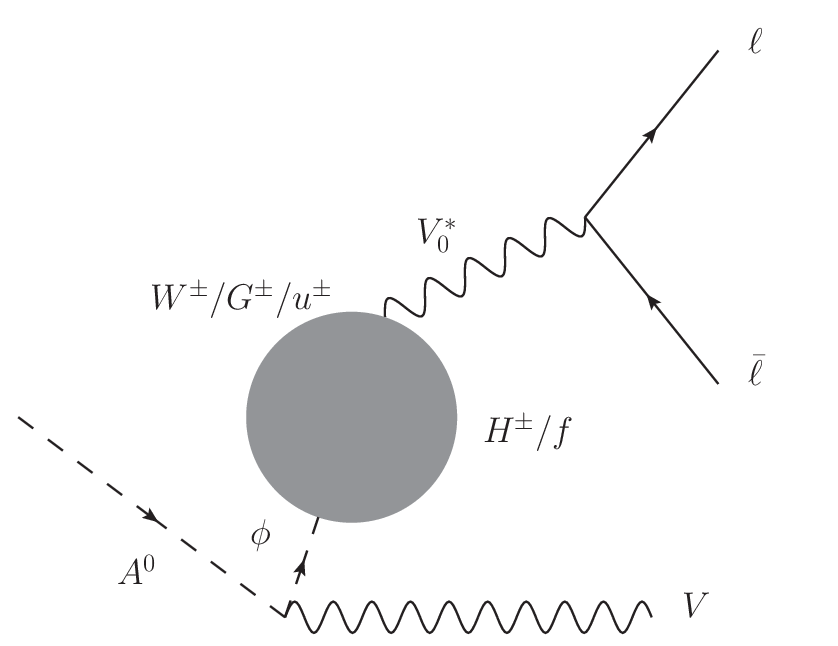}
\end{tabular}
\caption{\label{G6a} Group $5$
One-loop 
Feynman diagrams for 
the mixing of $V$ with $\phi$.}
\end{figure}

%%%%%%%%%%%%%%%%%%%%%%%%%%%%%
We next consider the mixing of $V$
with $\phi$. All self-energy diagrams
contributing to the mixing of $V$
with $\phi$ are shown in
Fig.~\ref{G6a}.
%%%%%%%%%%%%%%%%%%%%%%%%%%%%%%%%
%%%%%%%%%%%%%%%%%%%%%%%%%%
Self-energy Feynman diagrams for
one-loop CP-even Higgs boson
$\phi$ mixing with external
vector boson $V$, the amplitude
is decomposed as follows:
\begin{eqnarray}
\mathcal{A}_{G_5}^V
&=&
\sum \limits_{\phi = h^0, H_i}
\sum \limits_{V_0 = \gamma, Z}
u (q_1)
\Big(
\boldsymbol{a}_{V_0^* \ell \bar{\ell}}
\,
\gamma_5
\Big)
\bar{v} (q_2)
\times
\frac{\Big[
\sum \limits_{f / W^\pm / H^\pm}
F_{G_5,f / W^\pm / H^\pm}^{(V, \phi)}
\Big]
( p \cdot
\epsilon^* (q_3) )
}{
\big(
M_V^2
- 
M_{\phi}^2 
\big)
\Big[
\big(
q_{12} - M_{V_0}^2 
\big)
+ i M_{V_0} \Gamma_{V_0}
\Big]
}.
\end{eqnarray}
One-loop 
form factors are expressed
as follows
\begin{eqnarray}
F_{G_5,f}^{(V, \phi)}
&=&
\frac{-i\; 
N^C_f (m_{\ell}\; m_f^2)
}{\pi^2}
g_{\phi A^0 V_0^*} 
\times \\
&&\times 
\Big[
(
\boldsymbol{a}_{\phi f \bar{f}}
\,
\boldsymbol{a}_{V f \bar{f}}
-
\boldsymbol{v}_{\phi f \bar{f}}
\,
\boldsymbol{v}_{V f \bar{f}}
)
B_0 (M_V^2,m_f^2,m_f^2)
%\nonumber
%\\
%&&
-2 \,
\boldsymbol{v}_{\phi f \bar{f}}
\,
\boldsymbol{v}_{V f \bar{f}}
\,
B_1 (M_V^2,m_f^2,m_f^2)
\Big]
, \nonumber \\
%%%%%%%%%%%%%%%%%%%
&&
\nonumber\\
%%%%%%%%%%%%%%%%%%%%
F_{G_5,H^\pm}^{(V, \phi)}
&=&
\frac{-i\; m_{\ell}}{4\pi^2}
[ g_{\phi A^0 V_0^*}
\cdot
g_{\phi H^\pm H^\mp}
\cdot 
g_{V H^\pm H^\mp}
]
[
B_0 
+
2 B_1 
]
(M_V^2,M_{H^\pm}^2,M_{H^\pm}^2), 
\\
&\nonumber\\
%%%%%%%%%%%%%%%%%%%%%%%%%%%%%%%%%%
F_{G_5,W^\pm}^{(V, \phi)}
&=&
\frac{-i\; m_{\ell} 
}{4\pi^2}
g_{\phi A^0 V_0^*}
\Big\{ 
(
g_{\phi \, u^- \bar{u}^-}
\,
g_{V u^- \bar{u}^-}
+
g_{\phi \, u^+ \bar{u}^+}
\,
g_{V u^+ \bar{u}^+}
)
B_1 (M_V^2,M_W^2,M_W^2)
\\
&&\hspace{0.25cm}
+
[
g_{\phi \, W^- G^+}
\,
g_{V W^+ G^-}
+
g_{\phi \, W^+ G^-}
\,
g_{V W^- G^+}
]
[ B_0 - B_1]
(M_V^2,M_W^2,M_W^2)
\nonumber
\\
&&
\hspace{0.25cm}
+
[ g_{\phi \, G^\pm G^\mp}
\,
g_{V G^\pm G^\mp}
+
3 \, g_{\phi \, W^\pm W^\mp}
\,
g_{V W^\pm W^\mp}
]
[
B_0 
+
2
B_1 
]
(M_V^2,M_W^2,M_W^2)
\Big\}.
\nonumber
\end{eqnarray}

Noting that the form factor
$F_{G_5,f}^{(V, \phi)}$ are
simplified by applying
$\boldsymbol{a}_{\phi f \bar{f}}
= 0$. The result reads
\begin{eqnarray}
F_{G_5,f}^{(V, \phi)}
&=&
\frac{i(N^C_f \; m_{\ell}\; m_f^2)
}{\pi^2}
\cdot
g_{\phi A^0 V}
\,
\cdot
\boldsymbol{v}_{\phi f \bar{f}}
\,
\cdot
\boldsymbol{v}_{V_0^* f \bar{f}}
\cdot
[
B_0 
+
2
B_1 
]
(q_{12},m_f^2,m_f^2). 
\end{eqnarray}
The form factors become zero by
using PV-functions reduction
for $B_1$ as follows
\begin{eqnarray}
B_1 (k^2,m_1^2,m_2^2)
&=&
\frac{m_2^2 - m_1^2 - k^2}{2 k^2}
B_0 (k^2,m_1^2,m_2^2)
+
\frac{m_1^2 - m_2^2}{2 k^2}
B_0 (0,m_1^2,m_2^2). 
\label{B1reduction}
\end{eqnarray}
For the form factor
$F_{G_2,W^\pm}^{(V, \phi)}$,
applying the following relations
for the couplings as
\begin{eqnarray}
g_{\phi \,WW} &=&- s_{\beta - \alpha}
(- c_{\beta - \alpha})
\dfrac{e M_W}{s_W},
\\
g_{\phi \, G^\pm G^\mp} &=&
\dfrac{e}{2 M_W s_W}
M_\phi^2
\,
s_{\beta - \alpha}
(c_{\beta - \alpha})
=
- \dfrac{M_\phi^2}{2 M_W^2}
\,
g_{\phi \,WW},
\\
%%%%%%%%%%%%%%%%%%%%%%%%%%%%%%%%%%%%%
g_{\phi \, W^\pm G^\mp}
&=&
\pm \dfrac{e}{2 s_W}
s_{\beta - \alpha}
(c_{\beta - \alpha})
=
\mp \dfrac{1}{2 M_W}
\,
g_{\phi \, W^\pm W^\mp}, 
\\
%%%%%%%%%%%%%%%%%%%%%%%%%%%%%%%%%%%%
g_{\phi \, u^\pm \bar{u}^\pm}
&=&
\dfrac{e M_W}{2 s_W}
s_{\beta - \alpha}
(c_{\beta - \alpha})
=
\dfrac{1}{2}
\,
g_{\phi \, W^\pm W^\mp}. 
\end{eqnarray}
The factors $F_{G_2,W^\pm}^{(V, \phi)}$
become
\begin{eqnarray}
F_{G_2,W^\pm}^{(V, \phi)}
&=&
\frac{-i\; m_{\ell}}{4\pi^2}
g_{\phi A^0 V}
\; 
g_{\phi \, W^\pm W^\mp}
\Big\{
\frac{1}{2}
[
g_{V_0^* u^- \bar{u}^-}
+
g_{V_0^* u^+ \bar{u}^+}
]
B_1 (q_{12},M_W^2,M_W^2)
\nonumber
\\
&&\hspace{0.25cm}
+
\frac{1}{2 M_W}
(
g_{V_0^* W^+ G^-}
-
g_{V_0^* W^- G^+}
)
[
B_0 
-
B_1 
]
(q_{12},M_W^2,M_W^2)
\Big\},
\nonumber
\end{eqnarray}
These will be vanished once
$V_0^*$ being $Z$-boson,
the remaining couplings read
\begin{eqnarray}
g_{V_0^* u^\pm \bar{u}^\pm}
&=& \pm e\frac{c_W}{s_W},  \\
%%%%%%%%%%%%%%%%%%%%%%%%%%%%%
g_{V_0^* W^\pm G^\mp} &=& 
e M_W \frac{s_W}{c_W}.
\end{eqnarray}
In the case of $V_0^*$ being photon,
the related couplings are
\begin{eqnarray}
g_{V_0^* u^\pm \bar{u}^\pm}
&=& \pm e, \\
%%%%%%%%%%%%%%%%%%%%%%%%%%%%%
g_{V_0^* W^\pm G^\mp} &=& -e M_W.
\end{eqnarray}
We also verify that the factors
$F_{G_2,W^\pm}^{(V, \phi)}=0$ for both
$V_0^*=\gamma^*, Z^*$.

Similarly, the form factor 
$F_{G_2,H^\pm}^{(V, \phi)}$
is also eliminated by the
relation in Eq.~(\ref{B1reduction}).
%%%%%%%%%%%%%%%%%%%%%%%%%%%%%%%%%%%%%%%%%%

%%%%%%%%%%%%%%%%%%%%%%%%%%%%%%%%%%%%%%%%%%
\end{document}